\documentclass[a4paper,11pt]{article}
\usepackage[utf8]{inputenc}
\usepackage[english]{babel}
\usepackage{braket}
\usepackage{xspace}
\usepackage{bbm}
\usepackage{mathdots}
\usepackage{stackrel}
\usepackage{xcolor}
\usepackage{mathtools}
\usepackage{bbm}
\usepackage{subfigure}
\usepackage[T1]{fontenc}               %
\usepackage[left=3cm,
            right=2.5cm,
            top=2.5cm,
            bottom=3cm
            ]{geometry}                %
\usepackage{graphicx}
\usepackage{mathrsfs}
\usepackage{appendix}
\usepackage{fancyhdr}

\usepackage{amsmath,                   %
            amssymb,                   %
            amsthm}                    %
\usepackage{cite}
\usepackage{setspace} 
\usepackage[colorlinks=true,linkcolor=blue, citecolor=blue, bookmarks]{hyperref}
\usepackage[protrusion=true,expansion=true]{microtype}

\usepackage{authblk}

\usepackage{tikz}

\usetikzlibrary{decorations.pathmorphing}
\usetikzlibrary{shapes.geometric}

\theoremstyle{definition}

\theoremstyle{remark}

\def\RR{\mathbb{R}}
\def\ZZ{\mathbb{Z}}

\def\ii{\mathrm{i}}
\def\dd{\mathrm{d}}

\DeclareMathOperator{\Tr}{Tr}

\date{}\title{\bf Finite temperature negativity Hamiltonians\\ of the massless Dirac fermion}

\author[1]{Federico Rottoli}
\author[2,3]{Sara Murciano}
\author[1,4]{Pasquale Calabrese}

\affil[1]{SISSA and INFN Sezione di Trieste, 
    via Bonomea 265, 34136 Trieste, Italy.}
\affil[2]{Walter Burke Institute for Theoretical Physics, Caltech, 
    1200 E. California Bl., Pasadena, CA 91125, USA.}
\affil[3]{Department of Physics and IQIM, Caltech,
    1200 E. California Bl., Pasadena, CA 91125, USA.}
\affil[4]{International Centre for Theoretical Physics (ICTP),
    Strada Costiera 11, 34151 Trieste, Italy.}

\begin{document}
\maketitle

\begin{abstract}
The negativity Hamiltonian, defined as the logarithm of a 
partially transposed density matrix, provides  an operatorial characterisation of mixed-state entanglement. 
However, so far, it has only been studied for the mixed-state density matrices corresponding to subsystems of globally pure states. 
Here, we consider as a genuine example of a mixed state the one-dimensional massless Dirac fermions in a system at finite temperature and size. 
As subsystems, we consider an arbitrary set of disjoint intervals. 
The structure of the corresponding negativity Hamiltonian resembles the one for the entanglement Hamiltonian in the same geometry: in addition to a local term proportional to the stress-energy tensor, each
point is non-locally coupled to an infinite but discrete set of other points. However, when the lengths of the transposed and non-transposed intervals coincide, the structure remarkably simplifies and we retrieve the mild non-locality of the ground state negativity Hamiltonian. 
We also conjecture an exact expression for the negativity Hamiltonian associated to the twisted partial transpose, which is a Hermitian fermionic matrix.
We finally obtain the continuum limit of both the local and bi-local operators from exact numerical computations in free-fermionic chains.
\end{abstract}

\clearpage

\tableofcontents

\section{Introduction}
In the last few decades, the study of entanglement turned out to be an optimal tool to investigate quantum field theories, quantum
gravity models, condensed matter systems and quantum information theory \cite{nc-10,nrt-09,intro1,intro2,eisert-2010,intro3}. Several entanglement measures have been studied in order to better probe the different features of a system. For example, in the context of pure states, the most important entanglement measures are the von Neumann and the R\'enyi entropies. They are defined as follows.
Let us assume that our pure system is bipartite into $A \cup B$ and that the corresponding Hilbert space, $\mathcal{H}$, factorises as $\mathcal{H}_A \otimes \mathcal{H}_B$, where $\mathcal{H}_A$ ($\mathcal{H}_B$) is the Hilbert space containing the degrees of freedom of the subsystem $A$ ($B$).
The reduced density matrix of $A$ is then obtained by tracing over the degrees of freedom of $B$
\begin{equation}\label{eq:defReducedDensityMatr}
    \rho_A \equiv \Tr_B \rho\,,
\end{equation}
and the entropies of $\rho_A$, given by \cite{cc-04}
\begin{equation}\label{eq:defEntEntropies}
    S^{(n)} \equiv \frac{1}{1-n} \log \Tr \rho_A^n\,,\qquad S \equiv - \Tr \left [ \rho_A \log \rho_A \right ]= \lim_{n \to 1} S^{(n)}\,,
\end{equation}
are good entanglement monotones on pure states.
Despite the several successful applications of the entanglement entropy, they do not capture all of the properties of entanglement between the two subsystems. A more comprehensive measure of entanglement \cite{lh-08}  is provided by the entanglement Hamiltonian $K_A$, defined as the logarithm of the (appropriately normalised) reduced density matrix, i.e.
\begin{equation}
    \rho_A = \frac{1}{Z_A} e^{-2\pi K_A} , \qquad Z_A=\Tr e^{-2\pi K_A}\,.
\end{equation}
Being it an operator and not a scalar quantity, it is much more difficult to compute the entanglement Hamiltonian than the entropies, and it is known only for a limited number of cases.
One of them is provided by the Bisognano-Wichmann theorem. This extremely general result applies to all Lorentz invariant quantum field theories in any dimension $D+1$ and it states that, in the ground state, the entanglement Hamiltonian of the half-space $A = \left \{x \in \RR^{D+1} \big{|} x^1 > 0,  x^0 = t = 0 \right \}$ is the generator of the Lorentz boosts preserving the Rindler wedge\cite{bw-75, bw-76, haagbook, w-18}, i.e.
\begin{equation}\label{eq:BW}
    K_A = \int_{x^1>0} \dd^D x\, x^1\, T_{00}(x)\,, 
\end{equation}
where $T_{00}$ is the energy density.
This theorem provides a physical explanation of the Unruh effect \cite{Unruh1976} in terms of the entanglement of the vacuum.\\ Although computing $K_A$ is a challenging task, we mention here that several achievements have been gained, for example in Gaussian states arising in lattice models of statistical mechanics and in quantum field theories (see e.g. \cite{it-87,pkl-99,Gaussian, Chung2001,ct,ch-09}).
\\
If we shift our attention to mixed states, the entanglement entropies and Hamiltonian are not good entanglement measures, since they cannot distinguish between quantum and classical correlations. 
As a consequence, a plethora of alternative entanglement quantifiers have been proposed for dealing with the mixed states, even if in most cases these quantities are very difficult to compute even for few qubits (see e.g. \cite{NPcomplete,asm-22}).
A measure of entanglement in mixed states that has attracted a lot of interest is the \emph{logarithmic negativity} \cite{vidal,plenio-2005,eisert-thesis}, which is defined by doing a partial transpose operation on the reduced density matrix.
To fix the ideas, let us consider a further bipartition of our subsystem $A$ as $A=A_1 \cup A_2$ and let the Hilbert space be factorised as $\mathcal{H}_{A} = \mathcal{H}_{A_1} \otimes \mathcal{H}_{A_2}$.
Then, if $\ket{e^1_i}$, $\ket{e^2_j}$ are two arbitrary basis in, respectively, $\mathcal{H}_{A_1}$ and $\mathcal{H}_{A_2}$, the partial transposition in, e.g., $\mathcal{H}_{A_1}$ acts as 
\begin{equation}\label{eq:bosonicPT}
    \rho^{T_1}_A\, \equiv\sum_{i,j,k,l}\braket{e^1_k,e^2_j|\rho_A|e^1_i, e^2_l}\ket{e_i^1,e_j^2}\bra{e^1_k,e^2_l}.
\end{equation}
According to the Peres-Horodecki criterion \cite{p-96, s-00}, the presence of negative eigenvalues in the spectrum of $\rho^{T_1}_A$ is a sufficient, but not necessary, condition for the presence of entanglement between $A_1$ and $A_2$.
In light of this criterion, a quantifier of entanglement in mixed states is provided by the logarithmic negativity defined as \cite{vidal,plenio-2005,eisert-thesis}
\begin{equation}\label{eq:defLogNeg}
    \mathcal{E} \equiv \log \Tr \left |\rho_A^{T_1} \right |.
\end{equation}
Since the Peres-Horodecki criterion is only sufficient but not necessary, zero negativity does not imply the absence of entanglement.
Nevertheless, this quantifier presents several advantages with respect to other entanglement monotones.
On the one hand, from the quantum information point of view, as we have mentioned most measures of mixed state entanglement are very difficult to compute \cite{NPcomplete}, while the negativity is computationally more efficient.
On the other hand, from the field-theoretical perspective, the path integral construction of the partial transpose \cite{cct-neg-1,cct-neg-2,ctt-13,cct-neg-3} makes it possible to calculate the negativity in field theories using the same conformal field theory (CFT) techniques adopted in the computation of the entanglement entropies.\\
In the framework of the mixed state entanglement, in \cite{mvdc-22} the negativity Hamiltonian has been introduced as the logarithm of the partially transposed reduced density matrix
\begin{equation}\label{eq:defNegHam}
    \rho_A^{T_1} \equiv \frac{1}{Z_A}  e^{-2\pi \mathcal{N}_A}, \qquad Z_A=\Tr e^{-2\pi \mathcal{N}_A}\,.
\end{equation}
In particular, in \cite{mvdc-22, rmtc-23} the negativity Hamiltonian of the ground state of the massless Dirac fermion has been explicitly computed for tripartite configurations, both in the presence and in the absence of boundaries, but a truly global mixed state has never been studied. 
In this work, we fill this gap computing the negativity Hamiltonian of several disjoint intervals at finite temperature, both on a finite size system and on the infinite line. This computation requires the knowledge of the entanglement Hamiltonian in the same geometry. 
Thus, Sec.~\ref{sec:EH} of the manuscript is devoted to a summary of the known results for the entanglement Hamiltonian of the massless Dirac field theory at finite temperature.
This will set the stage for Sec.~\ref{sec:NH}, where we will present the main results of this manuscript: we evaluate the negativity Hamiltonian of the Dirac fermion on the torus, i.e. at finite temperature on the circle, showing explicit results for a tripartite and a bipartite geometry.
Our analytical findings will be checked in Sec.~\ref{sec:lattice}, where the continuum limit of the entanglement and negativity Hamiltonians is explicitly worked out starting from exact lattice numerical results.
This allows us to check our predictions for the underlying Dirac field theory. In the same section we also propose an alternative definition for the negativity Hamiltonian corresponding to an Hermitian partial transpose reduced density matrix. After the discussion in section \ref{concl}, we conclude the manuscript with two appendices containing computational details and additional results.

\section{Finite temperature entanglement Hamiltonian}\label{sec:EH}

In order to derive the expression of the negativity Hamiltonian, we use the construction introduced in \cite{mvdc-22, rmtc-23}, where it was argued that the effect of the partial transposition amounts to exchange the extrema of the transposed interval in the expression of the entanglement Hamiltonian, taking into account that the fermionic field picks up an imaginary phase if it belongs to the transposed interval. Therefore, in this section, we present the known results for the finite temperature entanglement Hamiltonian of the free massless Dirac fermion in a multi-component region $A$, underlying the major differences with respect to the ground state. 

\subsection{Entanglement Hamiltonian on the torus}\label{sec:torusEH}
Let us consider a free massless Dirac fermion on a circle of circumference $L$ at finite temperature $1/\beta$, i.e., on a torus. In the imaginary time direction we impose anti-periodic (also called Neveu-Schwarz) boundary conditions, while in the spatial direction we choose either anti-periodic or periodic (Ramond) ones. Then, in a subsystem $A = [a_1, b_1] \cup \ldots \cup [a_n, b_n]$ composed of $n$ intervals, the entanglement Hamiltonian is \cite{fr-19, bpn-19, bgpn-19} 
\begin{equation}\label{eq:EHtorus}\begin{split}
    &K_A(\beta,L) = K_A^\text{loc}(\beta,L) + K_A^\text{nl}(\beta,L) \\
    =&\int_A \dd x\, \beta_\text{loc}(x;\beta,L)\, T_{00}(x) + \!\!\!\sum_{(p, k) \neq (0,0)}\!\!\! (\pm 1)^{k} \int_A \dd x\, \frac{\beta_\text{loc}(\tilde{x}_{kp}; \beta, L)}{\frac{\beta}{\pi} \sinh\!\left [ \frac{\pi}{\beta} \left ( x - \tilde{x}_{kp} + k L\right )\right ]}\, T^\text{bl}\!\left (x, \tilde{x}_{kp}, t= 0 \right ) ,
\end{split}\end{equation}
where the signs $+$ and $-$ correspond, respectively, to the Ramond and Neveu-Schwarz sectors, $p \in \{0,\ldots, n-1\}, k\in \mathbb{Z}$. The Hamiltonian \eqref{eq:EHtorus} presents a local part, $K_A^\text{loc}(\beta,L)$, proportional to the energy density $T_{00}$, defined as ($::$ denotes normal ordering of the fields)
\begin{equation}\label{eq:energydensity}\begin{split}
    T_{00}(x,t) = \frac{\ii}{2} & : \!\! \Big[ \left ( \partial_x \psi_R^\dagger(x-t) \psi_R(x-t) - \psi_R^\dagger(x-t) \partial_x \psi_R(x-t) \right ) \\
    &  \quad - \left ( \partial_x \psi_L^\dagger(x+t) \psi_L(x+t) - \psi_L^\dagger(x+t) \partial_x \psi_L(x+t)\right )  \Big]\!\!:\, ,
\end{split}\end{equation}
with a weight given by the local entanglement temperature $\beta_\text{loc}(x) = 1 / z'(x)$, where \cite{fr-19, bpn-19, bgpn-19}
\begin{equation}\label{eq:TorusZeta}\begin{split}
    z(x; \beta, L) &=\log\!\left [ - \prod_{i = 1}^{n} \frac{\vartheta_1\!\left ( \frac{\pi}{L} \left ( x - a_i \right ) \big | q\right )}{\vartheta_1\!\left ( \frac{\pi}{L} \left ( x - b_i \right ) \big | q\right )} \right ] + \frac{2 \pi \ell}{\beta L} x \\
    &=  \log\!\left [ - \prod_{i = 1}^{n} \frac{\sigma\!\left ( x - a_i\right )}{\sigma\!\left ( x - b_i\right )} \right ] - \frac{2 \ell}{\ii \beta} \zeta\!\left ( \ii \beta/2\right ) x +\mathrm{const.}
\end{split}\end{equation}
Here, $\ell = \sum_{i} b_i - a_i$ is the total lenght of the subsystem $A$ and the additive constant term is only a shift which does not depend on $x$ and, therefore, does not affect the expression for $\beta_{\text{loc}}(x)$, that we report explicitly
\begin{equation}\label{eq:TorusZetabeta}\begin{split}
   \beta_{\text{loc}}(x) &=  \left [ \frac{\pi}{L}\sum_{i=1}^n\left [ \frac{\vartheta'_1\!\left ( \frac{\pi}{L} \left (  x-a_i \right ) \big | q\right )}{\vartheta_1\!\left ( \frac{\pi}{L} \left ( x- a_i\right ) \big | q\right )} - \frac{\vartheta'_1\!\left ( \frac{\pi}{L} \left (  x-b_i \right ) \big | q\right )}{\vartheta_1\!\left ( \frac{\pi}{L} \left (  x-b_i \right ) \big | q\right )} \right ] + \frac{2 \pi \ell}{\beta L} \right ]^{-1} \\
    &=  \left [ \sum_{i = 1}^{n} (\zeta\!\left ( x-a_i\right )-\zeta\!\left ( x-b_i \right )) - \frac{2 \ell}{\ii \beta} \zeta\!\left ( \ii \beta/2\right ) \right]^{-1}.
\end{split}\end{equation}
In Eqs.~\eqref{eq:TorusZeta} and \eqref{eq:TorusZetabeta}, $\sigma$ and $\zeta$ denote respectively Weierstrass' sigma and zeta functions and $\vartheta_1$ is the Jacobi's elliptic theta function with nome $q = e^{\ii\pi \tau}, \tau = \ii \beta / L$ (see Appendix \ref{app:tools} for their definitions).  In particular, the expression in the first row of \eqref{eq:TorusZeta} is the result obtained in \cite{fr-19} while the one in the second row follows the conventions of Refs.~\cite{bpn-19, bgpn-19}. While it is not obvious that the two alternative expressions coincide, one can show they are identical by using the properties of Weierstrass functions reported in Appendix \ref{app:tools}. In the rest of this manuscript we will adopt the conventions of \cite{fr-19} in terms of elliptic theta functions.\\
Regarding the non-local part $K_A^\text{nl}(\beta,L)$ of Eq.~\eqref{eq:EHtorus}, even in the case of one interval, this contains infinite terms proportional to the bi-local operator \cite{ch-09,mt-21,mt2-21} 
\begin{equation}\label{eq:biloc}
\begin{split}
    T^\text{bl}(x, y, t) = \frac{\ii}{2} &:\!\!\left [ \left ( \psi^\dagger_R(x-t) \psi_R(y-t) - \psi_R^\dagger(y-t) \psi_R(x-t) \right ) \right . \\
    &\left . \quad - \left ( \psi^\dagger_L(x+t) \psi_L(y+t) - \psi_L^\dagger(y+t) \psi_L(x+t) \right ) \right ]\!\!:\, .
\end{split}\end{equation}
In particular, the bi-local operator in \eqref{eq:EHtorus} couples one point $x$ with a single other \emph{conjugate point} $\tilde{x}_{kp}$, given by the non-trivial solutions of the equations \cite{fr-19, bpn-19, bgpn-19}
\begin{equation}\label{eq:conjTorus}
    z(x;\beta,L) - z(\tilde{x}_{kp};\beta,L) +  \frac{2 \pi k \ell}{\beta} = 0  \, , \quad k \in \ZZ\, ,
\end{equation}
indexed by the integer $k$. One can see that for every fixed index $k$, Eq.~\eqref{eq:conjTorus} admits $n$ solutions, indexed by $p = 0, \ldots, n-1$, each belonging to a different interval. In the following, we will use the index $p=0$ to denote the solution of Eq.~\eqref{eq:conjTorus} such that $\tilde{x}_{k0}$ belongs to the same interval as $x$. With this convention, we see that for $k=0$ Eq.~\eqref{eq:conjTorus} presents the trivial solution $y = \tilde{x}_{00} = x$, which does not contribute to the non-local part $K_A^\text{nl}(\beta,L)$ (see Eq.~\eqref{eq:EHtorus}).

It is instructive to compare the entanglement Hamiltonian on the torus \eqref{eq:EHtorus} with the one on the plane, i.e, of $n$ intervals on the infinite line at zero temperature, given by \cite{ch-09, achp-18} 
\begin{equation}\label{eq:CHEH}
    K_A  = K_A^\text{loc} + K_A^\text{bl} =\int \dd x\, \beta_\text{loc}(x)\, T_{00}(x) + \sum_{p = 1}^{n-1} \int \dd x\, \frac{\beta_\text{loc}(\tilde{x}_p)}{x - \tilde{x}_p}\, T^\text{bl}(x, \tilde{x}_p, t = 0)\,.
\end{equation}
The local part of Eq.~\eqref{eq:CHEH} is in form analogous to the one of Eq.~\eqref{eq:EHtorus}, with entanglement temperature $\beta_\text{loc}(x) = 1/z'(x)$ equal to the inverse of the derivative of the function
\begin{equation}\label{eq:CHZeta}
    z(x) = \log\! \left [ - \prod_{i = 1}^n \frac{x- a_i}{x-b_i} \right ].
\end{equation}
The main qualitative difference of the Hamiltonian \eqref{eq:EHtorus} on the toric space-time with respect to Eq.~\eqref{eq:CHEH} is the structure of the non-local part $K_A^\text{nl}$. While in Eq.~\eqref{eq:EHtorus} the non-local part contains infinite terms, indexed by the integer $k$ in Eq.~\eqref{eq:conjTorus}, on the plane the bi-local part $K_A^\text{bl}$ only contains $n-1$ terms, calculated in the non-trivial solutions of the equation $z(x) = z(\tilde{x}_p)$. In particular, for a single interval $A = [0, \ell]$ the entanglement Hamiltonian \eqref{eq:CHEH} becomes completely local \cite{HislopLongo82, ch-09, Casini2011, KlichWong2013, ct}
\begin{equation}\label{eq:singleintervalEH}
    K_A=\int_0^\ell dx\, \frac{x \left (\ell - x \right )}{\ell}\, T_{00}(x)\, ,
\end{equation}
as expected since it is conformally equivalent to the Bisognano-Wichmann result \eqref{eq:BW}. This shows that in general the entanglement Hamiltonian on the torus \eqref{eq:EHtorus} is much more non-local than the analogous configuration on the plane \cite{fr-19, bpn-19, bgpn-19}.

\subsection{Finite temperature entanglement Hamiltonian on the infinite line}\label{sec:finTempEH}
We will now review the known results for the finite temperature entanglement Hamiltonian on the infinite line, i.e., on an infinite cylinder of circumference $\beta$ in the time direction. In \cite{fr-19, bpn-19, bgpn-19}, this Hamiltonian was obtained from the result on the torus \eqref{eq:EHtorus} by taking the limit $L \to \infty$.\\
Using the asymptotic expansion of the elliptic theta function $\vartheta_1$ for $q= e^{\ii \pi \tau}$, $\tau = \ii \beta/L \to 0$ (see Eq.~\eqref{eq:thetaLimit} of the Appendix) in the expression \eqref{eq:TorusZeta} for the function $z(x; \beta, L)$, we obtain
\begin{equation}\label{eq:limitEH}\begin{split}
    z(x; \beta, L) &\longrightarrow \log\!\left [ - \prod_{i=1}^n \frac{\sinh\!\frac{\pi (x-a_i)}{\beta}\, e^{\frac{\pi}{\beta L} (2 a_i x - a_i^2)}}{\sinh\!\frac{\pi (x-b_i)}{\beta}\, e^{\frac{\pi}{\beta L} (2 b_i x - b_i^2)}}\right ] + \frac{2 \pi \ell}{\beta L} x\\
    &= z(x;\beta) - \frac{\pi}{\beta L} \sum_{i = 1}^{n} \left ( 2 b_i x - 2 a_i x - b_i^2 + a_i^2 \right ) + \frac{2 \pi \ell}{\beta L} x = z(x; \beta) + \text{const}\,,
\end{split}\end{equation}
where, using $\ell = \sum_i \left ( b_i - a_i\right )$, the contributions proportional to $x$ cancel and we have introduced \cite{fr-19, bpn-19, bgpn-19}
\begin{equation}\label{eq:finTempZeta}
    z(x; \beta) = \log\! \left [ - \prod_{i=1}^n \frac{\sinh\!\frac{\pi (x - a_i)}{\beta}}{\sinh\!\frac{\pi (x - b_i)}{\beta}} \right ].
\end{equation}
The local term of Eq.~\eqref{eq:EHtorus} becomes proportional to the entanglement temperature \cite{fr-19, bpn-19, bgpn-19}
\begin{equation}\label{eq:finTempBeta}
    \beta_\text{loc}(x;\beta) = \frac{1}{z'(x;\beta)} = \frac{\beta}{\pi} \left [ \sum_{i = 1}^{n} \left ( \coth\!\frac{\pi (x - a_i)}{\beta} - \coth\!\frac{\pi (x - b_i)}{\beta} \right ) \right ]^{-1}.
\end{equation}
In the non-local component $K_A^\text{nl}(\beta,L)$ of Eq.~\eqref{eq:EHtorus}, instead, we can see that in this limit the denominator $\sinh(\pi (x - \tilde{x}_{kp} +k L) / \beta )$ diverges for all $k \neq 0$ \cite{fr-19, bpn-19, bgpn-19}. For this reason, the only conjugate points that contribute in this limit are the $n-1$ non-trivial solutions of the equation \cite{fr-19, bpn-19, bgpn-19}
\begin{equation}
    z(x;\beta) = z(\tilde{x}_p; \beta) \, ,
\end{equation}
obtained as the limit of Eq.~\eqref{eq:conjTorus} with $k = 0$. This was expected by the fact that the cylinder can be conformally mapped into the plane, where the entanglement Hamiltonian is written in Eq.~\eqref{eq:CHEH}, which only contains $n-1$ bi-local terms.

Putting all together, we find that the finite temperature entanglement Hamiltonian for a multi-component subsystem $A = [a_1, b_1] \cup \ldots \cup [a_n, b_n]$ on the infinite line is \cite{fr-19, bpn-19, bgpn-19}
\begin{equation}\label{eq:finTempEH}\begin{split}
    &K_A(\beta) = K_A^\text{loc}(\beta) + K_A^\text{bl}(\beta)\\
    & = \int_A \dd x\, \beta_\text{loc}(x;\beta)\, T_{00}(x) + \sum_{p = 1}^{n-1} \int_A \dd x\, \frac{\beta_\text{loc}(\tilde{x}_{p}; \beta)}{\frac{\beta}{\pi} \sinh\! \frac{\pi \left ( x - \tilde{x}_{p}\right )}{\beta} }\, T^\text{bl}\!\left (x, \tilde{x}_{p}, t= 0 \right ),
\end{split}\end{equation}
with entanglement temperature $\beta_\text{loc}(\tilde{x}_{p}; \beta)$ given by Eq.~\eqref{eq:finTempBeta}. When specialising to a subsystem $A$ made up of one interval, the entanglement Hamiltonian in Eq.~\eqref{eq:finTempEH} is purely local and in agreement with the result of \cite{KlichWong2013, ct}, which reads
\begin{equation}\label{eq:singleintervalThermalEH}
    K_A(\beta) = \int_a^b \dd x\, \frac{\beta}{\pi} \left [ \coth\!\frac{\pi (x-a)}{\beta} + \coth\!\frac{\pi (x-b)}{\beta}  \right ]^{-1} T_{00}(x).
\end{equation}
As we mentioned earlier, since the cylinder is conformally equivalent to the plane, an alternative derivation of the finite temperature entanglement Hamiltonian on the infinite line in Eqs.~\eqref{eq:finTempZeta}, \eqref{eq:finTempEH} consists in mapping the expressions \eqref{eq:CHEH}, \eqref{eq:CHZeta} on the plane to the cylinder. We find it worthwhile to also present this additional derivation as a non-trivial check of the correctness of Eq.~\eqref{eq:finTempEH} and because we will adapt a similar trick later in the manuscript. We first present how to map the entanglement Hamiltonian from the plane to a generic geometry and we later specialise this procedure to the cylinder.
Let us consider a multi-component subsystem $A = [a_1, b_1] \cup \ldots \cup [a_n, b_n]$ made up of $n$ intervals in a geometry conformally isomorphic to the plane. In order to map it to the plane, it is convenient to switch to imaginary time $w = x + \ii t$ and consider, for simplicity, only the holomorphic component. Let then $\xi(w)$ be the transformation from this geometry to the plane, with the subsystem $A$ being mapped on the real line. On the complex plane, the holomorphic part of the entanglement Hamiltonian is given by the analytic continuation of Eq.~\eqref{eq:CHEH}
\begin{equation}\label{eq:holomorphicCHEH}
    K_A = \int_{\xi(A)} \dd \xi \, \frac{T(\xi(w))}{\partial_\xi z(\xi(w))} + \sum_{i = 1}^{n} \int_{\xi(A)} \dd \xi \, \frac{1}{\xi(w) - \xi(\tilde{w}_p)}\frac{T^\text{bl}(\xi(w), \xi(\tilde{w}_p))}{ \partial_\xi z(\xi(\tilde{w}_p))}\,,
\end{equation}
where the function $z(w) = z(\xi(w))$ is Eq.~\eqref{eq:CHZeta} evaluated in $\xi(w)$, i.e. 
\begin{equation}\label{eq:genZeta}
    z(w) = z(\xi(w)) = \log \left [- \prod_{i = 1}^{n} \frac{\xi(w) - \xi(a_i)}{\xi(w) - \xi(b_i)} \right ],
\end{equation}
with $\xi(a_i), \xi(b_i)$ the extrema of the mapping $\xi(A)$ of the subsystem $A$ on the plane, and the conjugate points $\tilde{w}_p$ are the solutions of $z(w) = z(\tilde{w}_p)$.\\
We first consider the mapping of the local part.
Despite the fact that the holomorphic stress-energy tensor $T$ is not a primary field, its transformation law only involves an additional function of $w$ proportional to the Schwarzian derivative of $\xi(w)$.
When integrated in the entanglement Hamiltonian, this simply gives a constant factor which can be reabsorbed in the overall normalisation and can therefore be neglected.
Considering also the Jacobian, the holomorphic part transforms as
\begin{equation}\begin{split}
    \int_{\xi(A)} \dd \xi\, \frac{T(\xi(w))}{\partial_\xi z(\xi(w))} &= \int_A \xi'(w)\, \dd w\, \frac{\xi'(w)^{-2}\, T(w)}{\partial_\xi z(\xi(w))} + \text{const} \\
    &= \int_A \dd w\, \frac{T(w)}{z'(w)} + \text{const} \equiv \int_A \dd w\, \beta_\text{loc}(w)\, T(w) + \text{const} \,,
\end{split}\end{equation}
where we see that in the original geometry the entanglement temperature $\beta_\text{loc}(w)$ is given by the inverse of the derivative of Eq.~\eqref{eq:genZeta} with respect to $w$.\\
In order to find the transformation of the bi-local part, it is necessary to understand how the bi-local operator in Eq.~\eqref{eq:biloc} transforms under conformal mappings. In complex coordinates, the holomorphic bi-local operator takes the form
\begin{equation}\label{eq:bilocalholomorphic}\begin{split}
    T^\text{bl}(\xi, \zeta) = \frac{\ii}{2} :\!\!  \left [ \psi^\dagger (\xi) \psi(\zeta) - \psi^\dagger (\zeta) \psi(\xi) \right ] \!\! : .
\end{split}\end{equation}
Since the fermions $\psi, \psi^\dagger$ are primary fields of conformal dimension $(\frac{1}{2},0)$, under the conformal mapping $\xi(z)$ they transform as $\psi(z) = \left ( \frac{\partial \xi}{\partial z}\right )^{1/2} \psi(\xi(z))$ (and analogously for the anti-holomorphic part). Replacing this transformation in Eq.~\eqref{eq:bilocalholomorphic} of the bi-local field, we find that in the original geometry it becomes
\begin{equation}\label{eq:biloctrans}\begin{split}
    T^\text{bl}(z, w) = \xi'(z)^{1/2}\, \xi'(w)^{1/2}\, T^\text{bl}(\xi(z), \xi(w))\, .
\end{split}\end{equation}
Using the transformation of the bi-local operator in the entanglement Hamiltonian in Eq.~\eqref{eq:holomorphicCHEH}, we obtain for the holomorphic part 
\begin{equation}\begin{split}\label{eq:holomorphicConformalMapEH}
    &\int_{\xi(A)} \dd \xi \, \frac{1}{\xi(w) - \xi(\tilde{w}_p)}\frac{T^\text{bl}(\xi(w), \xi(\tilde{w}_p))}{ \partial_\xi z(\xi(\tilde{w}_p))} \\
    =& \int_A \xi'(w) \dd w\, \frac{1}{\xi(w) - \xi(\tilde{w}_p)}\frac{\xi'(w)^{-1/2}\, \xi'(\tilde{w}_p)^{-1/2}\,T^\text{bl}(w, \tilde{w}_p)}{\xi'(\tilde{w}_p)^{-1}\, z'(\tilde{w}_p)} \\
    =& \int_A \dd w\, \frac{\xi'(w)^{1/2}\, \xi'(\tilde{w}_p)^{1/2}}{\xi(w) - \xi(\tilde{w}_p)} \frac{T^\text{bl}(w, \tilde{w}_p)}{z'(\tilde{w}_p)} = \int_A \dd w\, \frac{\xi'(w)^{1/2}\, \xi'(\tilde{w}_p)^{1/2}}{\xi(w) - \xi(\tilde{w}_p)}\, \beta_\text{loc}(\tilde{w}_p)\,T^\text{bl}(w, \tilde{w}_p)\, .
\end{split}\end{equation}
Putting together both the local and the bi-local components, we find that the holomorphic entanglement Hamiltonian in the original geometry takes the form
\begin{equation}\label{eq:holomorphicGenericEH}
    K_A = \int_A \dd w\, \beta_\text{loc}(w)\, T(w) + \int_A \dd w\, \frac{\xi'(w)^{1/2}\, \xi'(\tilde{w}_p)^{1/2}}{\xi(w) - \xi(\tilde{w}_p)}\, \beta_\text{loc}(\tilde{w}_p)\,T^\text{bl}(w, \tilde{w}_p),
\end{equation}
and an analogous result can be also derived for the anti-holomorphic component. We remark that the expression of $K_A$ in Eq.~\eqref{eq:holomorphicGenericEH} for multiple intervals under a generic conformal mapping to the complex plane is a novel result of this manuscript.
\begin{figure}
    \centering
    \begin{tikzpicture}
        \draw   (-6.5,1)   --  (-2.5,1);
        \draw   (-6.5,-1)  --  (-2.5,-1);
        %\draw[dashed]   (0,-2)  arc (-90:90:1cm and 2cm);
        \draw   (-6.5,-1)  arc (-90:90:-.5 and 1);
        \draw   (-2.5,0)  ellipse (.5 and 1);
        \filldraw[thick]    (-6.5,0)   circle  (.02)   --  (-5,0)   circle  (.02);
        \draw   (-6.5,-.3) node    {$a_1$};
        \draw   (-5,-.3) node    {$b_1$};
        \filldraw[thick]    (-4.5,0)   circle  (.02)   --  (-3.5,0)   circle  (.02);
        \draw   (-4.5,-.3) node    {$a_2$};
        \draw   (-3.5,-.3) node    {$b_2$};
        \draw[->]   (-7,-1)    arc (-90:90:-.5 and 1);
        \draw   (-7.8,0)    node    {$\beta$};
        \draw[->,thick] (-1.5,0) --  (1.5,0);
        \draw   (0,-.5)  node    {$\xi(w) = e^{\frac{2 \pi}{\beta} w}$};
        \node[trapezium,
            draw,
            minimum width = 6cm,
            trapezium left angle = -135,
            trapezium right angle = -45] (t) at (4.5,0) {};
        \filldraw[thick]    (3,0)   circle  (.02)   --  (4,0)   circle  (.02);
        \filldraw[thick]    (4.5,0)   circle  (.02)   --  (6,0)   circle  (.02);
    \end{tikzpicture}
    \caption{Conformal mapping from the infinite cylinder of circumference $\beta$ described by the coordinate $w$ to the plane, $\xi$, using the transformation $\xi(w)=e^{\frac{2\pi w}{\beta}}$. The segments $[a_1,b_1], [a_2,b_2]$ are mapped to the branch cuts on the left figure.}
    \label{fig:mapping}
\end{figure}
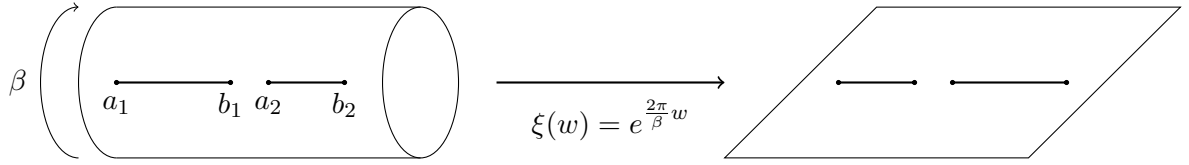
In order to use the result of Eq.~\eqref{eq:holomorphicGenericEH} for the finite temperature case, we recall that the cylinder is mapped into the plane under the transformation $\xi(w) = e^{\frac{2 \pi}{\beta} w}$, as shown in Fig.~\ref{fig:mapping}.
In particular, at the time $t = 0$ in which we are interested in, the holomorphic and anti-holomorphic coordinate $w$ coincides and the holomorphic and anti-holomorphic parts differ only in the operator.
Substituting this mapping in Eq.~\eqref{eq:genZeta}, we reproduce the expression for $z(x;\beta)$ at finite temperature reported in Eq.~\eqref{eq:finTempZeta}, which gives the entanglement temperature $\beta_\text{loc}(x;\beta)$ in Eq.~\eqref{eq:finTempBeta}. Regarding the bi-local part in Eq.~\eqref{eq:holomorphicConformalMapEH}, the weight function becomes
\begin{equation}
    \frac{\xi'(w)^{1/2}\, \xi'(\tilde{w}_p)^{1/2}}{\xi(w) - \xi(\tilde{w}_p)}\, \beta_\text{loc}(\tilde{w}_p) = \frac{\frac{2\pi}{\beta} e^{\frac{\pi}{\beta} ( w- \tilde{w}_p)}}{e^{\frac{2 \pi}{\beta} w} - e^{\frac{2 \pi}{\beta} \tilde{w}_p}}\, \beta_\text{loc}(\tilde{w}_p) = \frac{\beta_\text{loc}(\tilde{w}_p)}{\frac{\beta}{\pi} \sinh\!\frac{\pi(w-\tilde{w}_p)}{\beta}}\,,
\end{equation}
which is also in agreement with Eq.~\eqref{eq:finTempEH}, as expected. Therefore, we have used an alternative path to provide the results for the entanglement Hamiltonian of a disjoint set of intervals on the infinite cylinder. We stress that we find instructive to give this derivation here because we will use it also to evaluate the thermal twisted negativity Hamiltonian, which we introduce in the following section.

\section{Finite temperature negativity Hamiltonian}\label{sec:NH}

In this section, we present the main analytical result of this paper, which is the field-theoretical prediction for the negativity Hamiltonian on a torus. After recalling the definition of the negativity Hamiltonian for fermionic systems, we review the construction introduced in \cite{mvdc-22,rmtc-23} for its computation and we then extend it to the finite temperature case, showing two explicit examples. In particular, we find that in some cases the structure of the negativity Hamiltonian is more local than the one of the corresponding entanglement Hamiltonian.
\subsection{General definitions}
Let us consider a subsystem $A = A_1 \cup A_2$. As mentioned in the introduction, the negativity Hamiltonian $\mathcal{N}_A$ in \eqref{eq:defNegHam} is defined as the logarithm of the partially transposed reduced density matrix $\rho_A^{T_1}$ in \eqref{eq:bosonicPT}, where we perform a transpose operation only in $A_1$. The definition for such operation reported in Eq.~\eqref{eq:bosonicPT} is appropriate for bosonic systems, but it turns out to be ill-suited for fermions: while the partial transposition of Gaussian bosonic states is still a Gaussian state, due to the anti-commutation relation, this is not the case for a fermionic one \cite{ez-15,Eisert2018}, and this makes the computation difficult even for Gaussian states \cite{ctc-15,ctc-15b, ctc-15c}. For this reason, in \cite{ssr-17,ssr1-17,sr-19,shapourian-19,ryu} a more appropriate definition for fermionic systems has been introduced and the computational advantage one can gain is so remarkable that it has been employed in several contexts (see e.g. \cite{ac-18b,ge-20,mbc-21,pbc-22,fg-22,cmc-22,fmc-23}). 
In order to motivate this definition, let us remark that, in a bosonic system, the partial transposition is equivalent to a partial time-reversal or a mirror reflection in phase space \cite{s-00}. To see that this is indeed the case, we can consider a bosonic coherent state $\ket{\alpha} = e^{\alpha a^\dagger} \ket{0}$. On this state, the time-reversal transformation acts simply as the conjugation $\ket{\alpha} \to \ket{\alpha^*}$ \cite{s-00}, therefore the relative density matrix goes into its own transpose
\begin{equation}\label{eq:timeRevBoson}
    \ket{\alpha^*} \bra{\alpha} \longrightarrow \ket{\alpha} \bra{\alpha^*} = \left (\ket{\alpha^*} \bra{\alpha} \right )^T .
\end{equation}
For fermionic systems, the two transformations are not equivalent anymore. Under time-reversal, a fermionic coherent state $\ket{\xi} = e^{-\xi c^\dagger} \ket{0}, \bra{\bar{\xi}} = \bra{0} e^{- c \bar{\xi}}$ transforms as \cite{ssr-17}
\begin{equation}\label{eq:timeRevFermion}
    \ket{\xi}\bra{\bar{\xi}} \longrightarrow \ket{\ii \bar{\xi}}\bra{\ii \xi} ,
\end{equation}
which is different from the transposed density matrix because of the imaginary factor $\ii$. In light of this, one can define the partially time-reversed reduced density matrix $\rho_A^{R_1}$, obtained by acting with Eq.~\eqref{eq:timeRevFermion} only in $A_1$. This operation provides the fermionic logarithmic negativity $\mathcal{E}$ as
\begin{equation}\label{eq:defLogNegFermion}
    \mathcal{E} \equiv \log \Tr \left |\rho_A^{R_1} \right | = \log \Tr \sqrt{\rho_A^{R_1 \dagger} \rho_A^{R_1} }\,,
\end{equation}
although the spectrum of $\rho_A^{R_1}$ is not real in general \cite{shapourian-19}. \\An alternative definition for the fermionic partial transpose, called twisted partial transpose, has been introduced in~\cite{shapourian-19} as
\begin{equation}\label{eq:rhotilde}
    \rho_A^{\widetilde{R}_{1}}=\rho_A^{R_{1}}(-1)^{F_{A_1}},
\end{equation}
where $F_{A_1}=\sum_{j\in A_1}n_j$ is the number of fermions in the transposed subsystem $A_1$. This new operator has only real eigenvalues and the logarithmic negativity given by \cite{shapourian-19}
\begin{equation}\label{eq:lognegdef}
    \mathcal{E}=\log \mathrm{Tr}|\rho_A^{\widetilde{R}_1}|,
\end{equation}
is a measure of the negativeness of the eigenvalues, exactly as for the bosonic partial transpose. In this sense, the twisted fermionic partial transpose has a more transparent interpretation of the fermionic negativity and allows for the measure of mixed-state entanglement also from its moments, in full analogy with the bosonic partial transpose \cite{ekh-20,ncv-21}.\\
Following the definition of negativity in Eq.~\eqref{eq:defLogNegFermion}, in \cite{mvdc-22} the fermionic negativity Hamiltonian $\mathcal{N}_A$ has been defined as the logarithm of the partially time-reversed reduced density matrix $\rho_A^{R_1}$, with an appropriate normalisation
\begin{equation}\label{eq:defNegHamFermion}
    \rho_A^{R_1} \equiv \frac{e^{-2\pi \mathcal{N}_A}}{Z_A}  \,.
\end{equation}
In order to compute this operator, in \cite{mvdc-22} it was introduced a physically motivated procedure to construct the negativity Hamiltonian \eqref{eq:defNegHamFermion} from the knowledge of the entanglement Hamiltonian, as we will review for the case of the ground state on the infinite line, i.e., on the plane. Moreover, in Appendix 
\ref{app:resolvent} we also apply the resolvent method to rigorously justify the construction of this operator.\\
We can also define the twisted negativity Hamiltonian starting from Eq.~\eqref{eq:rhotilde} as
\begin{equation}\label{eq:defNegHamFermiontilte}
    \rho_A^{\widetilde{R}_1} \equiv \frac{e^{-\widetilde{\mathcal{N}}_A}}{Z_A},
\end{equation}
but in this section we focus only on Eq.~\eqref{eq:defNegHamFermion}. We will come back to $\widetilde{\mathcal{N}}_A$ in Sec.~\ref{sec:lattice}.\\
Before ending this section, we review the result for $\mathcal{N}_A$ obtained in Ref.~\cite{mvdc-22} (and we refer to the Appendix \ref{app:resolvent} for more details). Let us consider a multi-component subsystem $A = [a_1, b_1] \cup \ldots \cup [a_n, b_n]$ composed of $n$ intervals and, to fix the ideas, let us reverse only one interval $A_1 = [a_j, b_j]$; this case can be straightforwardly generalised to multiple reversed intervals. For this configuration, the entanglement Hamiltonian $K_A$ is given by Eqs.~\eqref{eq:CHEH}, \eqref{eq:CHZeta} \cite{ch-09, achp-18}. Under the path-integral construction of \cite{cct-neg-1, cct-neg-2}, the partial transpose has the net effect of applying a spatial reversal in the transposed interval. This can be understood in terms of CPT symmetry, since the time-reversal operation of Eqs.~\eqref{eq:timeRevBoson}, \eqref{eq:timeRevFermion} is equivalent to a parity transformation followed by a charge conjugation. This is implemented by exchanging the extrema $a_j, b_j$ of the reversed interval in the expression of the entanglement Hamiltonian \cite{mvdc-22}. Under this procedure, the function \eqref{eq:CHZeta} becomes
\begin{equation}\label{eq:CHZetaRev}
    z^R(x) = \log \! \left [ - \frac{x-b_j}{x-a_j} \prod_{i \neq j} \frac{x-a_i}{x-b_i} \right ].
\end{equation}
Moreover, the effect of the partial transposition on the Dirac spinor $\psi=\begin{pmatrix}
        \psi_R\\
        \psi_L
    \end{pmatrix}$ 
is simply $\psi(x)\to \ii \psi(x)$, $\psi^{\dagger}(x)\to \ii \psi^{\dagger}(x)$ for $x \in [a_j,b_j]$ \cite{rmtc-23} (see also \cite{ssr-17} and Appendix \ref{app:resolvent}). In the following section, we will show how this construction can be applied to compute the negativity Hamiltonian on the torus.
\subsection{Negativity Hamiltonian on the torus}\label{sec:NHtorus}
Starting from the result for the entanglement Hamiltonian in Eqs.~\eqref{eq:EHtorus}, \eqref{eq:TorusZeta}, by exchanging the extrema $a_j$, $b_j$ we can obtain the negativity Hamiltonian on the torus.
We remind that in the function $z(x;\beta,L)$ in Eq.~\eqref{eq:TorusZeta}, it appears a term proportional to $x$ and to the total length $\ell$ of the subsystem \cite{fr-19, bpn-19, bgpn-19}. 
It is useful to write the subsystem length $\ell$ as $\ell = \sum_i (b_i - a_i)$, since in order to obtain the correct negativity Hamiltonian it is necessary to exchange the endpoints of the reversed interval also in this expression. If we call $\ell_1 = \sum_{j \in A_1} (b_j - a_j)$ the total length of the partially reversed subsystem $A_1$ (for us, $\ell_1 = b_j - a_j$) and $\ell_2 = \sum_{i \in A_2} (b_i - a_i)$ the total length of $A_2$, this procedure gives 
\begin{equation}\label{eq:TorusZetaRev}\begin{split}
    z^R(x; \beta, L) = \log\left [ -\frac{\vartheta_1\!\left ( \frac{\pi}{L} \left ( x - b_j \right ) \big | q\right )}{\vartheta_1\!\left ( \frac{\pi}{L} \left ( x - a_j \right ) \big | q\right )} \prod_{i \neq j} \frac{\vartheta_1\!\left ( \frac{\pi}{L} \left ( x - a_i \right ) \big | q\right )}{\vartheta_1\!\left ( \frac{\pi}{L} \left ( x - b_i \right ) \big | q\right )} \right ] + \frac{2 \pi x}{\beta L}  \left ( \ell_2 - \ell_1\right  ) \,.
\end{split}\end{equation} 
Analogously, Eq.~\eqref{eq:conjTorus}, that determines the position of the conjugate points, becomes
\begin{equation}\label{eq:conjTorusRev}
    z^R(x;\beta,L) - z^R(\tilde{x}^R_{kp};\beta,L) +  \frac{2 \pi k}{\beta} \left ( \ell_2 - \ell_1\right  ) = 0 \, , \quad k \in \ZZ\, ,
\end{equation}
where again we have exchanged $\ell$ with $\ell_2 - \ell_1$.
For $\ell_1 \neq \ell_2$, the negativity Hamiltonian on the torus has a non-local structure analogous to the one of the corresponding entanglement Hamiltonian in Eq.~\eqref{eq:EHtorus}, containing infinite terms coupling different points 
\begin{equation}\label{eq:NHtorus}\begin{split}
    &\mathcal{N}_A(\beta,L) = \mathcal{N}_A^\text{loc}(\beta,L) + \mathcal{N}_A^\text{nl}(\beta,L) \\
    =&\int \dd x\, \beta_\text{loc}^R(x;\beta,L)\, T_{00}(x) \\
    &\hspace{2cm}+ \!\!\!\sum_{(p, k) \neq (0,0)}\!\!\! (\pm 1)^{k}  \int_A \dd x\, \frac{\beta_\text{loc}^R(\tilde{x}_{kp}^R; \beta, L)\, \ii^{\Theta_1(x)} (-\ii)^{\Theta_1(\tilde{x}^R_{kp})}}{\frac{\beta}{\pi} \sinh\!\left [ \frac{\pi}{\beta} \left ( x - \tilde{x}_{kp}^R + k L\right )\right ]}\, T^\text{bl}\!\left (x, \tilde{x}_{kp}^R, t= 0 \right ) ,
\end{split}\end{equation}
where $p\in \{0,\ldots,n-1\}, k\in \ZZ$ and the function $\Theta_1(x)$ is equal to 1 only for $x\in A_1$, 0 otherwise
\begin{equation}\label{eq:theta1}
    \Theta_1(x) = \begin{cases}
        1,  &x\in A_1,\\
        0,  &x\notin A_1 .
    \end{cases}
\end{equation}
In Eq.~\eqref{eq:NHtorus} we have introduced the ``negativity temperature'' 
\begin{equation}\label{eq:inversenegtemperature}
    \beta_\text{loc}^R(x;\beta,L) = \frac{1}{(z^{R}(x;\beta,L))'},
\end{equation} 
and, analogously to Eq.~\eqref{eq:EHtorus}, the signs $+$ and $-$ correspond respectively to the Ramond and to the Neveu-Schwarz sectors.
On the other hand, when $\ell_1=\ell_2$, the dependence on the integer index $k$ in Eq.~\eqref{eq:conjTorusRev} cancels out exactly and the solutions $\tilde{x}^R_{kp}$ with different $k$ collapse on each another, giving a striking qualitative difference with respect to the entanglement Hamiltonian in Eq.~\eqref{eq:EHtorus}. 
In Eq.~\eqref{eq:NHtorus}, the bi-local terms with different $k$ and same $p$ are then calculated in the same conjugate point $\tilde{x}_{p}^R$, leading to a bi-local structure with only $n-1$ bi-local terms
\begin{equation}\label{eq:NHtorusEqual}\begin{split}
    &\mathcal{N}_A(\beta,L) = \mathcal{N}_A^\text{loc}(\beta,L) + \mathcal{N}_A^\text{nl}(\beta,L) \\
    =&\int \dd x\, \beta_\text{loc}^R(x;\beta,L)\, T_{00}(x) \\
    &\hspace{2cm}+ \sum_{p = 1}^{n} \int_A \dd x\, \beta_\text{loc}^R(\tilde{x}_{p}^R; \beta, L)\, \frac{g^R_\pm( x - \tilde{x}_{p}^R )}{L}\,\ii^{\Theta_1(x)} (-\ii)^{\Theta_1(\tilde{x}^R_{kp})}\, T^\text{bl}\!\left (x, \tilde{x}_{p}^R, t= 0 \right ) ,
\end{split}\end{equation}
where we have introduced the (dimensionless) functions $g_{\pm}^R(z)$ defined by the infinite series 
\begin{equation}\label{eq:g}
    g_{\pm}^R(z; \beta,L) = \frac{\pi L}{\beta} \sum_{k = -\infty}^{+\infty} \frac{(\pm 1)^{k}}{ \sinh\!\left [ \frac{\pi}{\beta} \left ( z + k L\right )\right ]}\,.
\end{equation}
In the Ramond sector ($+$ sign), Eq.~\eqref{eq:g} can be resummed to give
\begin{equation}\label{eq:ramond} \begin{split}
    &g_{+}^R(z; \beta,L) = \psi_{q}\!\left(-\frac{z}{L}\right) - \psi_{q}\!\left(1 + \frac{z}{L}\right)  + \psi_{q}\!\left(1 + \frac{z - \ii \beta}{L} \right) - \psi_{q}\!\left(-\frac{z + \ii \beta}{L} \right) , \quad q = e^{\frac{\pi L }{\beta}},
\end{split}\end{equation}
where $\psi_{q}$ denotes the $q$-digamma function  (see Appendix \ref{app:tools} for its definition), while in the Neveu-Schwarz sector ($-$ sign) it reads
\begin{equation}\label{eq:NS}\begin{split}
    &g_{-}^R(z; \beta,L) = \frac{1}{2} \Bigg [ \psi_{q^2}\!\left(-\frac{z}{2 L}\right) -\psi_{q^2}\!\left( \frac{L-z}{2 L}\right) + \psi_{q^2}\!\left(\frac{L+z}{2
   L}\right) - \psi_{q^2}\!\left(\frac{z}{2 L}+1\right) \\&+ \psi_{q^2}\!\left(\frac{L-z-\ii \beta }{2 L}\right) - \psi_{q^2}\!\left(\frac{L+z-\ii \beta
   }{2 L}\right) + \psi_{q^2}\!\left(\frac{2 L+z-i \beta }{2 L}\right)-\psi_{q^2}\!\left(-\frac{z+i \beta
   }{2 L}\right) \Bigg ].
\end{split}\end{equation}
To summarise, when the length of the reversed intervals is equal to the non-reversed one, the negativity Hamiltonian recovers a mild non-local structure given by a finite number of bi-local terms, while such a simplification does not arise in the entanglement Hamiltonian. \\
In the following we specialise to the case of $n$ intervals lying on an infinite line at finite temperature (i.e. the space-time is a cylinder), and then we 
present explicit examples for the case of two intervals.

\subsection{Finite temperature negativity Hamiltonian on the infinite line}

The finite temperature negativity Hamiltonian on the infinite line can be obtained either by directly exchanging the extrema of the reversed interval in the related entanglement Hamiltonian reported in Eqs.~\eqref{eq:finTempEH}, \eqref{eq:finTempZeta} or by taking the $L \to \infty$ limit of the negativity Hamiltonian in Eq.~\eqref{eq:TorusZetaRev}, similarly to the limit reported in Eq.~\eqref{eq:limitEH}. By applying the exchanging procedure to Eq.~\eqref{eq:finTempEH}, we find that the function $z(x;\beta)$ in Eq.~\eqref{eq:finTempZeta} reduces to 
\begin{equation}\label{eq:finTempZetaRev}
    z^R(x;\beta) = \log \left [ - \frac{\sinh\frac{\pi(x - b_j)}{\beta}}{\sinh\frac{\pi (x - a_j)}{\beta}} \prod_{i\neq j} \frac{\sinh\frac{\pi(x - a_i)}{\beta}}{\sinh\frac{\pi (x - b_i)}{\beta}} \right ],
\end{equation}
and the $n - 1$ conjugate points $\tilde{x}_p^R$ are found to be the non-trivial solutions of $z^R(x;\beta) = z^R(\tilde{x}_p^R;\beta)$.  
Thus, the finite temperature negativity Hamiltonian on the infinite line is
\begin{equation}\label{eq:finTempNH}\begin{split}
    &\mathcal{N}_A(\beta) = \mathcal{N}_A^\text{loc}(\beta) + \mathcal{N}_A^\text{bl}(\beta) \\
    & = \int_A \dd x\, \beta_\text{loc}^R(x;\beta)\, T_{00}(x) + \sum_{p = 1}^{n-1} \int \dd x\, \frac{\beta_\text{loc}^R(\tilde{x}_{p}; \beta)\, \ii^{\Theta_1(x)} (-\ii)^{\Theta_1(\tilde{x}^R_{p})}}{\frac{\beta}{\pi} \sinh\! \frac{\pi \left ( x - \tilde{x}_{p}^R\right )}{\beta} }\, T^\text{bl}\!\left (x, \tilde{x}_{p}^R, t= 0 \right )\,,
\end{split}\end{equation}
where the negativity temperature $\beta_\text{loc}^R(x;\beta)$ is given by 
\begin{equation}\begin{split}
    \beta_\text{loc}^R(x;\beta) = \frac{1}{{z^R}'(x;\beta)} = \frac{\beta}{\pi} \Bigg [ &\coth\!\frac{\pi (x - b_j)}{\beta} - \coth\!\frac{\pi (x - a_j)}{\beta} \\
    &+ \sum_{i\neq j} \left ( \coth\!\frac{\pi (x - a_i)}{\beta} - \coth\!\frac{\pi (x - b_i)}{\beta} \right ) \Bigg ]^{-1}, 
\end{split}\end{equation}
and the bi-local terms are calculated in the $n-1$ conjugate points obtained as the non-trivial solutions of $z^R(x;\beta) = z^R(\tilde{x}_p^R;\beta)$. As we also commented for the entanglement Hamiltonian, the negativity Hamiltonian only contains $n-1$ bi-local terms. 

\subsection{Tripartite geometry}\label{sec:tripartite}

As a first explicit example regarding the negativity Hamiltonian on the torus, we consider a tripartite geometry made up of two intervals $A_1 = [a_1, b_1]$, $A_2 = [a_2, b_2]$. Let us call $\ell_1 = b_1 - a_1$ the length of $A_1$ and $\ell_2 = b_2 - a_2$ the one of $A_2$, and let us reverse the interval $A_1$. Then, specialising Eq.~\eqref{eq:TorusZetaRev} to this configuration we find
\begin{equation}\label{eq:TorusZetaRevTripartite}\begin{split}
    z^R(x; \beta, L) &= \log\left [ -\frac{\vartheta_1\!\left ( \frac{\pi}{L} \left ( x - b_1 \right ) \big | q\right ) \vartheta_1\!\left ( \frac{\pi}{L} \left ( x - a_2 \right ) \big | q\right )}{\vartheta_1\!\left ( \frac{\pi}{L} \left ( x - a_1 \right ) \big | q\right ) \vartheta_1\!\left ( \frac{\pi}{L} \left ( x - b_2 \right ) \big | q\right ) }   \right ] + \frac{2 \pi x}{\beta L}  \left ( \ell_2 - \ell_1\right ),
\end{split}\end{equation}
while the conjugate point equation in Eq.~\eqref{eq:conjTorusRev} becomes
\begin{equation}\label{eq:conjTorusRevTripartite}
    z^R(x; \beta, L) - z^R(\tilde{x}_{k}^R; \beta, L) + \frac{2 \pi k}{\beta} \left( \ell_2 - \ell_1\right ) = 0 \, , \quad k \in \ZZ\, .
\end{equation}
We stress again that for $\ell_1=\ell_2$, the non-local structure of the negativity Hamiltonian drastically simplifies since the solutions of Eq.~\eqref{eq:conjTorusRevTripartite} do not depend on the index $k$, leading to a single bi-local term.
We can now also consider some interesting limits of Eq.~\eqref{eq:TorusZetaRevTripartite}.
\paragraph{Finite temperature on the infinite line:}
If the two intervals $A_1 = [a_1, b_1]$ and $A_2 = [a_2, b_2]$ lie on the infinite line, the function $z^R(x;\beta)$ in Eq.~\eqref{eq:TorusZetaRevTripartite} becomes 
\begin{equation}\label{eq:finTempZetaRevTripartite}\begin{split}
    z^R(x;\beta) = \log\left[ \frac{\sinh\!\frac{\pi(x-b_1)}{\beta}\,\sinh\!\frac{\pi(x-a_2)}{\beta}}{\sinh\!\frac{\pi (x-a_1)}{\beta}\,\sinh\!\frac{\pi(b_2-x)}{\beta}} \right],
\end{split}\end{equation}
which gives the negativity temperature
\begin{equation}\label{eq:betatripinfinite}
    \beta_\text{loc}^R(x;\beta) = \frac{\beta}{\pi}\left [ - \coth\!\frac{\pi (x-a_1)}{\beta} + \coth\!\frac{\pi (x-b_1)}{\beta} + \coth\!\frac{\pi (x-a_2)}{\beta} - \coth\!\frac{\pi (x-b_2)}{\beta}\right ]^{-1} .
\end{equation}
There is a single bi-local term, calculated in the conjugate point $\tilde{x}^R$
\begin{equation}\label{eq:tildeTripartitoThermal}
    \tilde{x}^R = \frac{\beta}{2 \pi} \log \! \left [ \frac{2\,e^{\frac{2 \pi}{\beta}x}\, \sinh\!\frac{\pi(\ell_2-\ell_1)}{\beta}  + \left (e^{\frac{2 \pi}{\beta}a_1} + e^{\frac{2 \pi}{\beta}b_2}\right ) e^{\frac{\pi}{\beta}(\ell_1 - \ell_2)} - \left (e^{\frac{2 \pi}{\beta}b_1} + e^{\frac{2 \pi}{\beta}a_2}\right) e^{\frac{\pi}{\beta}(\ell_2 - \ell_1)}}{e^{- \frac{\pi}{\beta}(a_1 + b_1 + a_2 + b_2)}\left (e^{\frac{2 \pi}{\beta}a_1} - e^{\frac{2 \pi}{\beta}b_1} + e^{\frac{2 \pi}{\beta}b_2} - e^{\frac{2 \pi}{\beta}a_2}\right ) e^{\frac{2 \pi}{\beta}x} -2 \sinh\!\frac{\pi(\ell_2-\ell_1)}{\beta}} \right ]
\end{equation}
which is the only non-trivial solution of $z^R(x;\beta) = z^R(\tilde{x}^R;\beta)$. In particular, for $\ell_1 = \ell_2$ Eq.~\eqref{eq:tildeTripartitoThermal} reduces simply to $\tilde{x}^R = a_1 + b_2 - x$.
The weight function of the bi-local operator reads 
\begin{equation}\label{eq:tildeTripartitoThermalbilocl}
    \beta^R_{\mathrm{bl}}(x;\beta)=\frac{\beta_\text{loc}^R(\tilde{x}^R;\beta)}{\frac{\beta}{\pi}\sinh(\frac{\pi}{\beta}(x-\tilde{x}^R))},
\end{equation}
As a further cross-check of our result, it is interesting to consider the zero-temperature limit $\beta \to \infty$ of the negativity Hamiltonian. In this regime, we expect to retrieve the result for the tripartite configuration in the ground state, which was obtained in \cite{mvdc-22} by directly applying the exchanging procedure to the result in Eqs.~\eqref{eq:CHEH}, \eqref{eq:CHZeta}. Indeed, we see that taking the limit $\beta \to \infty$ of $z^R(x;\beta)$ in Eq.~\eqref{eq:finTempZetaRevTripartite}, we reproduce the function on the plane found in \cite{mvdc-22} 
\begin{equation}\label{eq:plane2ZetaRev}
    z^R(x) = \log\!\left [ \frac{(x - b_1) (x - a_2)}{(x - a_1) (b_2 - x)}\right ].
\end{equation}
Regarding the conjugate points, again we see that they are given by the single non-trivial solution of $z^R(x) = z^R(\tilde{x})$, with $z^R(x)$ in Eq.~\eqref{eq:plane2ZetaRev}, finding the same conjugate point of \cite{mvdc-22}
\begin{equation}\label{eq:conjnegTzero}
    \tilde{x}^R=\frac{(a_1 b_2 - b_1 a_2) x + (a_1 + b_2) b_1  a_2 - (b_1 + a_2) a_1 b_2}{(a_1 - b_1 + b_2 - a_2) x + b_1 a_2 - a_1 b_2},
\end{equation}
as expected. For this geometry, i.e. two intervals on the plane, we provide a rigorous derivation of the result in Appendix \ref{app:resolvent}.

\subsection{Bipartite geometry}\label{sec:bipartite}

We now study a bipartite geometry on the torus where $A_1 = [0,\ell_1]$ and the rest of the system is $A_2 = [\ell_1, L]$.
Notice that, differently from the case studied above, now the union $A = A_1 \cup A_2$ of the reversed interval $A_1$ and $A_2$ is not a proper subset of the circle, but it covers all the system.
Such a geometry can be obtained from the tripartite case of Sec.~\ref{sec:tripartite} by choosing $a_1 = 0$, $b_1 = a_2 = \ell_1$ and $b_2 = L$.
Taking this limits in the function $z^R$ in Eq.~\eqref{eq:TorusZetaRevTripartite}, we obtain
\begin{equation}\label{eq:TorusZetaRevBipartite}\begin{split}
    z^R(x; \beta, L) &= \log\left [ -\frac{\vartheta_1\!\left ( \frac{\pi}{L} \left ( x - \ell_1 \right ) \big | q\right )^2}{\vartheta_1\!\left ( \frac{\pi}{L} x \big | q\right ) \vartheta_1\!\left ( \frac{\pi}{L} \left ( x - L \right ) \big | q\right ) }   \right ] + \frac{2 \pi x}{\beta}  \left ( 1 - \frac{2 \ell_1}{L}\right ) \\
    &= 2 \log\left | \frac{\vartheta_1\!\left ( \frac{\pi}{L} \left ( x - \ell_1 \right ) \big | q\right )}{\vartheta_1\!\left ( \frac{\pi}{L} x \big | q\right ) }   \right | + \frac{2 \pi x}{\beta}  \left ( 1 - \frac{2 \ell_1}{L}\right ) \, ,
\end{split}\end{equation}
where we have used the periodicity of the theta function $\vartheta_1(z-\pi | q ) = - \vartheta_1(z|q)$, while Eq.~\eqref{eq:conjTorusRevTripartite} for the conjugate points becomes 
\begin{equation}\label{eq:conjTorusRevBipartite}
    z^R(x;\beta,L) - z^R(\tilde{x}^R_{k};\beta,L) + \frac{2 \pi k}{\beta} \left( L - 2 \ell_1\right ) = 0 \, , \quad k \in \ZZ\, .
\end{equation}
The corresponding negativity temperature is provided by 
\begin{equation}\label{eq:localTorusRevBipartite}
   \beta^R_{\text{loc}}(x;\beta,L) =\frac{1}{z^R(x;\beta,L)'}.
\end{equation}
We again remark that for $\ell_1=L/2$ the dependence in $k$ drops out from Eq.~\eqref{eq:conjTorusRevBipartite}, and therefore all the infinite non-local solutions collapse into a single bi-local term with weight given by
\begin{equation}\label{eq:bilocalTorusRevBipartite}
   \beta^R_{\mathrm{bl}}(x;\beta,L)=\beta^R_{\text{loc}}(\tilde{x}^R;\beta,L) \frac{g^R_{\pm}(x-\tilde{x}^R)}{L},
\end{equation}
where $g^R_{\pm}$ are given in Eqs.~\eqref{eq:ramond} and \eqref{eq:NS}, respectively.
Let us stress that this represents an important result of this manuscript, since a bipartite system at finite temperature is a neat example of mixed state: in this case, the negativity is a genuine entanglement measure, differently from the entanglement entropy which mixes both quantum and thermal correlations. Therefore, the result for the negativity Hamiltonian provides the first operatorial characterisation of a thermal state. 
Let us now consider some interesting limits also for this bipartite geometry.
\paragraph{Finite temperature on the infinite line:}
Finding the theoretical prediction for the bipartite negativity Hamiltonian on the infinite line is more subtle than in the tripartite case of Sec.~\ref{sec:tripartite} because now $A_1$ and $A_2$ cover the full infinite line.
The geometry of interest is $A_1 = [0, \ell_1]$, $A_2 = [-\infty, 0] \cup [\ell_1, + \infty]$ and we reverse the interval $A_1$.
We can obtain this geometry from a three interval configuration on the infinite line $A_1 = [0, \ell_1]$, $A_2 = [-L/2, 0] \cup [\ell_1, \ell_1 + L/2]$, taking then the limit $L \to \infty$ \cite{ryu}. Specialising the function $z^R(x;\beta)$ in Eq.~\eqref{eq:finTempZetaRev} to this geometry and taking the $L \to \infty$ limit we find (up to $x$-independent terms)
\begin{equation}\begin{split}
    \log\left [ \left ( \frac{\sinh\!\frac{\pi(\ell_1-x)}{\beta}}{\sinh\!\frac{\pi x}{\beta}} \right )^{\!\!2} \frac{\sinh\!\frac{\pi(x+L/2)}{\beta}}{\sinh\!\frac{\pi (L/2+\ell_1-x)}{\beta}} \right ] \longrightarrow & \log\left [ \left ( \frac{\sinh\!\frac{\pi(\ell_1-x)}{\beta}}{\sinh\!\frac{\pi x}{\beta}} \right )^{\!\!2} \frac{e^{\frac{\pi}{\beta}x}}{e^{\frac{\pi}{\beta}(\ell_1-x)}} \right ]\\
    &= z^R(x; \beta) + \text{const.}\,,    
\end{split}\end{equation}
where now $z^R(x; \beta)$ reads 
\begin{equation}\label{eq:zRbeta}
    z^R(x; \beta) = 2 \log \!\left | \frac{\sinh\!\frac{\pi (\ell_1-x)}{\beta}}{\sinh\!\frac{\pi x}{\beta}} \right | + \frac{2 \pi x}{\beta}\,.
\end{equation}
This form differs from the one in Eq.~\eqref{eq:finTempZetaRevTripartite} for the tripartite geometry, since now we find a term proportional to $x$. 
From this result we see that the negativity temperature is
\begin{equation}\label{eq:betaloc_bipartiteNH}
    \beta_\text{loc}^R (x;\beta) = \frac{1}{z'(x)} = \frac{\beta}{2 \pi} \left [ 1 + \coth\!\frac{\pi (x -\ell_1)}{\beta} -  \coth\!\frac{\pi x}{\beta} \right ]^{-1}.
\end{equation}
Since the geometry is made of three intervals, the equation for the conjugate points obtained from Eq.~\eqref{eq:finTempZetaRev}, with $z^R(x; \beta)=z^R(y; \beta)$ is a polynomial of third order in $y$ and one has the trivial solution $y = x$ and also two non-trivial solutions $y=\tilde{x}_{\pm}^R$, that in the limit $L\to\infty$ read 
\begin{equation}
\begin{split}\label{eq:conjugate}
    \tilde{x}_+^R =&\,\frac{\beta}{2\pi}\log\Bigg[\frac{1}{8} \left(-4 e^{\frac{2 \pi \ell_1}{\beta }}+e^{\frac{4 \pi \ell_1}{\beta }}+\left(e^{\frac{2 \pi 
   \ell_1}{\beta }}-1\right) \sqrt{-6 e^{\frac{2 \pi  \ell_1}{\beta }}+e^{\frac{4 \pi \ell_1}{\beta }}+4
   e^{\frac{2 \pi  (\ell_1-x)}{\beta }}+4 e^{\frac{2 \pi x}{\beta }}-3}\right.\\&\left.+2 e^{\frac{2 \pi 
   (\ell_1-x)}{\beta }}+2 e^{\frac{2 \pi  x}{\beta }}-1 \right) \text{csch}^2\!\left(\frac{\pi 
   x}{\beta }\right)\Bigg],\\
   \tilde{x}_-^R =&\, \frac{\beta}{2\pi}\log\Bigg[\frac{1}{8} \left(-4 e^{\frac{2 \pi  \ell_1}{\beta }}+e^{\frac{4 \pi  \ell_1}{\beta }}-\left(e^{\frac{2 \pi 
  \ell_1}{\beta }}-1\right)  \sqrt{-6 e^{\frac{2 \pi  \ell_1}{\beta }}+e^{\frac{4 \pi \ell_1}{\beta }}+4
   e^{\frac{2 \pi  (\ell_1-x)}{\beta }}+4 e^{\frac{2 \pi x}{\beta }}-3}\right.\\&\left.+2 e^{\frac{2 \pi  (\ell_1-x)}{\beta }}+2 e^{\frac{2 \pi  x}{\beta }}-1\right)
   \text{csch}^2\!\left(\frac{\pi  x}{\beta }\right)\Bigg].
   \end{split}
\end{equation}
The bi-local inverse temperature corresponding to each conjugate point $\tilde{x}^R_{\pm}$ is  
\begin{equation}\label{eq:bipartiteThermalbilocl}
    \beta^R_{\mathrm{bl}}(\tilde{x}^R_{\pm};\beta)=\frac{\beta_\text{loc}^R(\tilde{x}^R_{\pm};\beta)}{\frac{\beta}{\pi}\sinh(\frac{\pi}{\beta}(x-\tilde{x}^R_{\pm}))}\,.
\end{equation}
Another interesting limit we can study is when $\beta \to \infty$, i.e. the zero temperature case, in which the state becomes pure. From Eq.~\eqref{eq:betaloc_bipartiteNH}, the negativity temperature is given by 
\begin{equation}
   \beta_\text{loc}^R (x;\infty) =\frac{(x-\ell_1)x}{2\ell_1}\,,  
\end{equation}
which is half of the weight function of the entanglement Hamiltonian for one single interval in the ground state in Eq.~\eqref{eq:singleintervalEH}. The limit of Eq.~\eqref{eq:zRbeta} is 
\begin{equation}\label{eq:zetaRevPure}
    z^R(x;\infty)=2\log\Big |1-\frac{\ell_1}{x}\Big |\,,
\end{equation}
and the two conjugate point in Eq.~\eqref{eq:conjugate} are 
\begin{equation}\label{eq:conjugatePlusMinLimit}
\begin{split}
    \tilde{x}_+^R=
    \begin{cases}
    \frac{\beta}{\pi}  \log \left |\frac{\ell_1-x}{x}\right|,\quad &x< \ell_1/2\,,\\
    \frac{x\ell_1}{2x-\ell_1}, \quad & x> \ell_1/2\,,
    \end{cases}\\
     \tilde{x}_-^R=
    \begin{cases}
    \frac{x\ell_1}{2x-\ell_1}, \quad & x< \ell_1/2\,,\\
    \frac{\beta}{\pi}  \log \left|\frac{\ell_1-x}{x}\right|,\quad &x> \ell_1/2\,.
    \end{cases}
    \end{split}
\end{equation}
In the limit $\beta\to \infty$, the conjugate point $\tilde{x}_+^R$ ($\tilde{x}_-^R$) diverges as $\mathcal{O}(\beta)$ for $x< \ell_1/2$ ($x> \ell_1/2$), and the bi-local operators
calculated in this point do not contribute because the fermionic field $\psi(x)$ vanish as $x\to \infty$ \cite{mt-21,rmtc-23}. In the other regions, instead, $\tilde{x}_+^R$ and $\tilde{x}_-^R$ are joined together to give the conjugate point $\tilde{x}^R = x\ell_1 / (2x-\ell_1)$ in which the fermion does not vanish. Notice that, as expected, this conjugate point is precisely the only non-trivial solution of $z^R(x;\infty) = z^R(\tilde{x}^R;\infty)$ with $z^R(x;\infty)$ in Eq.~\eqref{eq:zetaRevPure}. We can explicitly compute the weight functions of the bi-local operators as
\begin{equation}
\begin{split}
    &\frac{\beta_\text{loc}^R(\tilde{x}_{+}^R;\beta)}{\frac{\beta}{\pi}\sinh\Big[\frac{\pi}{\beta}(x-\tilde{x}_{+}^R)\Big]}=\begin{cases}
        \frac{|x|(\ell_1-x)}{\ell_1^2(2x-\ell_1)},\quad & x<\ell_1/2\,,\\
        \frac{\ell_1  }{4 (\ell_1-2 x)},\quad &x >\ell_1/2\,, 
    \end{cases}\\
    &\frac{\beta_\text{loc}^R(\tilde{x}_{-}^R;\beta)}{\frac{\beta}{\pi}\sinh\Big[\frac{\pi}{\beta}(x-\tilde{x}_{-}^R)\Big]}=\begin{cases}
        \frac{\ell_1  }{4 (\ell_1-2 x)},\quad &x <\ell_1/2\,, \\
        \frac{x|\ell_1-x|}{\ell_1^2(2x-\ell_1)},\quad & x>\ell_1/2\,.
    \end{cases}
\end{split}
\end{equation}
As we can see, considering only the bi-local weights calculated in the region in which the conjugate points in Eq.~\eqref{eq:conjugatePlusMinLimit} remain finite, the bipartite negativity Hamiltonian at zero temperature is
\begin{equation}\label{eq:NHbipartitePure}\begin{split}
    &\mathcal{N}_A = \mathcal{N}_A^\text{loc} + \mathcal{N}_A^\text{loc} \\
    &= \int_{-\infty}^{+\infty} \dd x\, \frac{(x-\ell_1)x}{2\ell_1} \, T_{00}(x) - \ii \left ( \int_{-\infty}^0 - \int_0^{\ell_1} + \int_{\ell_1}^{+\infty} \right ) \dd x\, \frac{\ell_1  }{4 (\ell_1-2 x)}\, T^\text{bl}(x, \frac{x\ell_1}{2x-\ell_1})\,.
\end{split}\end{equation}
We remark that, although one of the imaginary bi-local operators of the negativity Hamiltonian does not vanish, as $\beta\to\infty$ the state becomes pure and $[\rho_A^{R_1},(\rho_A^{R_1})^{\dagger}]=0$ \cite{ez-15,ssr-17}. As a consequence, we find
\begin{equation}
    \sqrt{\rho_A^{R_1}(\rho_A^{R_1})^{\dagger}}=|\rho_A^{R_1}|=\frac{1}{Z}e^{-\pi(\mathcal{N}_A+\mathcal{N}^{\dagger}_A)}=\frac{1}{Z}e^{-2\pi \mathcal{N}_A^{\text{loc}}}.
\end{equation}
 The local part of the negativity Hamiltonian can be also rewritten as 
 \begin{equation}
 \mathcal{N}^{\text{loc}}_A = \frac{1}{2}\left ( \mathbb{I}_{A_1} \otimes K_{A_2} - K_{A_1} \otimes \mathbb{I}_{A_2} \right ),
 \end{equation}
 where $\mathbb{I}_{A_2}$ and $\mathbb{I}_{A_2}$ denote the identity operators on $A_1$ and $A_2$, respectively, and 
\begin{equation}
    K_{A_{1}}=\displaystyle \int_{0}^{\ell_1}dx\frac{x(\ell_1-x)}{\ell_1}T_{00}(x), \qquad K_{A_{2}}=\displaystyle \int_{-\infty}^0dx\frac{x(x-\ell_1)}{\ell}T_{00}(x)+\displaystyle \int_{\ell_1}^{\infty}dx\frac{x(x-\ell_1)}{\ell_1}T_{00}(x),
\end{equation}
are the entanglement Hamiltonians of the interval $A_1=[0,\ell_1]$ ($K_{A_{1}}$) and of its complement ($K_{A_{2}}$).
This result does not come as a surprise since a bipartite geometry at zero temperature is a pure state and one recovers that \cite{ssr-17}
\begin{equation}
    \mathrm{Tr}|\rho_A^{R_1}|=\mathrm{Tr}(\rho_{A_1}^{1/2})^2.
\end{equation}
In other words, for a pure state the
logarithmic negativity is equal to the R\'enyi entropy of order $1/2$ defined in Eq.~\eqref{eq:defEntEntropies}.
\section{Numerical analysis}\label{sec:lattice}

In this section we present exact numerical calculations on the lattice in order to compare them with our field-theoretical predictions.
For Gaussian states, as those we are considering in this manuscript, it is possible to compute both the entanglement and the negativity Hamiltonian from the knowledge of the two-point correlation matrix restricted to the subsystem $A$, $C_A$ \cite{pkl-99, Peschel2009, Chung2001, Peschel2003, Peschel2004, Peschel2012,ep-17,ep-18}. For lattice fermions at finite temperature on the circle, $C_A$ is known both for periodic and for anti-periodic boundary conditions \cite{vidal2,jiaju}. However, we stress that, as it has been studied in the literature \cite{abch-16,gmcd-18,ETP19,ETP22,EH-1,EH-2,EH-3,jt-21,defv-22,fsc-22,rmtc-23,zcdr-20}, comparing the lattice entanglement and negativity Hamiltonians with the analytical results is highly non-trivial.
Indeed, while the terms of the field-theoretical predictions are localised around certain points (see, e.g. Eq.~\eqref{eq:CHEH}), on the lattice the Hamiltonians are more delocalised, and this requires taking the continuum limit carefully.
In this section, we first review how to recover the lattice entanglement and negativity Hamiltonians and we explain how to take the continuum limit of the lattice results.
We then present numerical lattice computations, showing their good agreement with our predictions.

\subsection{Lattice entanglement and negativity Hamiltonians for free fermions}

On a circle of $L$ sites, let us consider the tight-binding Hamiltonian 
\begin{equation}\label{eq:tightBinding}
    H = -\sum_i \left [ c_{i}^\dagger c_{i + 1} + c_{i + 1}^\dagger c_{i} \right ] ,
\end{equation}
where the lattice fermions satisfy the canonical anti-commutation relations
\begin{equation}
    \big\{ c_i, c_j^\dagger \big\}  = \delta_{ij}, 
    \qquad 
    \big\{ c_i, c_j \big\} = \big\{ c_i^\dagger, c_j^\dagger \big\} = 0\,,
\end{equation}
and we impose either anti-periodic boundary conditions $c_{L+1} = - c_1,\, c_{L+1}^\dagger = - c_1^\dagger$ or periodic ones $c_{L+1} = c_1,\, c_{L+1}^\dagger = c_1^\dagger$.
We can write down the Hamiltonian \eqref{eq:tightBinding} in the Fourier modes $c_k, c_k^\dagger$ and the dispersion relation of the tight-binding model \eqref{eq:tightBinding} reads
\begin{equation}\label{eq:tightBindingDispers}
    H = \sum_k \varepsilon(k)\, c_{k}^\dagger c_{k}\,, \qquad \varepsilon(k) = - \cos\!\frac{2 \pi k}{L}\,,
\end{equation}
where the allowed momenta $k$ depend on the boundary conditions, i.e., in the Neveu-Schwarz sector, the momenta are semi-integer
\begin{equation}
    k = - \frac{L}{2} + \frac{1}{2}, \ldots, -\frac{1}{2}, \frac{1}{2}, \ldots, \frac{L}{2} - \frac{1}{2}\,, \quad \text{(NS)},
\end{equation}
while they are integer in the Ramond one
\begin{equation}
    k =  - \frac{L}{2} + 1, \ldots, -1, 0, 1, \ldots, \frac{L}{2}\,, \quad \text{(R)}.
\end{equation}
Notice that, when $L$ is divisible by 4, in the Ramond sector there are two zero-modes corresponding to the momenta $k = \pm \frac{L}{4}$. As discussed in \cite{kvw-17, kvw-18, fr-19, bpn-19, bgpn-19}, their presence is responsible for a non-local term in the ground state entanglement Hamiltonian. Choosing $L=(2 \mod 4)$ (i.e. divisible by 2 but not by 4), there are no zero-modes in the Ramond sector, while $k=\pm L/2$ correspond to two zero-modes in the Neveu-Schwarz sector. To simplify the discussion, in the following we will focus on the case in which $L$ is a multiple integer of 4.\\
In terms of the energy $\varepsilon(k)$ in Eq.~\eqref{eq:tightBindingDispers}, the two-point correlation matrix takes the form \cite{ryu}
\begin{equation}\label{eq:NSCorrMatrix}
    C_{i,j} = \frac{1}{L} \sum_{k = - \frac{L}{2} + \frac{1}{2}}^{\frac{L}{2} - \frac{1}{2}} \frac{e^{2\pi \ii k r / L}}{1 + e^{\beta \varepsilon(k)}}\quad \text{(NS)},\qquad    C_{i,j} = \frac{1}{L} \sum_{k = - \frac{L}{2} + 1}^{\frac{L}{2} } \frac{e^{2\pi \ii k r / L}}{1 + e^{\beta \varepsilon(k)}}\quad \text{(R)}.
\end{equation}
Since the finite temperature state of free fermions is Gaussian, we can write its reduced density matrix in the subsystem $A$ as \cite{Peschel2003}
\begin{equation}\label{eq:latticeEH}
    \rho_A = \frac{1}{Z_A} e^{-2\pi K_A} =  \frac{1}{Z_A} \exp\!\bigg \{- \sum_{i,j} c_i^\dagger h_{i,j} c_j \bigg \},
\end{equation}
where $h_{i,j}$ plays the role of the matrix kernel of the entanglement Hamiltonian $2 \pi K_A$.
For Gaussian density matrices, $h_{i,j}$ is related to the two-point correlation matrix restricted to the subsystem $A$, $C_A$, via Peschel's formula \cite{pkl-99, Peschel2009, Chung2001, Peschel2003, Peschel2004, Peschel2012}
\begin{equation}\label{eq:Peschel}
    \frac{1}{\mathbb{I}_A + e^{h}} = C_A\,,\qquad h = \log\!\left [ \frac{\mathbb{I}_A - C_A}{C_A} \right ] = \log\!\left [ \frac{\mathbb{I}_A + \Gamma_A}{\mathbb{I}_A - \Gamma_A} \right ] ,
\end{equation}
where $\mathbb{I}_A$ denotes the identity matrix in $A$ and we have introduced the covariance matrix restricted to $A$ 
\begin{equation}\label{eq:defCovariance}
    \Gamma_A = \mathbb{I}_A - 2 C_A\,.
\end{equation}
By numerically computing Eq.~\eqref{eq:Peschel} using the correlation matrices in Eq.~\eqref{eq:NSCorrMatrix}, we obtain the lattice entanglement Hamiltonian in both the Neveu-Schwarz and Ramond sector. \\
We stress that, due to the presence of the logarithm, the numerical computation of the formula \eqref{eq:Peschel} requires that the eigenvalues of $C_A$ are strictly included in $(0, 1)$. For this reason, the numerical computation must be performed at high precision; in our study we used both the software \emph{Mathematica} and the python library mpmath \cite{mpmath}, keeping up to 300 digits.\\
Peschel's formula \eqref{eq:Peschel} can be generalised to compute the negativity Hamiltonian, too. As we explained at the beginning of Sec.~\ref{sec:NH}, the partial time-reversal $\rho_A^{R_1}$ of a Gaussian density matrix is still Gaussian, and this allows us to express the lattice negativity Hamiltonian as
\begin{equation}\label{eq:latticeNH}
    \rho_A^{R_1} = \frac{1}{Z_A} e^{-2\pi \mathcal{N}_A} = \frac{1}{Z_A} \exp\!\bigg \{- \sum_{i,j} c_i^\dagger \eta_{i,j} c_j \bigg \},
\end{equation}
where now the kernel $\eta_{i,j}$ is non-hermitian. We consider a two-interval configuration $A_1 = [a_1, b_1], A_2 = [a_2, b_2]$, even though the generalisation to a multi-interval geometry is straightforward. Following \cite{ssr-17}, we write the blocks of the covariance matrix \eqref{eq:defCovariance} as
\begin{equation}\label{eq:CovarianceBlocks}
    \Gamma_A = \begin{pmatrix}
        \Gamma_A^{(1,1)}&   \Gamma_A^{(1,2)}\\
        \vspace{-.25cm}\\
        \Gamma_A^{(2,1)}&   \Gamma_A^{(2,2)}
    \end{pmatrix},
\end{equation}
where $\left(\Gamma_A^{(\sigma, \zeta)}\right)_{i, j}$ denotes $i \in A_\sigma, j \in A_\zeta$. Under partial time-reversal in, e.g., the interval $A_1$, the covariance matrix changes simply because of an imaginary factor for every index belonging to the reversed interval \cite{ssr-17}
\begin{equation}\label{eq:defCovarianceRev}
    \Gamma_A^{R_1} = \begin{pmatrix}
        -\Gamma_A^{(1,1)}&      \ii\,\Gamma_A^{(1,2)}\\
        \vspace{-.25cm}\\
        \ii\,\Gamma_A^{(2,1)}&  \Gamma_A^{(2,2)}
    \end{pmatrix}, \qquad \left (\Gamma_A^{R_1}\right )^{\dagger} = \begin{pmatrix}
        -\Gamma_A^{(1,1)}&      -\ii\,\Gamma_A^{(1,2)}\\
        \vspace{-.25cm}\\
        -\ii\,\Gamma_A^{(2,1)}& \Gamma_A^{(2,2)}
    \end{pmatrix}.
\end{equation}
The kernel of the negativity Hamiltonian $\eta_{i,j}$ in Eq.~\eqref{eq:latticeNH} is then related to the reversed covariance matrix $\Gamma_A^{R_1}$ in Eq.~\eqref{eq:defCovarianceRev} in a way analogous to Peschel's formula \eqref{eq:Peschel} for the entanglement Hamiltonian \cite{shapourian-19}
\begin{equation}\label{eq:PeschelNeg}
    \eta = \log\!\left [ \frac{\mathbb{I}_A + \Gamma_A^{R_1}}{\mathbb{I}_A - \Gamma_A^{R_1}} \right ].
\end{equation}
Since we are dealing with Gaussian states, also the relation between the twisted partial transpose $\rho^{\widetilde{R}_{1}}_A$ in Eq.~\eqref{eq:rhotilde} and $\rho^{R_{1}}_A$ can be written as
\begin{equation}    e^{\widetilde{\eta}}=\frac{\mathbb{I}_A + \Gamma_{A}^{R_1}}{\mathbb{I}_A - \Gamma_{A}^{R_1}}U_{A_1},
\end{equation}
where the matrix $U_A=-\mathbb{I}_{A_1} \oplus \mathbb{I}_{A_2}$ is related to the transformation $(-1)^{F_{A_1}}$. Therefore, the kernel of the twisted negativity Hamiltonian, $\widetilde{\eta}$, is given by
\begin{equation}\label{eq:PeschelTwist}
    \widetilde{\eta} = \log\!\left [ \frac{\mathbb{I}_A + \Gamma_A^{R_1}}{\mathbb{I}_A - \Gamma_A^{R_1}} U_{A_1} \right ].
\end{equation}
As we mentioned, a proper comparison between the field-theoretical prediction and the lattice results requires a careful limiting procedure.
To fix the ideas, we consider a configuration in which the field-theoretical entanglement Hamiltonian is local, such as a single interval on the plane in the ground state reported in Eq.~\eqref{eq:singleintervalEH} or at finite temperature in Eq.~\eqref{eq:singleintervalThermalEH}.
The Hamiltonians in Eqs.~\eqref{eq:singleintervalEH} and \eqref{eq:singleintervalThermalEH} are completely local and proportional to the energy density $T_{00}$.
In light of this, one could naively expect that the corresponding entanglement Hamiltonian on the lattice would only have non-zero contribution on the first sub-diagonals $h_{i, i+1}$, $h_{i+1, i}$. However, as studied for the first time in \cite{abch-16,ETP19}, this is not the case:
On the lattice, higher hopping terms which couple fermions at longer distances are non-negligible and they must be included in order to recover the continuum limit.
In the next subsection we review how to carry over this limit, showing why more care is needed for the case on the torus.

\subsection{Continuum limit of the entanglement Hamiltonian}\label{sec:EHlatticeLim}

The key idea to take the continuum limit is to linearise the fluctuations of the lattice fermion $c_i$ around the Fermi points $\pm k_F$ according to \cite{ETP19, giamarchibook, vds-98}
\begin{equation}\label{eq:linearFermion}
    c_i \sim \sqrt{s}\left [ e^{-\ii k_F x} \psi_L(x) + e^{\ii k_F x} \psi_R(x) \right ],
\end{equation}
where $s$ is the lattice spacing (which will be put $s=1$ in our numerical calculations) and we have introduced the continuous coordinate $x = i s$. To lighten the notation, from now on, we redefine the momenta $k$ introduced in Eq.~\eqref{eq:tightBindingDispers} as $k\to k = 2k'\pi/(L s)$, such that $k_F=\pi/(2s)$ and we restore the lattice spacing $s$.
The fields $\psi_L, \psi_R$ are respectively the left- and right-moving components of a massless Dirac fermion, which describes the scaling limit of the tight-binding model in Eq.~\eqref{eq:tightBinding}.\\
Let us divide the entanglement Hamiltonian kernel $h_{i,j}$ in Eq.~\eqref{eq:latticeEH} in matrix blocks $\left (h^{(\sigma, \zeta)} \right )_{i,j}$ such that $i \in A_\sigma, j \in A_\zeta$.
In order to recover the local term of the entanglement Hamiltonian, one needs to consider the diagonal blocks $h^{(\sigma, \sigma)}$. Following \cite{ETP19}, one substitutes the linearisation in Eq.~\eqref{eq:linearFermion} in the expression of the lattice entanglement Hamiltonian in Eq.~\eqref{eq:latticeEH}, obtaining
\begin{equation}\label{eq:locLimitDeriv1}\begin{split}
    h_{i, i+r}^{(\sigma, \sigma)} \Big[ c_i^\dagger c_{i+r} + c_{i+r}^\dagger c_{i} \Big] 
    \sim 
    &\, s\, h_{x, x+rs}^{(\sigma, \sigma)} \left [ e^{- \ii k_F r s}  \psi_L^\dagger(x) \, \psi_L(x+rs) + e^{\ii k_F r s} \,\psi_R^\dagger(x) \,\psi_R(x+rs)  \right .
    \\
    &
    \hspace{-.4cm}
    \left . +\, e^{\ii k_F (2x + rs)} \psi_L^\dagger(x) \,\psi_R(x+rs) + e^{-\ii k_F (2x + rs)} \psi_L^\dagger(x)\, \psi_R(x+rs) + \text{h.c.} \right ],
\end{split}\end{equation}
where we expressed also the matrix element $h_{x, x+rs}^{(\sigma, \sigma)}$ as a function of the continuous variable $x$. Since the massless Dirac field-theory presents conformal symmetry, one expects that in the continuum limit $s \to 0$ the right- and left-moving fermions $\psi_R, \psi_L$ will decouple. 
From Eq.~\eqref{eq:locLimitDeriv1} we can understand that the decoupling mechanism is due to the phases: the terms proportional to the product of left- and right-movers are multiplied by a strongly oscillating phase $e^{\pm \ii k_F (2x + rs)}$ and in the limit $s \to 0$, these phases will average to zero, leading to the decoupling between $\psi_L$ and $\psi_R$ \cite{ETP19}.
Dropping the highly oscillating terms and expanding in powers of $s$ both the fields $\psi_L, \psi_R$ and the matrix element $h_{x, x+rs}^{(\sigma, \sigma)}$ we find 
\begin{equation}\label{eq:locLimitDeriv2}\begin{split}
    h_{i, i+r}^{(\sigma, \sigma)} &\left [ c_i^\dagger c_{i+r} + c_{i+r}^\dagger c_{i} \right ] \approx \\
   \approx& \, s \left ( h_{x-\frac{rs}{2}, x+\frac{r s}{2}}^{(\sigma, \sigma)} + \frac{r s}{2} \partial_x h \right )2\, \cos\!\left ( k_F r s \right ) \left ( \psi_L^\dagger(x) \psi_L(x) + \psi_R^\dagger(x) \psi_R(x) \right ) + \\
   &+ s\, h_{x-\frac{rs}{2}, x+\frac{r s}{2}}^{(\sigma, \sigma)} 
   \Big [ \cos\!\left ( k_F r s \right ) r\,  s \, \partial_x\!\left ( \psi_L^\dagger(x) \psi_L(x) + \psi_R^\dagger(x) \psi_R(x) \right ) + \\
   &\hspace{3cm}- \ii\, \sin\!\left ( k_F r s \right ) r\, s \left ( \psi_L^\dagger(x) \,\partial_x \psi_L(x) - \psi_R^\dagger(x) \,\partial_x \psi_R(x) \right )  + \text{h.c.} \Big ].
\end{split}\end{equation}
We now plug the expansion~\eqref{eq:locLimitDeriv2} into Eq.~\eqref{eq:latticeEH} and we promote the sum over the index $i$ to an integral over $x$.
Integrating by parts the operator in the third row of \eqref{eq:locLimitDeriv2}, this term cancels out with the one proportional to the derivative of the matrix element $\partial_x h$ in the second row. In the second row, we recognise the number operator $N(x)$
\begin{equation}\label{eq:number}
    N(x,t) =\, : \!\! \Big[ \psi_R^\dagger(x-t) \psi_R(x-t) + \psi_L^\dagger(x+t) \psi_L(x+t)  \Big]\!\!:\, ,
\end{equation}
while in the last row the energy density $T_{00}(x)$ defined in Eq.~\eqref{eq:energydensity}. Thus, at leading order in the lattice spacing, we find that the diagonal blocks of the entanglement Hamiltonian can be written as \cite{ETP19}
\begin{equation}\label{eq:localLatticeLimit}
    \sum_i h_{i, i+r}^{(\sigma, \sigma)} \left [ c_i^\dagger c_{i+r} + c_{i+r}^\dagger c_{i} \right ] \sim \int \dd x \left [ \mathcal{S}^{\text{loc}}(x)\, T_{00} (x) + \mathcal{C}^{\text{loc}}(x)\, N(x) \right ],
\end{equation}
where 
we have introduced the weighted sums over the matrix elements \cite{ETP19}
\begin{equation}\label{eq:limSinLoc}
    \mathcal{S}^{\text{loc}}(x)\equiv
     - \,2\, s \sum_{r\geq 1} r\, \sin\!\left ( k_F r s \right ) h_{i-\frac{r}{2}, i+\frac{r}{2}}^{(\sigma, \sigma)}\,,
\end{equation}
\begin{equation}\label{eq:limCosLoc}
    \mathcal{C}^{\text{loc}}(x)\equiv h_{i, i}^{(\sigma, \sigma)}
     + \,2 \sum_{r\geq 1} \cos\!\left ( k_F r s \right ) h_{i-\frac{r}{2}, i+\frac{r}{2}}^{(\sigma, \sigma)}\,.
\end{equation}
Let us compare Eq.~\eqref{eq:localLatticeLimit} with the field-theoretical predictions for the entanglement Hamiltonian on the plane in Eq.~\eqref{eq:CHEH} or at finite temperature in Eq.~\eqref{eq:finTempEH}.
Identifying the terms proportional to the energy density $T_{00}(x)$, in \cite{ETP19} it was verified that in the case of a single interval, the sum $\mathcal{S}^{\text{loc}}(x)$ in Eq.~\eqref{eq:limSinLoc} correctly reproduces the prediction for the local entanglement temperature $\beta_\text{loc}(x)$ both in the ground state (Eq.~\eqref{eq:singleintervalEH}) and at finite temperature (Eq.~\eqref{eq:singleintervalThermalEH}).

\noindent On the other hand, the term proportional to the number operator $N(x)$ in Eq.~\eqref{eq:localLatticeLimit} is expected to vanish at half-filling  
$k_F=\frac{\pi}{2 s}$.
In this case, because of the particle-hole symmetry, the correlation matrix presents a checkerboard structure, inherited by the lattice entanglement Hamiltonian, which implies that Eq.~\eqref{eq:limCosLoc} is identically zero.
\begin{figure}[t!]
    \centering
    {\includegraphics[width=.7\textwidth]{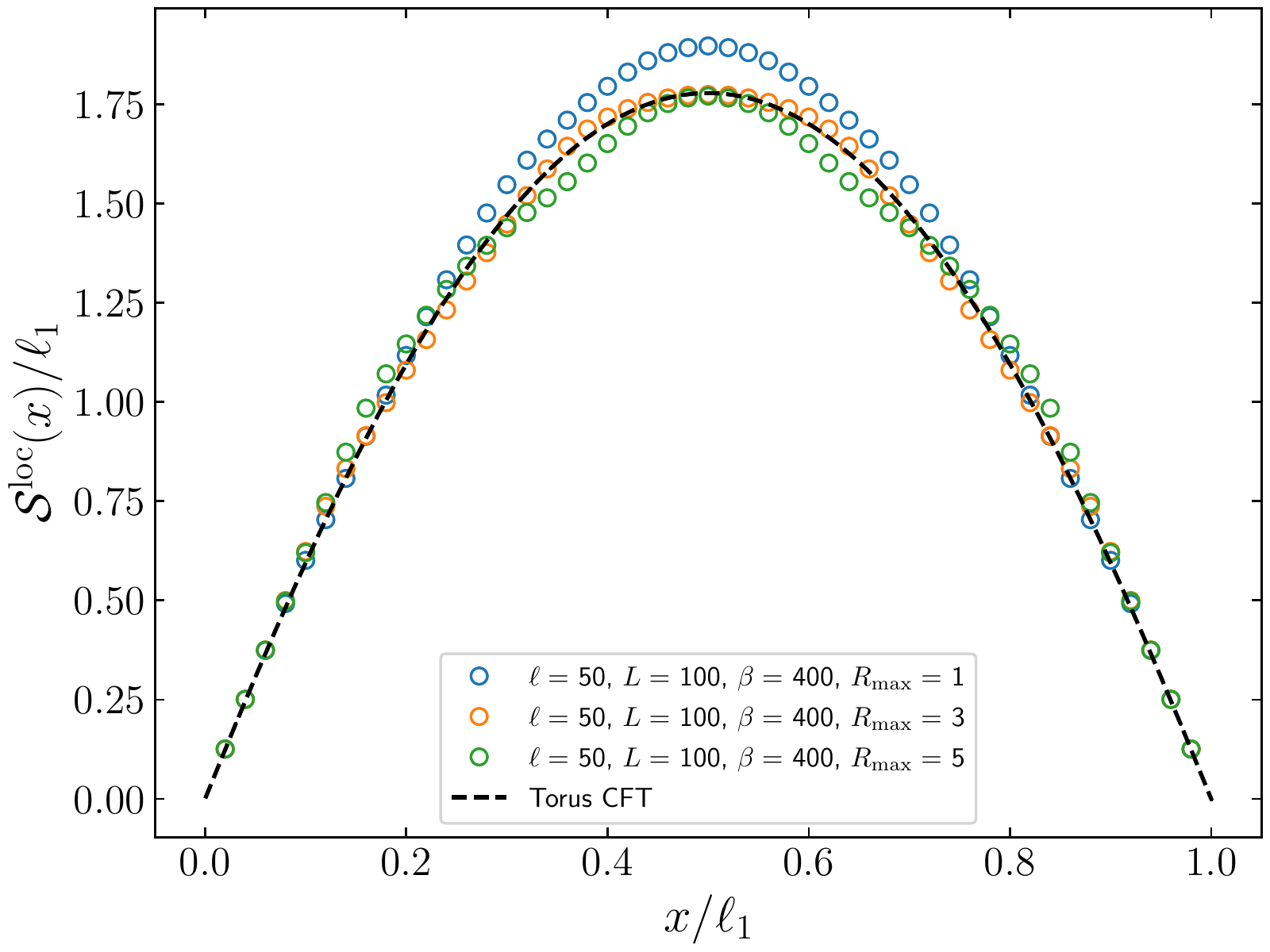}}
    \caption{Local effective temperature of the entanglement Hamiltonian for one single interval of length $\ell=50$ in a system of size $L=100$ at finite temperature $\beta=400$. Due to the presence of non-local terms, we introduce a cut-off, $R_{\mathrm{max}}$ in Eq.~\eqref{eq:limSinLoc} to recover the continuum limit (symbols). The best agreement with the theoretical prediction in Eq.~\eqref{eq:EHtorus} is obtained for $R_{\mathrm{max}}=3$. 
    }
    \label{fig:LocEntTorus}
\end{figure}

\noindent We are now interested in extending the analysis above to study free fermions on a torus, i.e. at finite temperature and size.
The derivation of \cite{ETP19} reviewed in Eqs.~\eqref{eq:locLimitDeriv1}, \eqref{eq:locLimitDeriv2}, \eqref{eq:localLatticeLimit} relies on the fact that all the matrix elements in the diagonal blocks contribute to the local term of the field-theoretical entanglement Hamiltonian \eqref{eq:localLatticeLimit}.
However, we have observed that the field-theoretical entanglement Hamiltonian in Eq.~\eqref{eq:EHtorus} contains infinite bi-local terms, even in the case of a single interval.
This implies that summing over all matrix elements $\mathcal{S}^{\text{loc}}(x)$ of Eq.~\eqref{eq:limSinLoc} gives the wrong continuum limit, since we would be also including contributions that reproduce the bi-local terms of the entanglement Hamiltonian.
\begin{figure}[t!]
    \centering
    {\includegraphics[width=.49\textwidth]{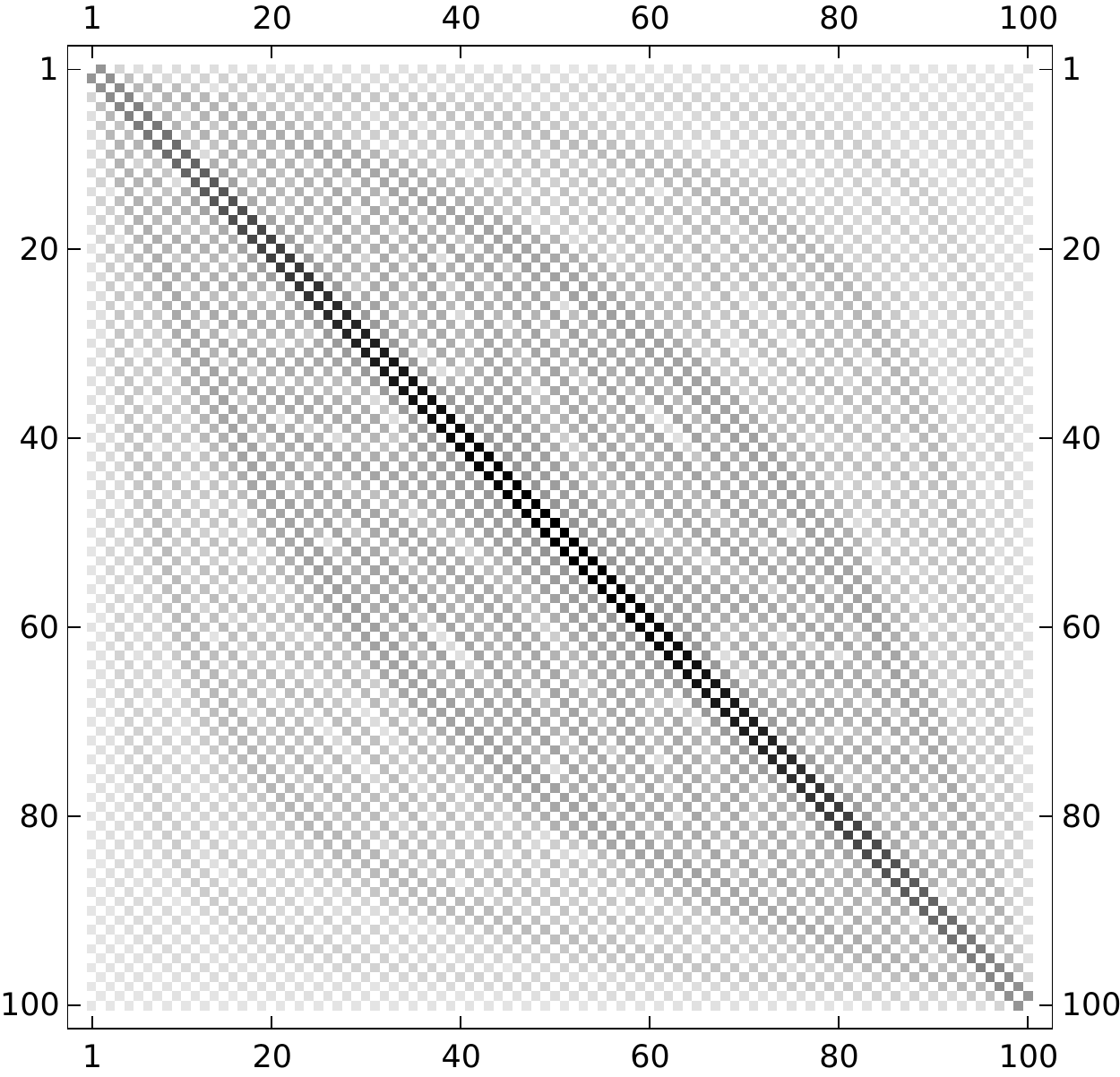}}
    \subfigure
    {\includegraphics[width=.49\textwidth]{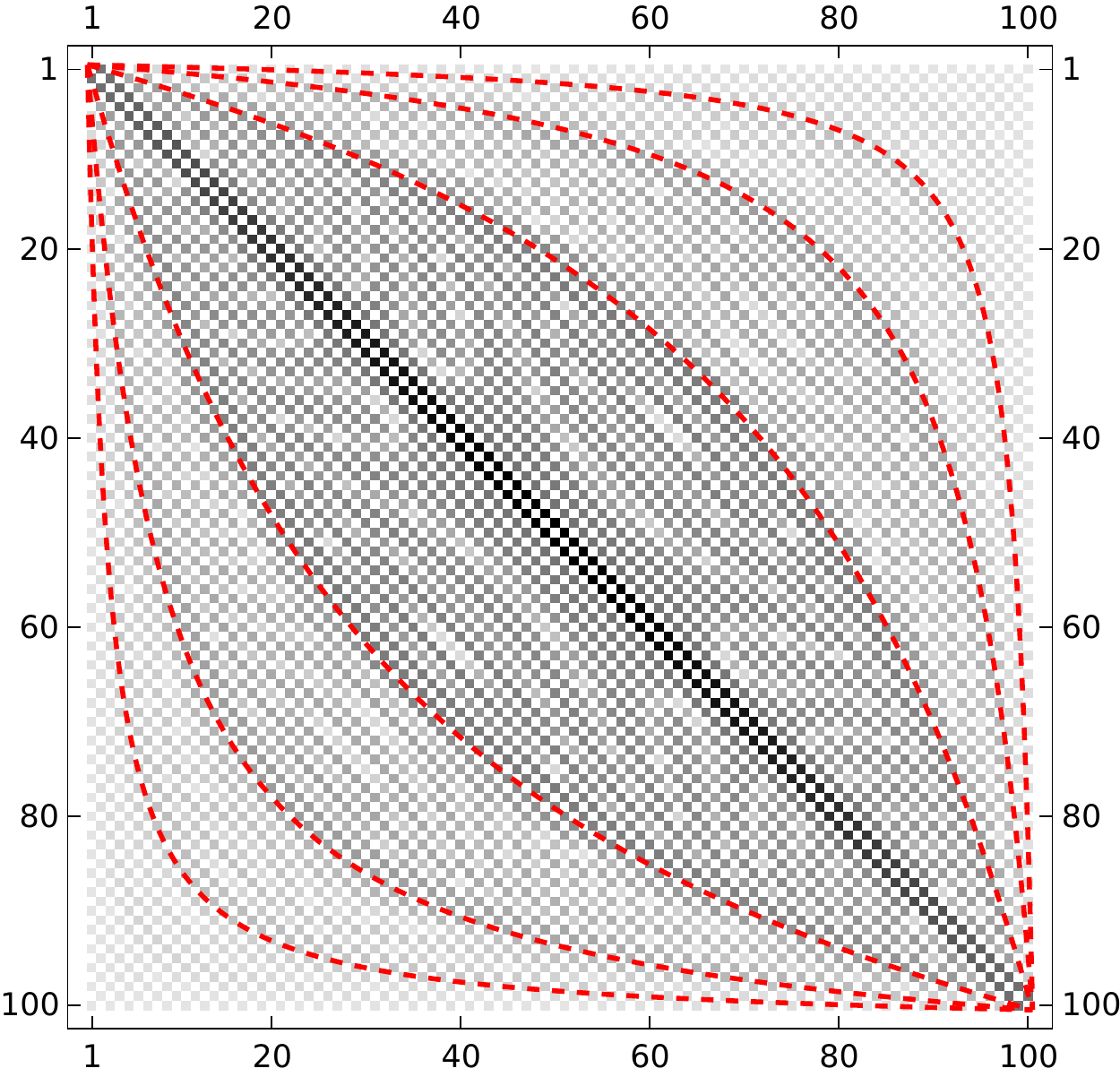}}
    \caption{Matrix plot of the kernel of the lattice entanglement Hamiltonian for one interval of length $\ell=100$ on a torus of length $L = 200$ and inverse temperature $\beta=500$ in the Neveu-Schwarz sector. We see that besides the local part around the first sub-diagonal, there are bi-local terms coupling different points, localised around the solutions $\tilde{x}_k$ of Eq.~\eqref{eq:conjTorus} (red dashed lines) for $k = \pm 1, \pm 2, \pm 3.$
    }
    \label{fig:torusEHOneMatrixplot}
\end{figure}
It is therefore necessary to introduce a maximum cut-off $R_\text{max}$ in the sum in Eq.~\eqref{eq:limSinLoc}, to only include the local contributions. We show this in Fig.~\ref{fig:LocEntTorus} for the local part of the entanglement Hamiltonian of one interval of length $\ell=50$ on the torus with $L=100$ and $\beta=400$. As we vary the cut-off $R_\text{max}$, the agreement between the lattice bi-local weight in Eq.~\eqref{eq:limSinLoc} and the theoretical prediction in Eq.~\eqref{eq:EHtorus} worsens.

\noindent This non-local behaviour is also visible in Fig.~\ref{fig:torusEHOneMatrixplot}, where we report the matrix plot of the entanglement Hamiltonian kernel $h$ obtained via Eq.~\eqref{eq:Peschel} for the case of a fermion on a torus at temperature $\beta = 500$ and system size $L = 200$ with anti-periodic boundary conditions. We see that besides the diagonal contributions, the matrix plot presents other terms located in the position of the conjugate points given by Eq.~\eqref{eq:conjTorus} for one interval.\\In \cite{ETP22}, the limit of the entanglement Hamiltonian was carried over also for the bi-local terms of a multi-interval entanglement Hamiltonian. Using Eq.~\eqref{eq:linearFermion} and again dropping the strongly oscillating contributions, we get
\begin{equation}\begin{split}\label{eq:expansionBilocal}
    c_i^\dagger h_{i,j}^{(1,2)} c_j\,
    \sim&\; s\, h_{i,j}^{(1,2)}\, \Big [e^{\ii k_F (i-j) s } \psi_L^\dagger(x) \, \psi_L(y) + e^{\ii  k_F (j-i) s } \psi_R^\dagger(x) \, \psi_R(y)  
    \\
    & \hspace{1.4cm} + e^{\ii k_F (i+j) s} \psi_L^\dagger(x)\, \psi_R(y) + e^{- \ii k_F (i+j) s} \psi_R^\dagger(x) \, \psi_L(y) \Big ]  
    \\
    \rule{0pt}{.8cm}
    =&\; \, \ii\, s\,   \sin\!\left ( k_F (j-i) s  \right ) h_{i,j}^{(1,2)} \left [ \psi_R^\dagger(x) \, \psi_R(y) - \psi_L^\dagger(x) \, \psi_L(y) \right ] 
    \\
    \rule{0pt}{.6cm}
    & \hspace{.2cm} + s\, \cos\!\left ( k_F (j-i) s \right ) h_{i,j}^{(1,2)}  \left [ \psi_R^\dagger(x) \, \psi_R(y) + \psi_L^\dagger(x) \, \psi_L(y) \right ].
\end{split}\end{equation}
In the second-to-last row we recognise the bi-local operator $T^\text{bl}(x,y)$ defined in Eq.~\eqref{eq:biloc}, while the term in the last row is proportional to a different operator $j^\text{bl}(x, y)=j^\text{bl}(x, y, 0)$ with
\begin{equation}\label{eq:bilocalNumber}\begin{split}
    j^\text{bl}(x, y, t) = \frac{1}{2} & : \!\! \Big[ \left ( \psi_R^\dagger(x-t) \psi_R(y-t) + \psi_R^\dagger(y-t) \psi_R(x-t) \right ) \\
    & \quad + \left ( \psi_L^\dagger(x+t) \psi_L(y+t) + \psi_L^\dagger(y+t) \psi_L(x+t) \right ) \Big]\!\!:\, ,
\end{split}\end{equation}
which was already identified in \cite{ETP22}.\\
In order to find the proper continuum limit, we now expand the field in position $y$ around the conjugate point $\tilde{x}_p$, keeping only the term at leading order in $s$, obtaining \cite{ETP22}
\begin{equation}\label{eq:bilocalLatticeLimit}
    \sum_i \sum_j c_i^\dagger h_{i,j}^{(1,2)} c_j \sim \int \dd x \left [ \mathcal{S}^\text{bl}(x)\, T^\text{bl}(x, \tilde{x}_p) + \mathcal{C}^\text{bl}(x)\, j^\text{bl}(x, \tilde{x}_p) \right ],
\end{equation}
where we have again promoted the sum over the row index $i$ to an integral over $x$ and we have introduced the sums \cite{ETP22}
\begin{equation}\label{eq:limSinBiloc}
    \mathcal{S}^\text{bl}(x) \equiv \sum_{j \in A_2} \sin\!\left ( k_F (j-i) s \right ) h_{i,j}^{(1,2)} ,
\end{equation}
\begin{equation}\label{eq:limCosBiloc}
    \mathcal{C}^\text{bl}(x) \equiv \sum_{j \in A_2} \cos\!\left ( k_F (j-i) s \right ) h_{i,j}^{(1,2)} .
\end{equation}
If we now compare the limit of the off-diagonal blocks in Eq.~\eqref{eq:bilocalLatticeLimit} with the bi-local terms of the field-theoretical entanglement Hamiltonian \eqref{eq:EHtorus}, we see that the sum $\mathcal{S}^\text{bl}(x)$ in Eq.~\eqref{eq:limSinBiloc} needs to reproduce the bi-local weight, since they are both proportional to the bi-local operator $T^\text{bl}$ \eqref{eq:biloc}.
Analogously to what happens in the local case, we expect that the sum $\mathcal{C}^\text{bl}(x)$ \eqref{eq:limCosBiloc} multiplying the new operator $j^\text{bl}(x)$ \eqref{eq:bilocalNumber} vanishes, since such an operator does not appear in the field-theoretical entanglement Hamiltonian \eqref{eq:EHtorus}.
Also in the off-diagonal blocks, at half-filling $k_F = \frac{\pi}{2 s}$, the checkerboard structure of the lattice entanglement kernel $h$ implies that Eq.~\eqref{eq:limCosBiloc} vanishes identically, simplifying the calculations.
\subsection{Negativity Hamiltonian}\label{sec:NHlatticeLim}
\begin{figure}[t!]
    \centering
    {\includegraphics[width=.49\textwidth]{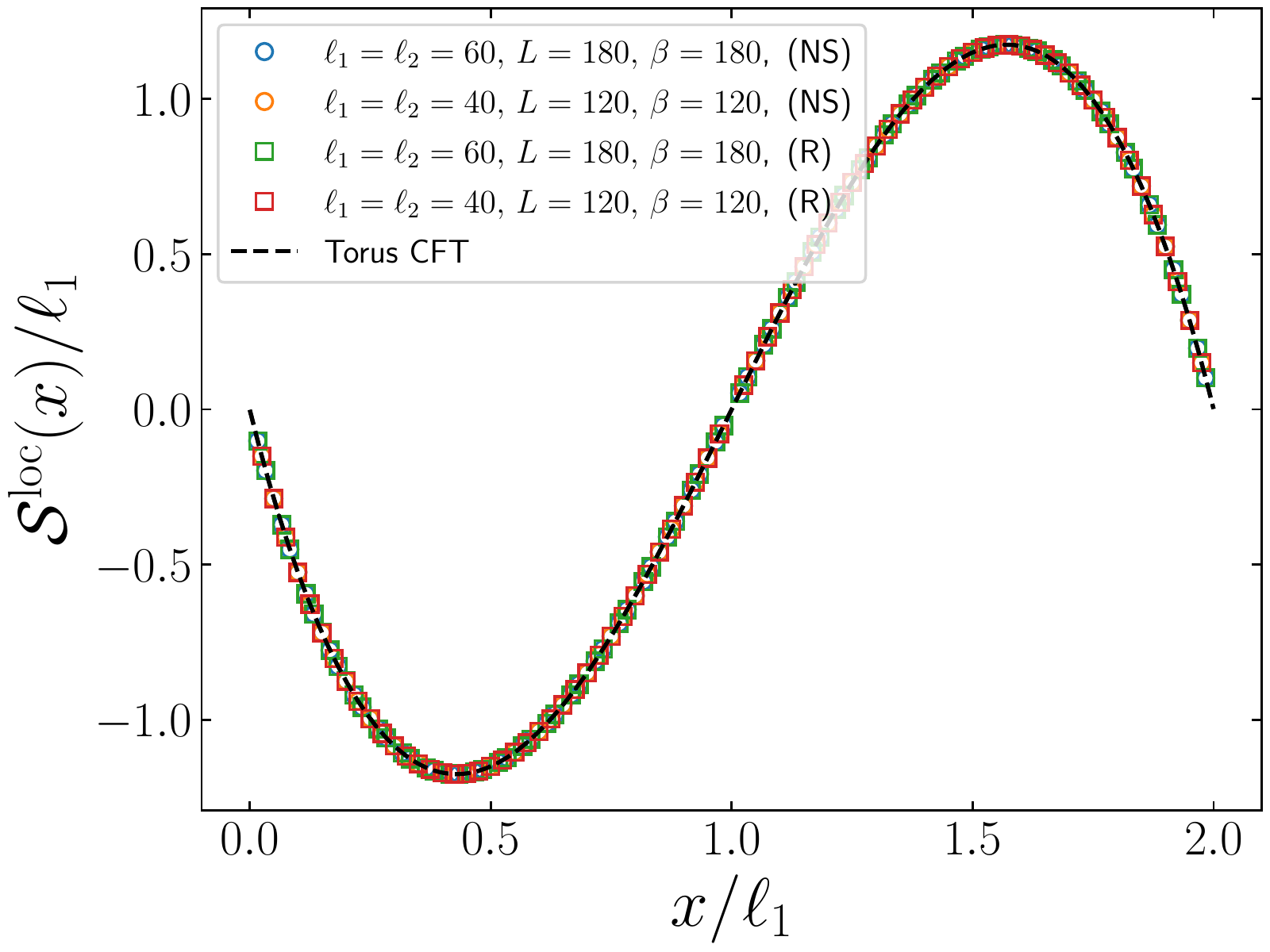}}
    \subfigure
    {\includegraphics[width=.49\textwidth]{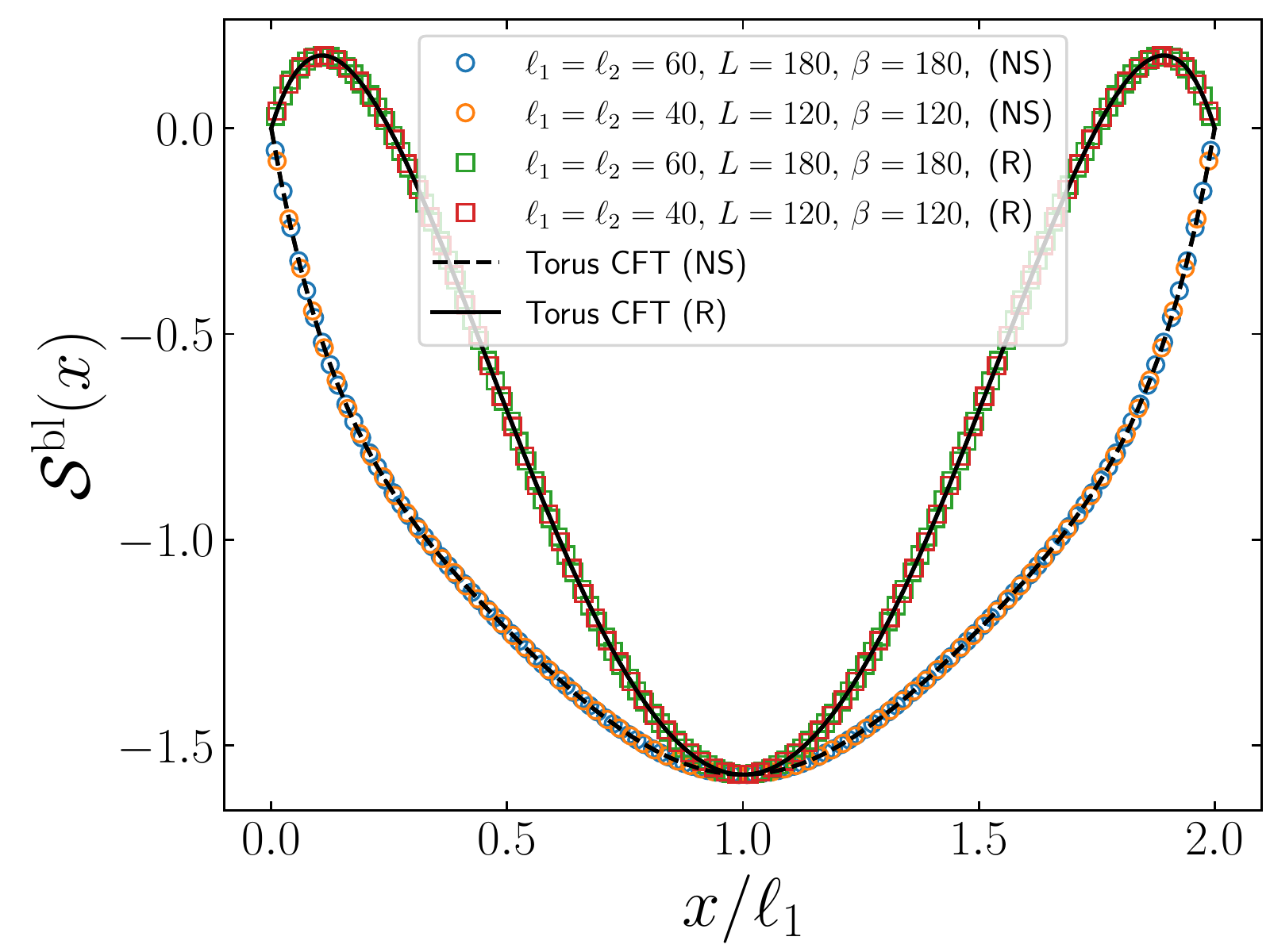}}
    \caption{Benchmark of the analytical prediction for the negativity Hamiltonian of adjacent blocks of equal length on the torus for a Dirac fermion. 
    In the left panels, the symbols are obtained from Eq.~\eqref{eq:limSinLoc} while the dashed lines correspond to Eq.~\eqref{eq:inversenegtemperature}, rescaled by $\ell_1$ in order to show the collapse for different sizes.  
    In the right panel, we perform the same analysis for the bi-local part of the negativity Hamiltonian in the same geometry. The symbols are obtained from Eq.~\eqref{eq:limitBilocDiagNeg} while the dashed line corresponds to 
    the weight function in the bi-local term in Eq.~\eqref{eq:ramond} and \eqref{eq:NS} for Ramond and Neveu-Schwarz boundary condition, respectively.}\label{fig:LocNegTorus_tripartito_intervalliuguali}
\end{figure}
In \cite{rmtc-23} it was argued that the limiting procedure of the lattice entanglement Hamiltonian $h_{i,j}$ reviewed in the previous section is almost identical to the one of the lattice negativity Hamiltonian $\eta_{i,j}$ of Eqs.~\eqref{eq:latticeNH}, \eqref{eq:PeschelNeg}.
Indeed, the limit only depends on the expansion of the lattice fermion of Eq.~\eqref{eq:linearFermion}, which is identical also for the negativity Hamiltonian.
The only difference is due to the presence of the imaginary factors $\ii^{\Theta_1(x)} (-\ii)^{\Theta_1(\tilde{x}_p)}$ in Eq.~\eqref{eq:limSinBiloc}.\\
This allows us to extract the negativity temperature from the lattice, in order to check our predictions of Sec.~\ref{sec:NH}. The weight function of the local term can be read from Eq.~\eqref{eq:limSinLoc}, while the bi-local terms take different signs and imaginary factors in different intervals. In order to compare the continuum limit of the lattice negativity Hamiltonian, in the special case of two intervals, Eq.~\eqref{eq:limSinBiloc} must be modified as follows
\begin{equation}\label{eq:limitBilocDiagNeg}
    \mathcal{S}^\text{bl}(x) = \begin{cases}
       \;  - \ii \sum_{j } \sin\!\left ( k_F (j-i) s \right ) \eta_{i,j}^{(1,2)},   \;\;\; &x \in [a_1, b_1],
        \\
        \rule{0pt}{.8cm}
       \;  \ii \sum_{j } \sin\!\left ( k_F (j-i) s \right ) \eta_{i,j}^{(2,1)},    &x \in [a_2, b_2]\,.
    \end{cases}
\end{equation}
Also for the negativity, at half-filling Eqs.~\eqref{eq:limCosLoc}, \eqref{eq:limCosBiloc} vanish identically. Now we can study the continuum limit of Eqs.~\eqref{eq:limSinLoc} and \eqref{eq:limitBilocDiagNeg} 
to check the field theory predictions for the negativity Hamiltonian, 
Eq.~\eqref{eq:NHtorus}, for two disjoint intervals at finite temperature and size, in different regimes and both in a tripartite and bipartite geometry.
\begin{figure}[t!]
    \centering
    {\includegraphics[width=.49\textwidth]{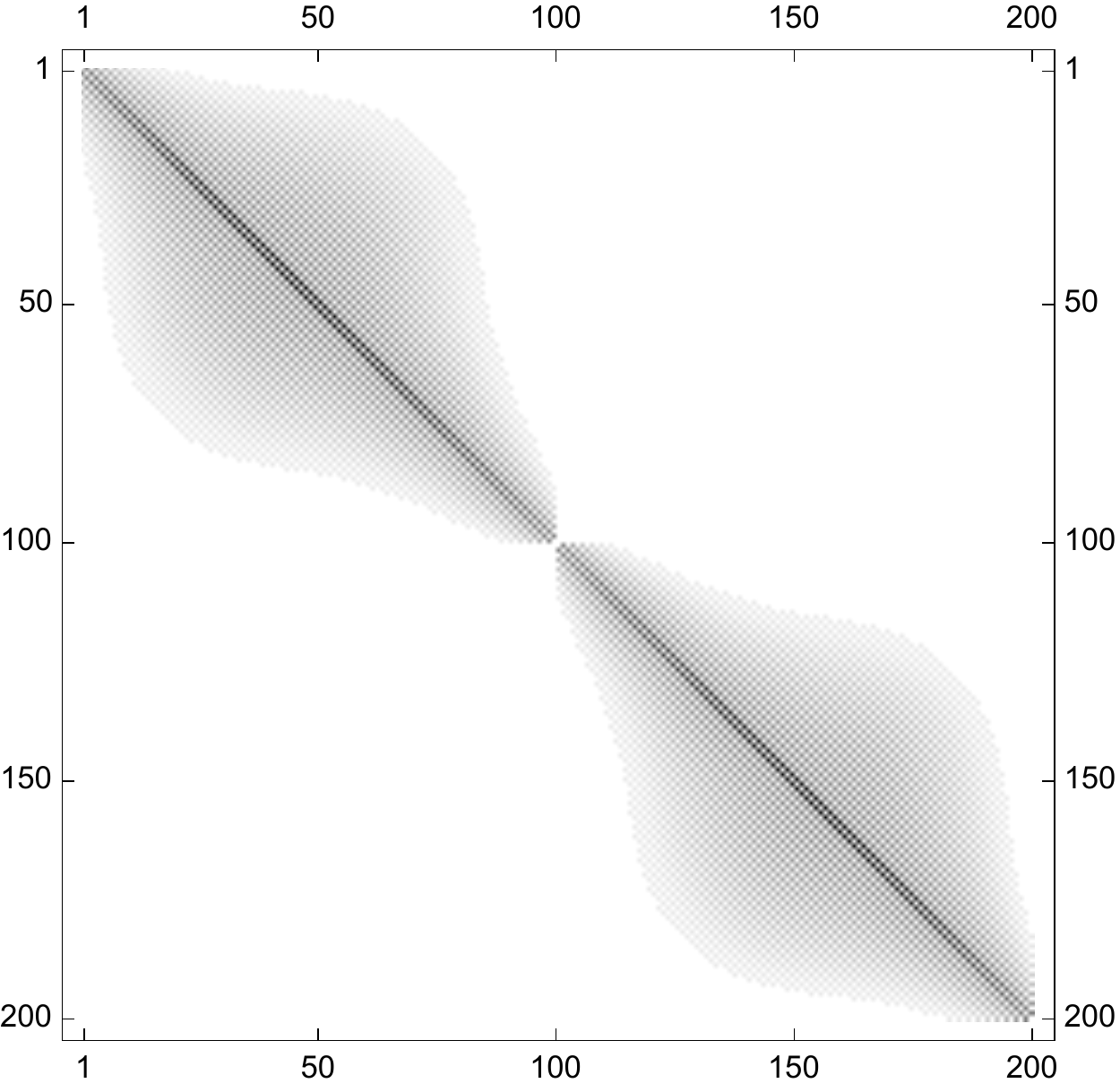}}
    \subfigure
    {\includegraphics[width=.49\textwidth]{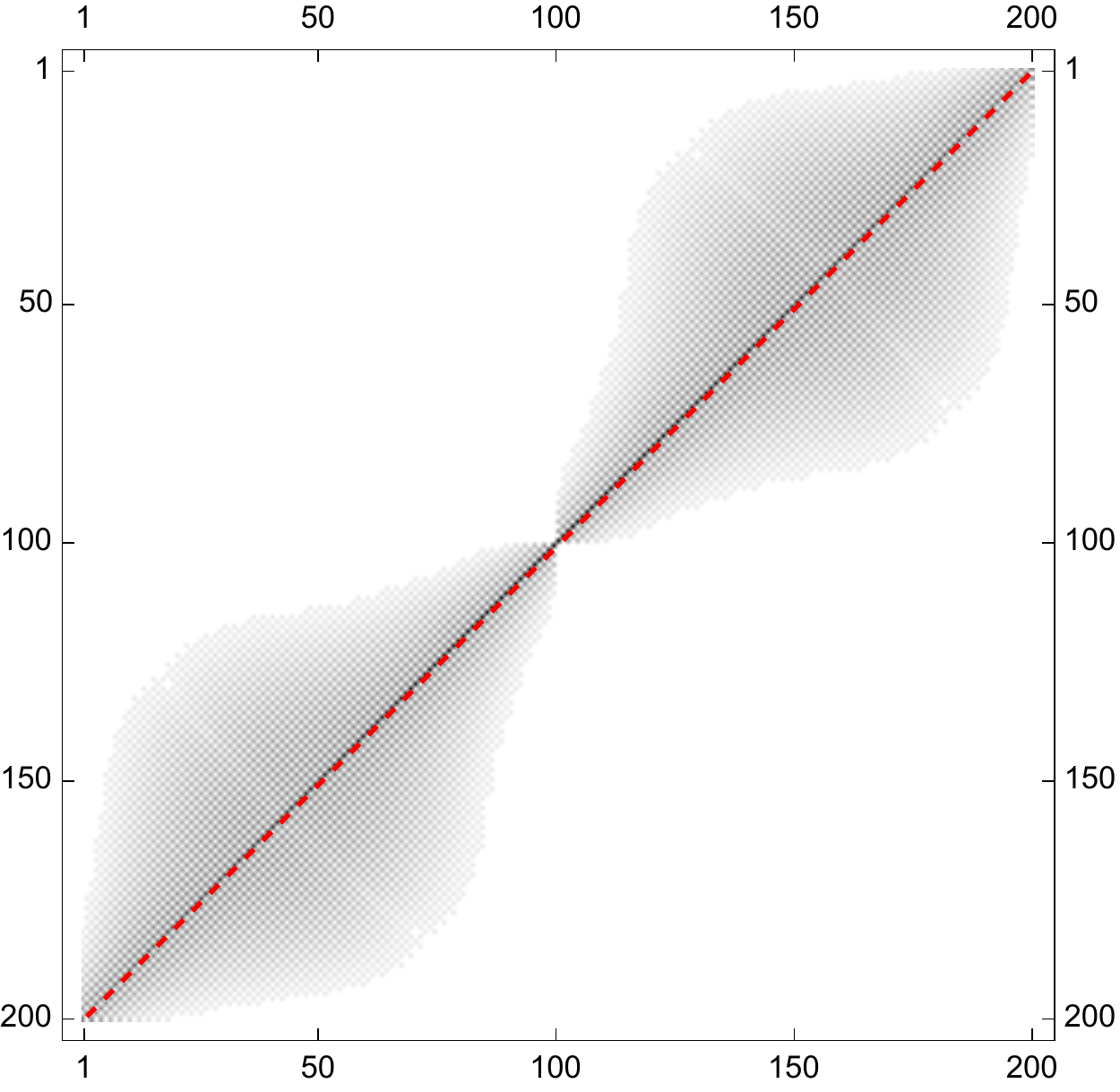}}
    \caption{Matrix elements of the negativity Hamiltonian kernel $\eta$ for two adjacent intervals of equal size, $\ell_1=\ell_2=100$, in a system of size $L=300$ and (inverse) temperature $\beta=300$ with Neveu-Schwarz boundary conditions. The left panel corresponds to the real local part, while the right panel is the bi-local contribution.
    The dashed lines correspond to the only conjugate point obtained by solving Eq.~\eqref{eq:conjTorusRev}.
    }
    \label{fig:torusNHMatrixplot}
\end{figure}

\noindent In Fig.~\ref{fig:LocNegTorus_tripartito_intervalliuguali} we consider two adjacent intervals of equal length $\ell_1=\ell_2$, for several values of $\ell_1$ and system size $L$ and for different values of $\beta$, both with NS and R boundary conditions.
In the left panel we find that the sum $\mathcal{S}^\text{loc}$ over the higher hoppings is in perfect agreement with the field-theoretical local effective inverse temperature in Eq.~\eqref{eq:TorusZetaRevTripartite}.
In the right panel, we report a similar analysis for the non-local term of the negativity Hamiltonian for the same geometry: we compare $\mathcal{S}^\text{bl}$ in Eq.~\eqref{eq:limitBilocDiagNeg} with the field-theoretical weight function occurring in the bi-local term of the negativity Hamiltonian in Eq.~\eqref{eq:NHtorusEqual}, finding a good agreement. We stress that this geometry is quite interesting because the infinite non-local terms of the negativity Hamiltonian collapse on each other
and we recover a bi-local structure, as we discussed in Sec.~\ref{sec:NHtorus}.
This is also clear by studying the matrix plot of the kernel of the negativity Hamiltonian in Fig.~\ref{fig:torusNHMatrixplot} for two intervals of equal length, $\ell_1=\ell_2=100$, $L=\beta=300$, where the left panel corresponds to the real local part of Eq.~\eqref{eq:NHtorusEqual} while the right panel describes the bi-local imaginary contribution. The structure differs from the one for the entanglement Hamiltonian shown in Fig.~\ref{fig:torusEHOneMatrixplot} and the dashed lines corresponds to the position of the single conjugate point.

\begin{figure}[t!]
    \centering
    {\includegraphics[width=.49\textwidth]{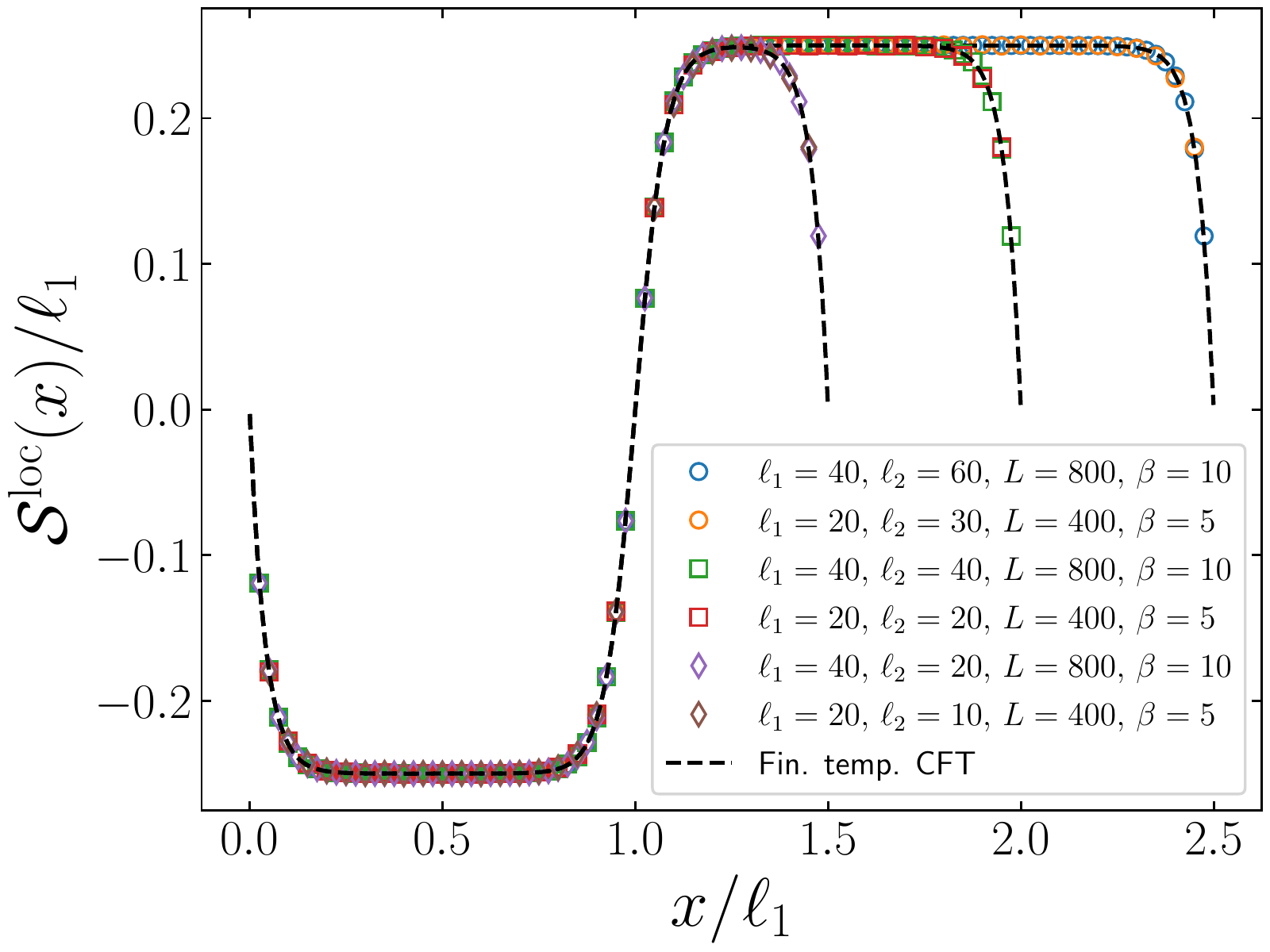}}
    \subfigure
     {\includegraphics[width=.49\textwidth]{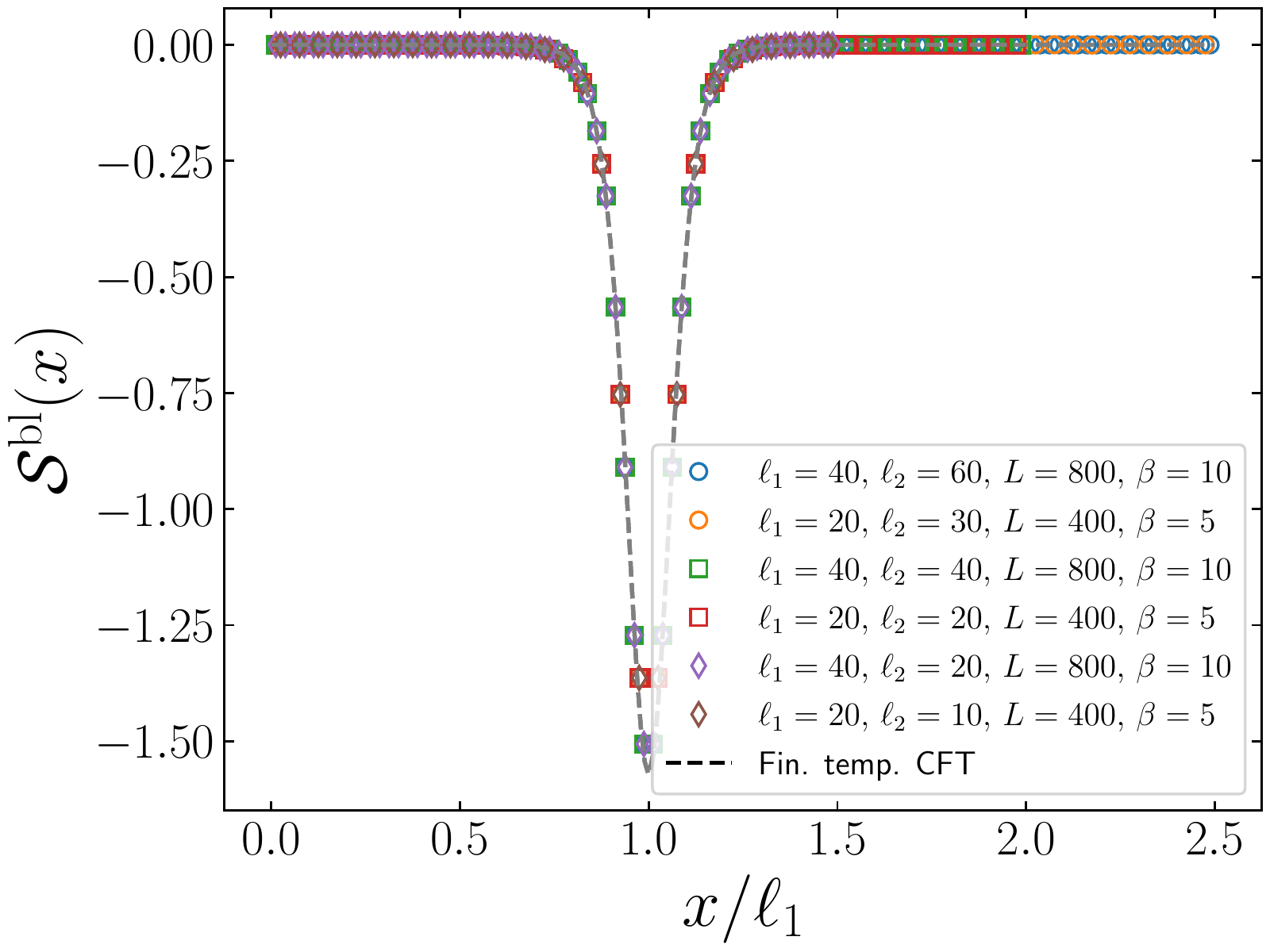}}
    \caption{Local (left) and bi-local (right) inverse effective temperature of the negativity Hamiltonian, rescaled with $\ell_1$ as a function of $x/\ell_1$. 
    The geometry we consider is $A=[1,\ell_1] \cup [\ell_1+1,\ell_1+\ell_2]$ for different values of the ratio $\ell_1/\ell_2 = 0.5, 1, 1.5$. Here we fix the system size as $L/\ell_1=20$ and we rescale the inverse temperature $\beta$ such that $\beta/\ell_1=1/4$.
    The data points are obtained by applying Eq.~\eqref{eq:limSinLoc} (Eq.~\eqref{eq:limitBilocDiagNeg}) in the left (right) panel while 
    the dashed curves correspond to the prediction in Eq.~\eqref{eq:betatripinfinite} (Eq.~\eqref{eq:tildeTripartitoThermalbilocl}).}    \label{fig:LocNegTh_tripartito_intervallidiversi_HighT}
\end{figure}

\noindent In Fig.~\ref{fig:LocNegTh_tripartito_intervallidiversi_HighT}, we consider again two intervals for different ratios of the length $\ell_2/\ell_1 = 0.5, 1, 1.5$, with $\beta/\ell_1=1/4$. Here the system size is $L=20\, \ell_1$, but since $L\gg \beta$, this amounts to study a thermal tripartite geometry on the infinite line, whose analytical predictions are reported in Eq.~\eqref{eq:finTempNH}. Indeed, both the left and the right panels confirm what we find analytically in Eqs.~\eqref{eq:betatripinfinite} and \eqref{eq:tildeTripartitoThermalbilocl} for the local and bi-local terms of the negativity Hamiltonian, respectively.

\begin{figure}[t!]
    \centering
    {\includegraphics[width=.49\textwidth]{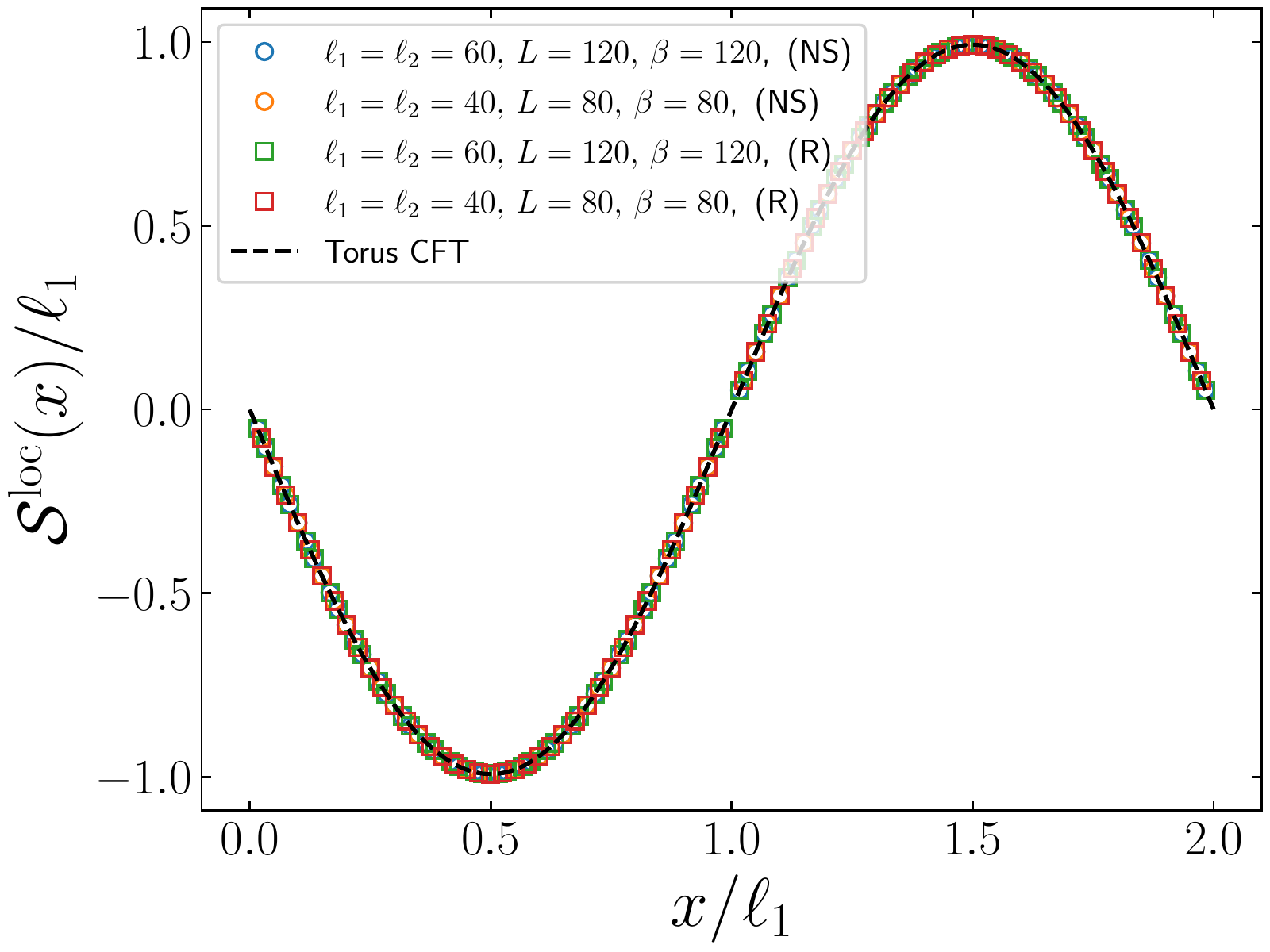}}
    \subfigure
    {\includegraphics[width=.49\textwidth]{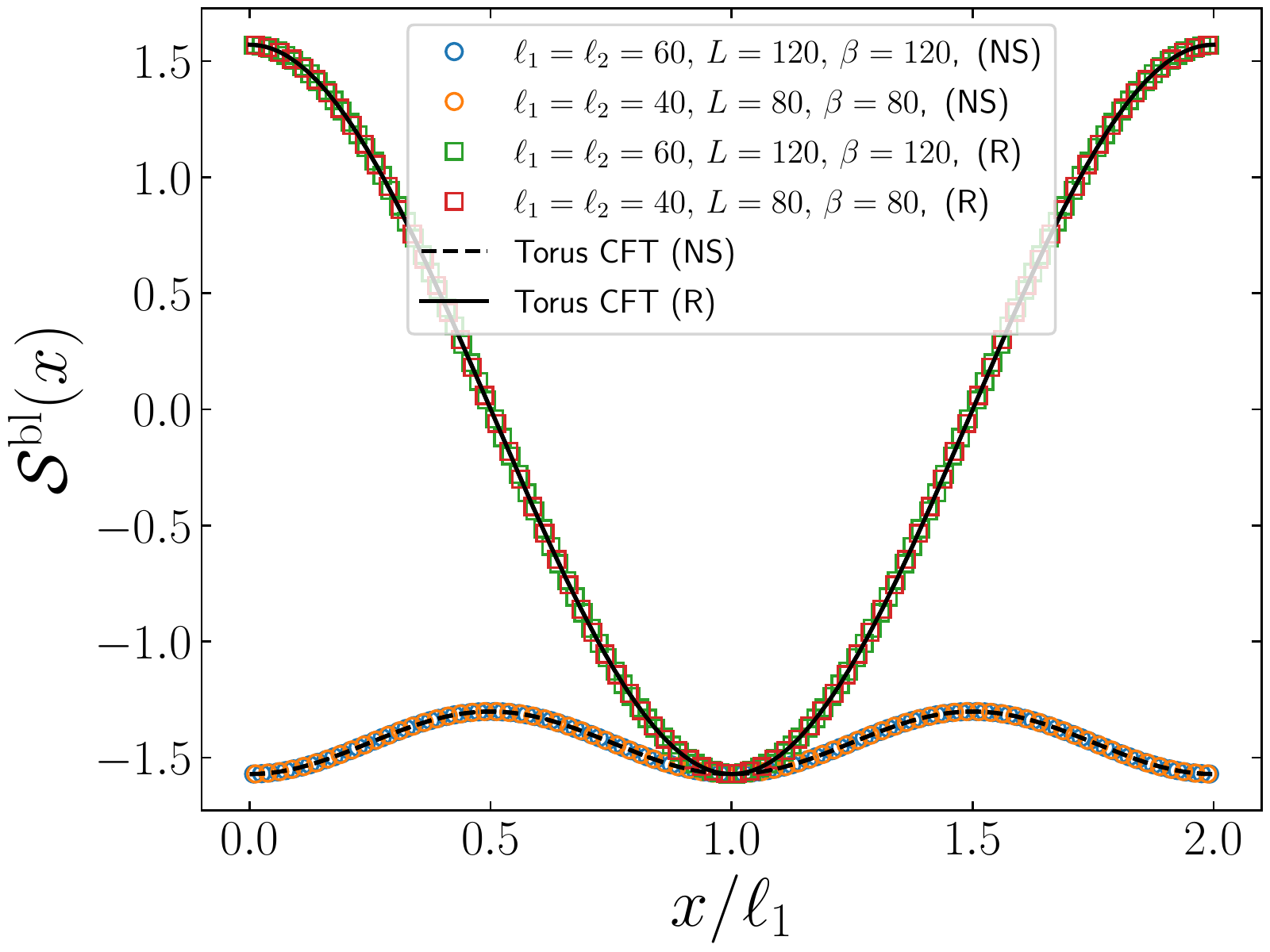}}
    \subfigure
    {\includegraphics[width=.49\textwidth]{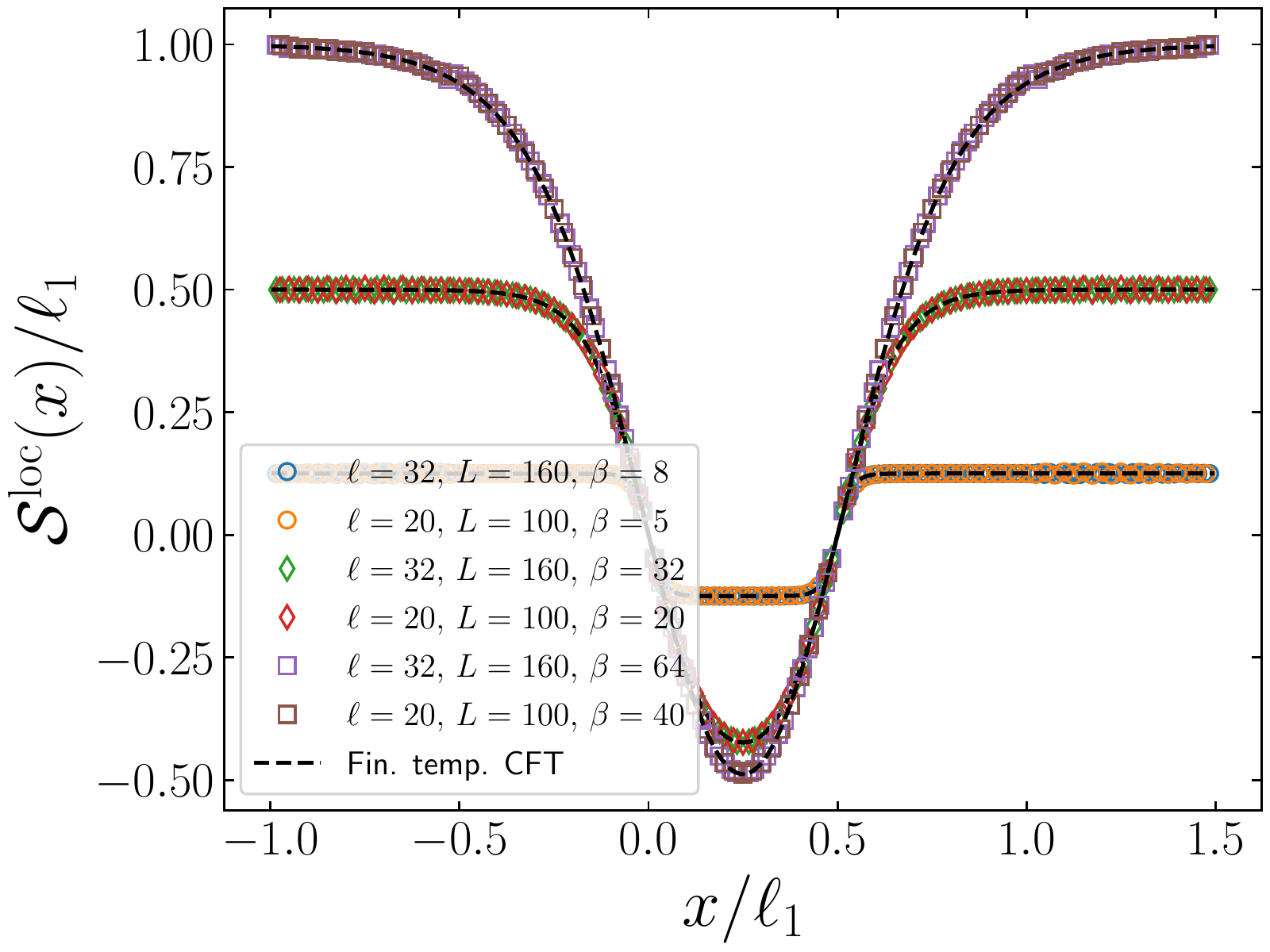}}
    \subfigure
    {\includegraphics[width=.49\textwidth]{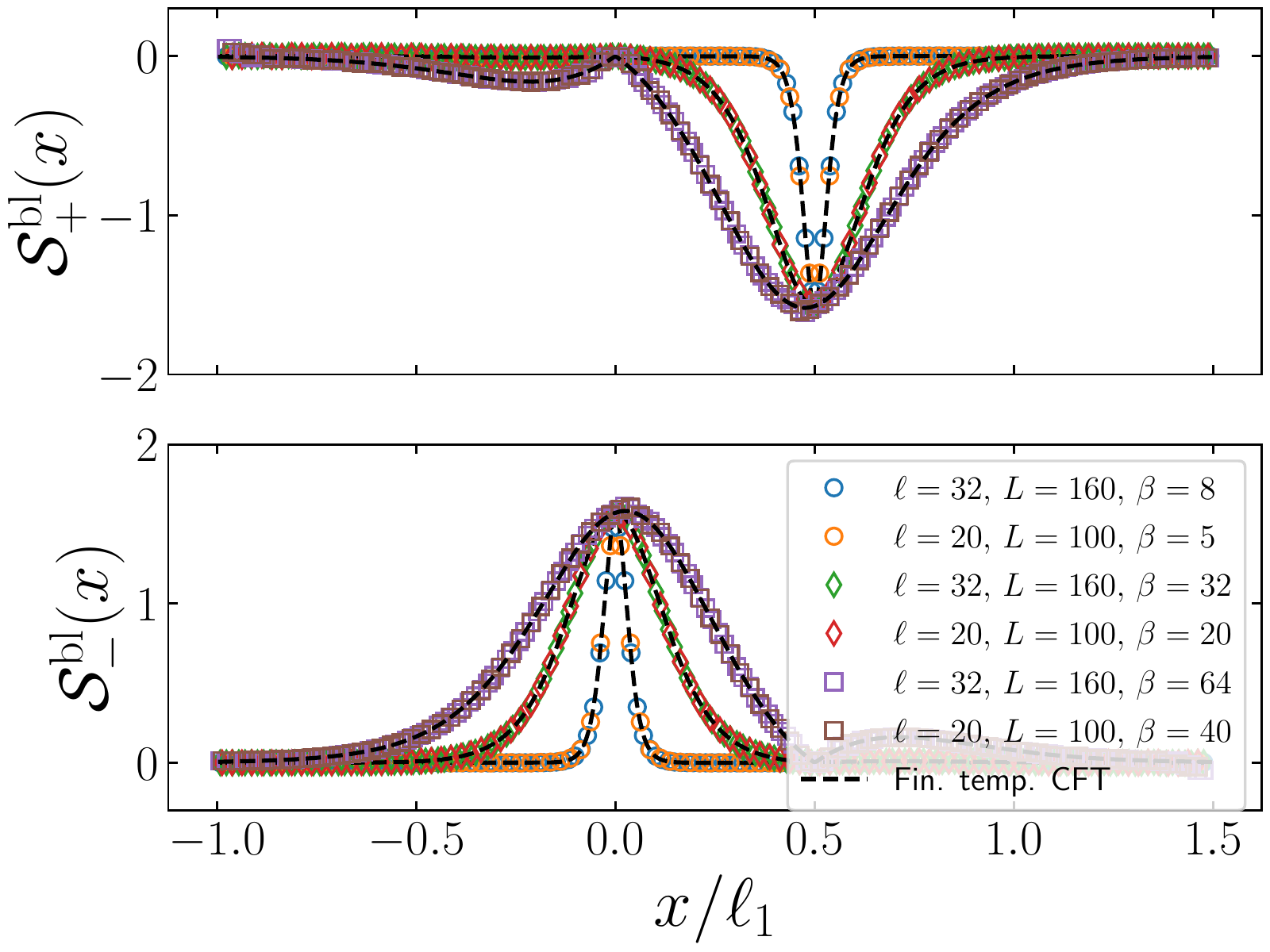}}
    \caption{Top panels: Local and bi-local weight functions of the negativity Hamiltonian in the left and right panel, respectively. The geometry we are considering is a bipartition of a system of size $L$ into two intervals of equal length, $\ell_1=\ell_2=L/2$, at inverse temperature $\beta=L$. The dashed line corresponds to Neveu-Schwarz boundary conditions, while the solid line describes a system with Ramond boundary conditions. The theoretical prediction are Eq.~\eqref{eq:localTorusRevBipartite} (left) and Eq.~\eqref{eq:bilocalTorusRevBipartite} (right). Bottom panels: same analysis as above, for the geometry $A=[-\ell_2/2,0]\cup[1,\ell_1] \cup [\ell_1+1,\ell_1+\ell_2/2]$, with $\ell_2=L-\ell_1$ and $A_2=[1,\ell_1]$. It corresponds to a bipartite case, where now we fix $L\gg \beta$, such that in the left panel we can use our theoretical prediction in Eq.~\eqref{eq:betaloc_bipartiteNH}(left) and Eq.~\eqref{eq:bipartiteThermalbilocl} (right).}   \label{fig:LocNegTorus_biipartito_intervalliuguali}
\end{figure}

\noindent Before concluding the section, we want to check also the results for a bipartite geometry found in Sec.~\ref{sec:bipartite}.
In the top panels of Fig.~\ref{fig:LocNegTorus_biipartito_intervalliuguali}, we consider a bipartition of a system of size $L$ into two intervals of equal length, $\ell_1=\ell_2=L/2$, at inverse temperature $\beta=L$. This choice is particularly convenient because from Eq.~\eqref{eq:conjTorusRevBipartite} we can deduce that the infinite non-local terms are suppressed.
Both the local and the bi-local component of the negativity Hamiltonian are in good agreement with Eq.~\eqref{eq:localTorusRevBipartite} and Eq.~\eqref{eq:bilocalTorusRevBipartite}, respectively.
In the bottom panels, we consider a different geometry, $A=[-\ell_2/2,0]\cup[1,\ell_1] \cup [\ell_1+1,\ell_1+\ell_2/2]=A_1\cup A_2\cup A_3$, with $\ell_2=L-\ell_1$ and we perform a partial transpose operation with respect to the middle interval $A_2=[1,\ell_1]$. Since now $A$ consists of three intervals, in the limit $L\to \infty$, we have two conjugate points $\tilde{x}_{\pm}^R$  given by Eq.~\eqref{eq:conjugate}. We can find the continuum limit by studying 
\begin{equation}
\begin{split}
    &\mathcal{S}^\text{bl}_+(x) \equiv (-\ii)^{\delta_{\sigma,2}}(\ii)^{\delta_{\zeta,2}} \sum_{j \in A_{\zeta}} \sin\!\left ( k_F (j-i) s \right ) \eta_{i,j}^{(\sigma,\zeta)} , \quad (\sigma,\zeta) \in \{(1,3), (2,3), (3,2)\}, \\
     &\mathcal{S}^\text{bl}_-(x) \equiv (-\ii)^{\delta_{\sigma,2}}(\ii)^{\delta_{\zeta,2}} \sum_{j \in A_{\zeta}} \sin\!\left ( k_F (j-i) s \right ) \eta_{i,j}^{(\sigma,\zeta)}, \quad (\sigma,\zeta) \in \{(1,2), (2,1), (3,1)\}.\\
    \end{split}
\end{equation}
We observe a good agreement with Eq.~\eqref{eq:betaloc_bipartiteNH} for the local part (left) and Eq.~\eqref{eq:bipartiteThermalbilocl} for the bi-local weight (right).

\subsection{Twisted negativity Hamiltonian}

While for the entanglement and negativity Hamiltonians we presented both known and novel field-theoretical predictions and we could compare them with the continuum limit of the lattice results, for the twisted negativity Hamiltonian defined in Eq.~\eqref{eq:defNegHamFermiontilte}, there are no field theory results.
To avoid confusion with the notation, we stress that we define the negativity Hamiltonian related to $\rho_A^{R_{1}}$ as $\mathcal{N}_A$  and the one related to $\rho_A^{\widetilde{R}_{1}}$ as $\widetilde{\mathcal N}_A$. The advantage of studying $\rho_A^{\widetilde{R}_{1}}$ is that it is an Hermitian operator, so the logarithmic negativity recovers its original meaning of measure of the negativeness of the eigenvalues.
Although we do not manage to derive its form theoretically, we perform a numerical study on the lattice using the limiting procedure described in Sec.~\ref{sec:EHlatticeLim}.
This allows us to identify which operators appear in the continuum limit of the lattice twisted negativity Hamiltonian and we can formulate a conjecture for the local weight functions in the case of two identical intervals on the plane. We comment that this approach allows us to identify all the operators appearing in $\widetilde{\mathcal N}_A$, contrarily to the analysis done in \cite{mvdc-22}, where only the nearest neighbour
negativity Hamiltonian has been considered. 

\subsubsection{Twisted negativity Hamiltonian on the plane}

\begin{figure}[t!]
    \centering
    {\includegraphics[width=.49\textwidth]{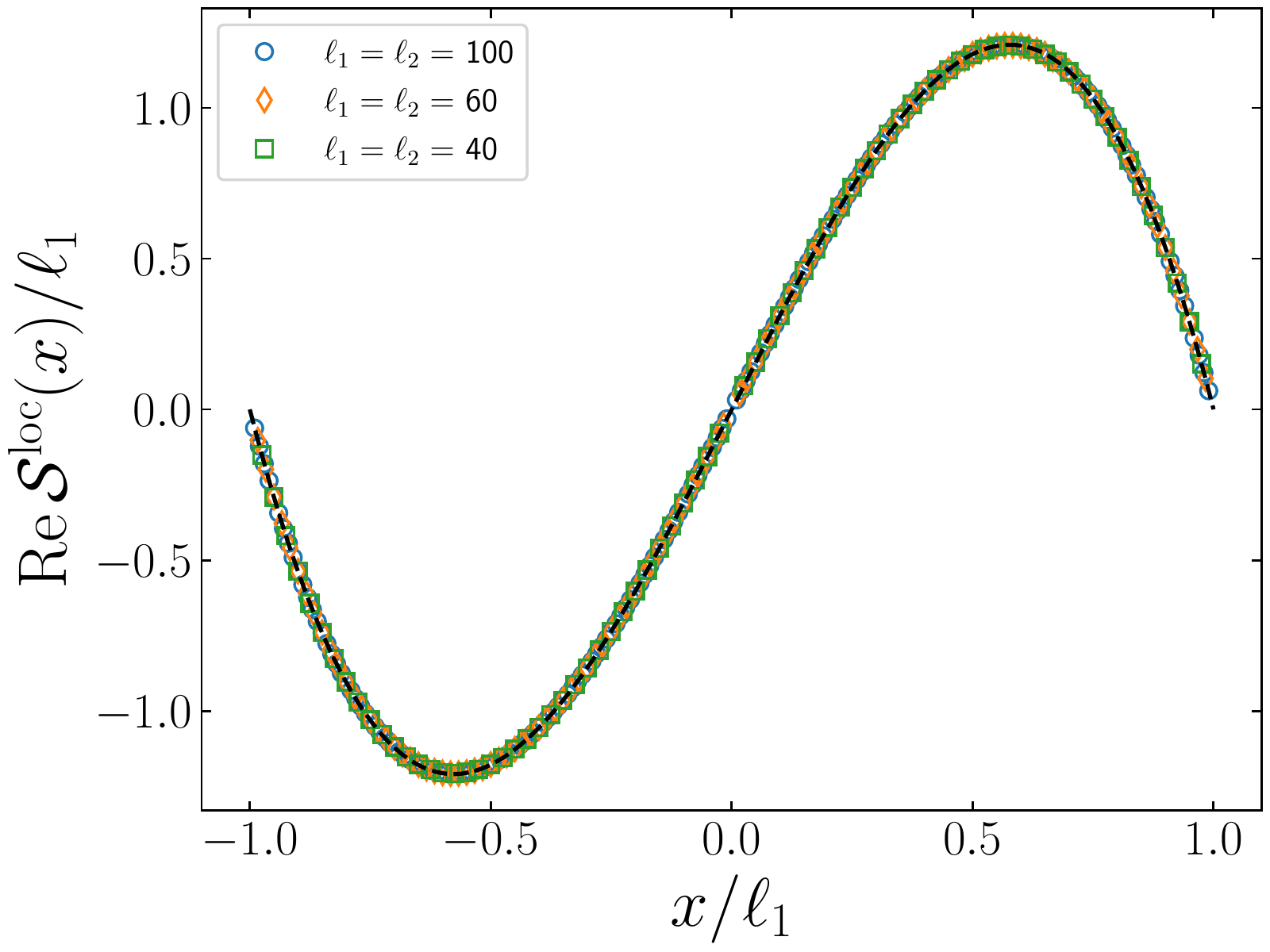}}
    \subfigure
    {\includegraphics[width=.49\textwidth]{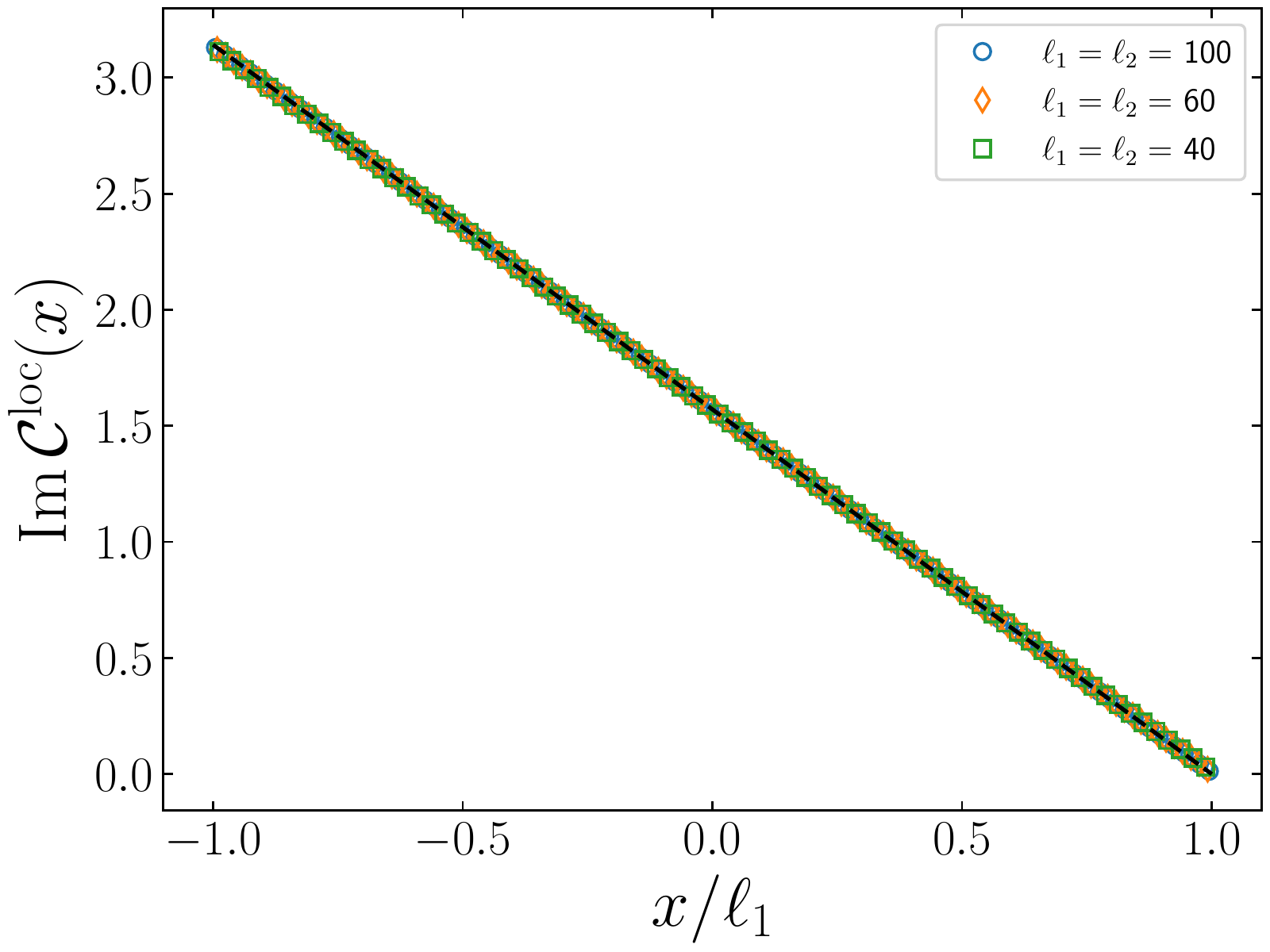}}
    \subfigure
    {\includegraphics[width=.49\textwidth]{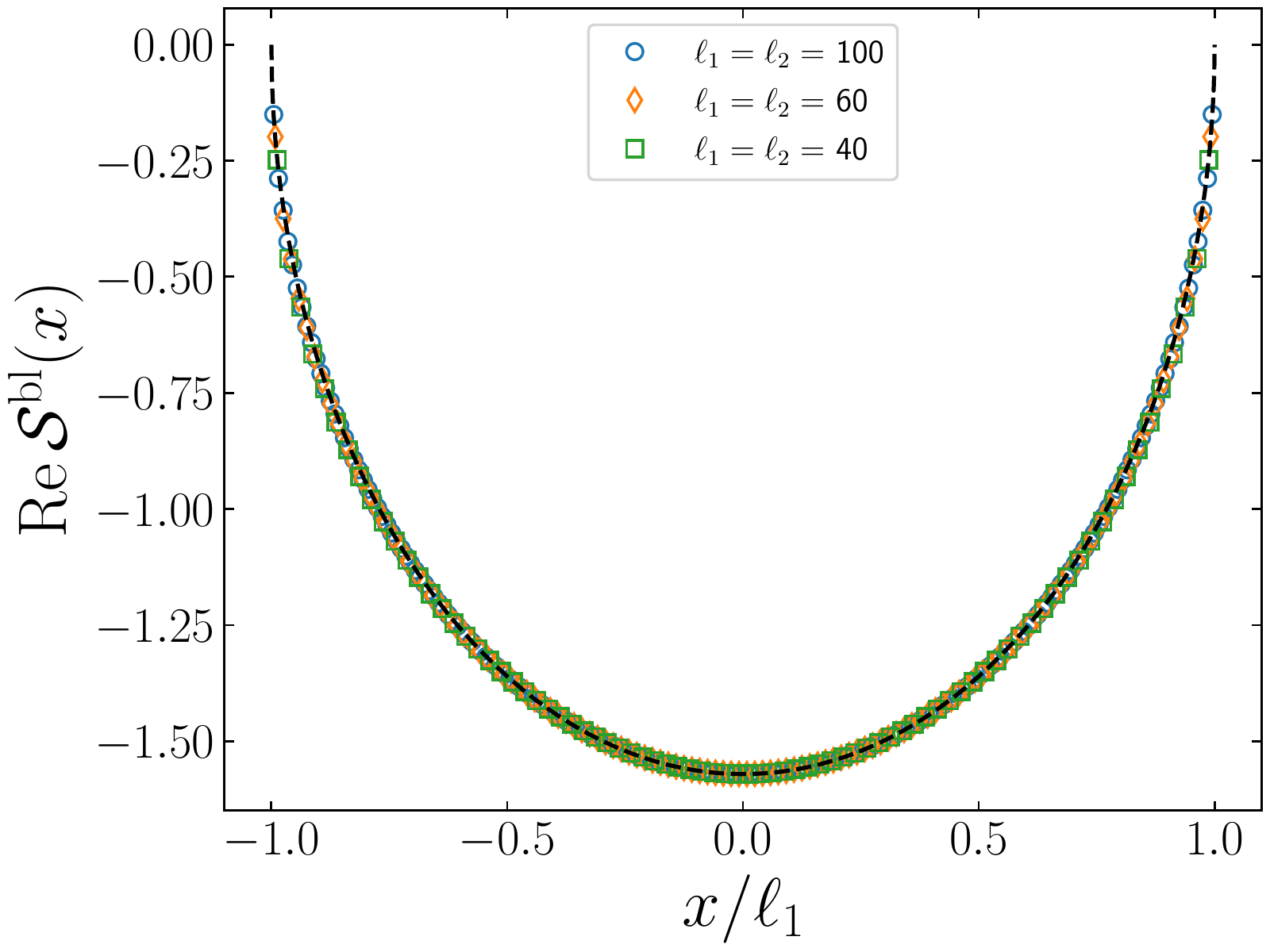}}
    \subfigure
    {\includegraphics[width=.49\textwidth]{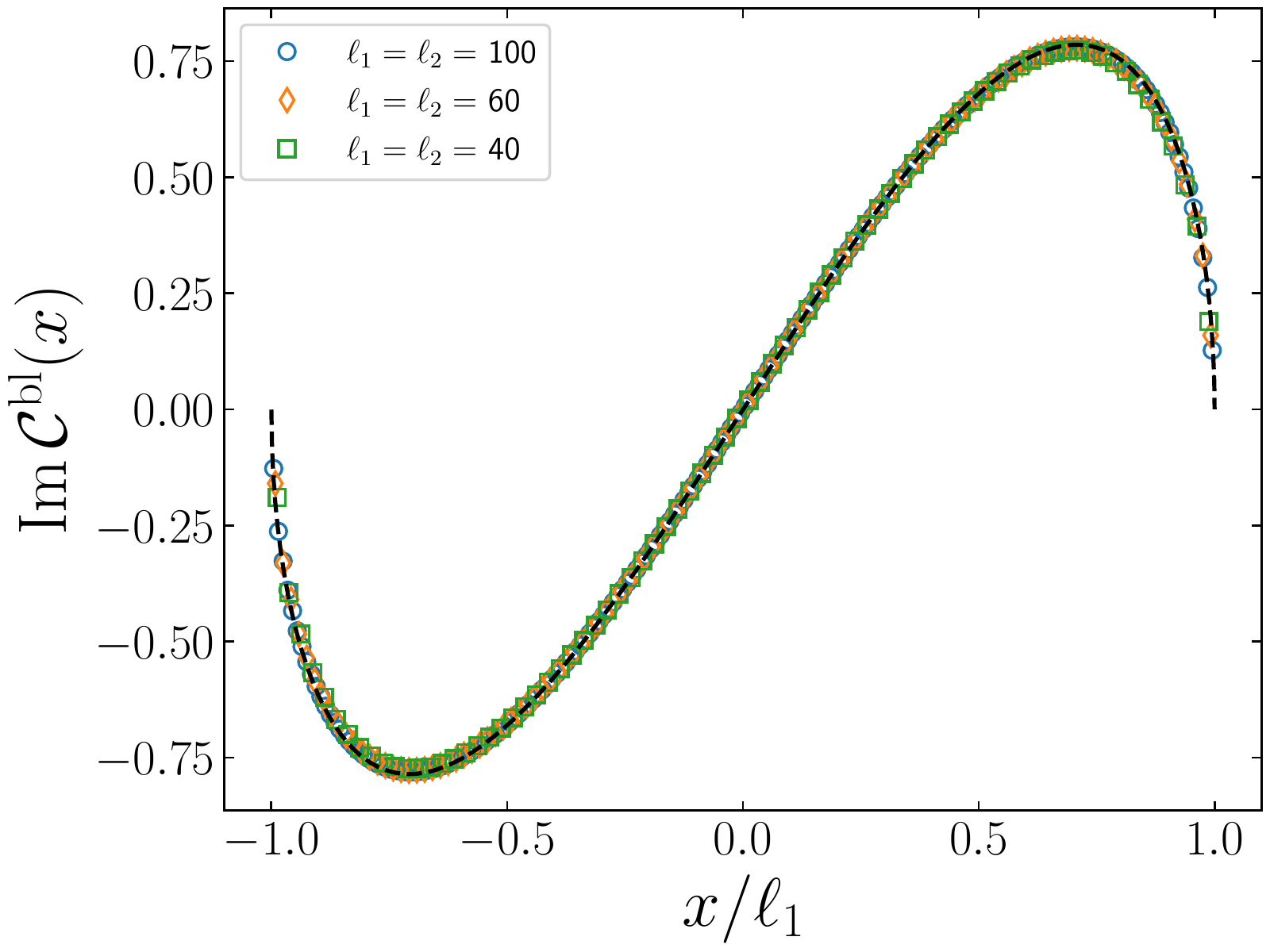}}
    \caption{Benchmark of the analytic prediction for the twisted negativity Hamiltonian $\widetilde{\mathcal{N}}_A$ for two adjacent intervals of equal length on the infinite line. The symbols correspond to the numerical
data obtained using Eqs.~\eqref{eq:limSinLoc} and \eqref{eq:limCosLoc} for the top left and right panel, respectively, and Eqs.~\eqref{eq:limSinBiloc} and \eqref{eq:limCosBiloc} for the bottom left and right. The solid lines are our analytical conjectures in Eqs.~\eqref{eq:betaloc_twisted} (top left) and \eqref{eq:betaloc_chem} (top right) for the local terms and in Eqs.~\eqref{eq:betabiloc_twisted} (bottom left) and \eqref{eq:betabiloc_chem} (bottom right) in the bi-local part.  }
    \label{fig:twisted_plane_intervalliuguali}
\end{figure}

\begin{figure}[t!]
    \centering
    {\includegraphics[width=.49\textwidth]{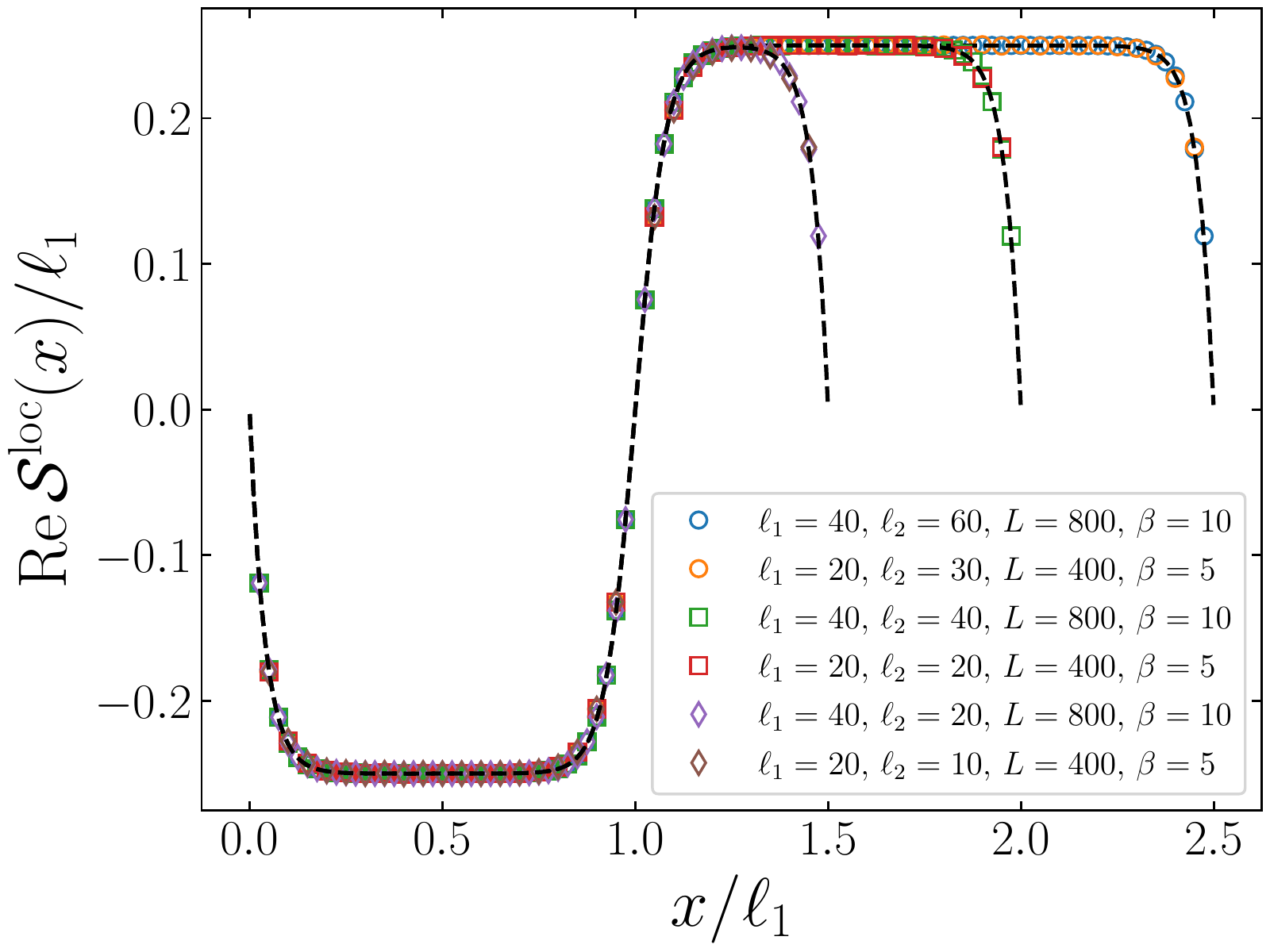}}
    \subfigure
    {\includegraphics[width=.49\textwidth]{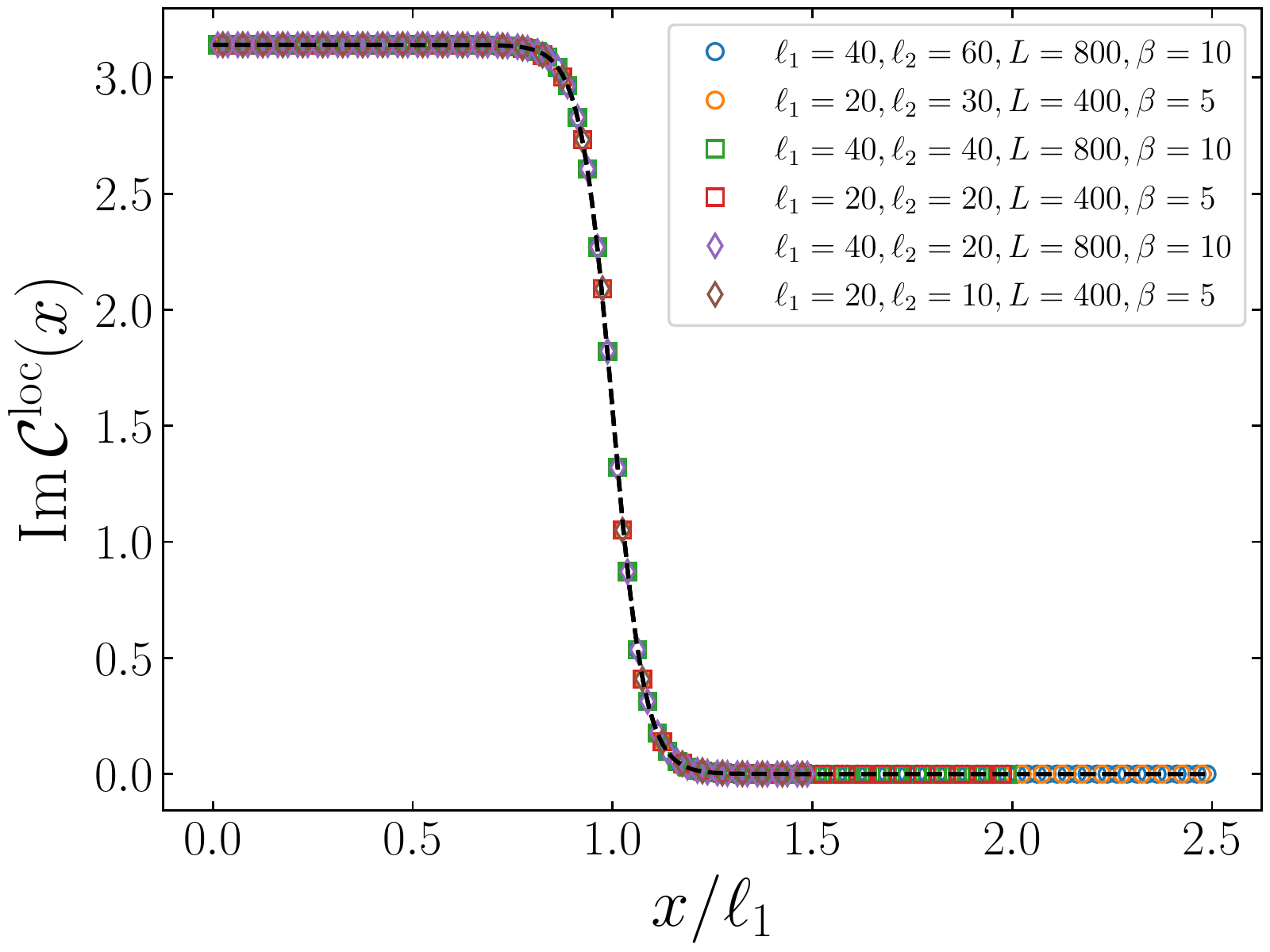}}
    \subfigure
    {\includegraphics[width=.49\textwidth]{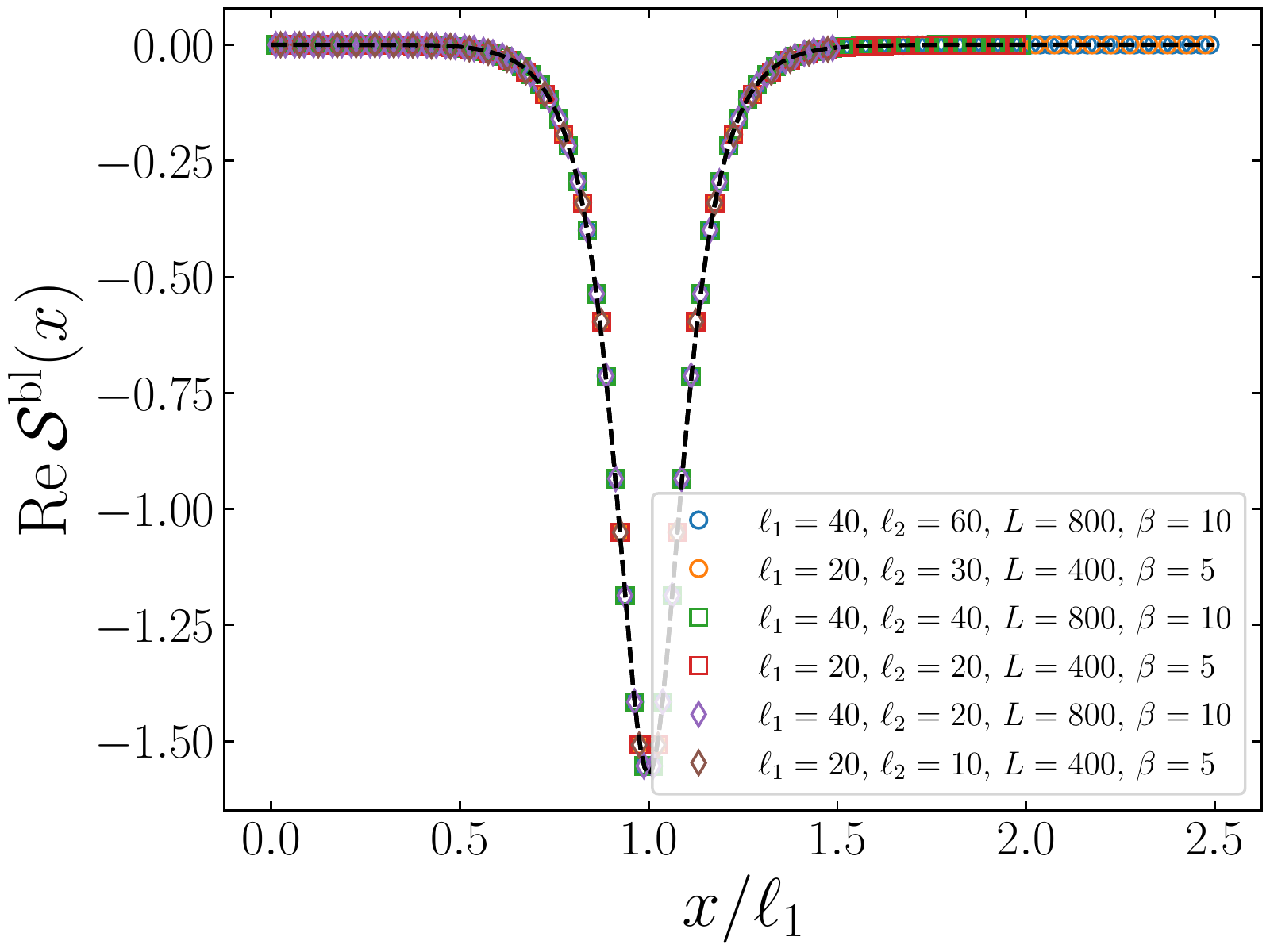}}
    \subfigure
    {\includegraphics[width=.49\textwidth]{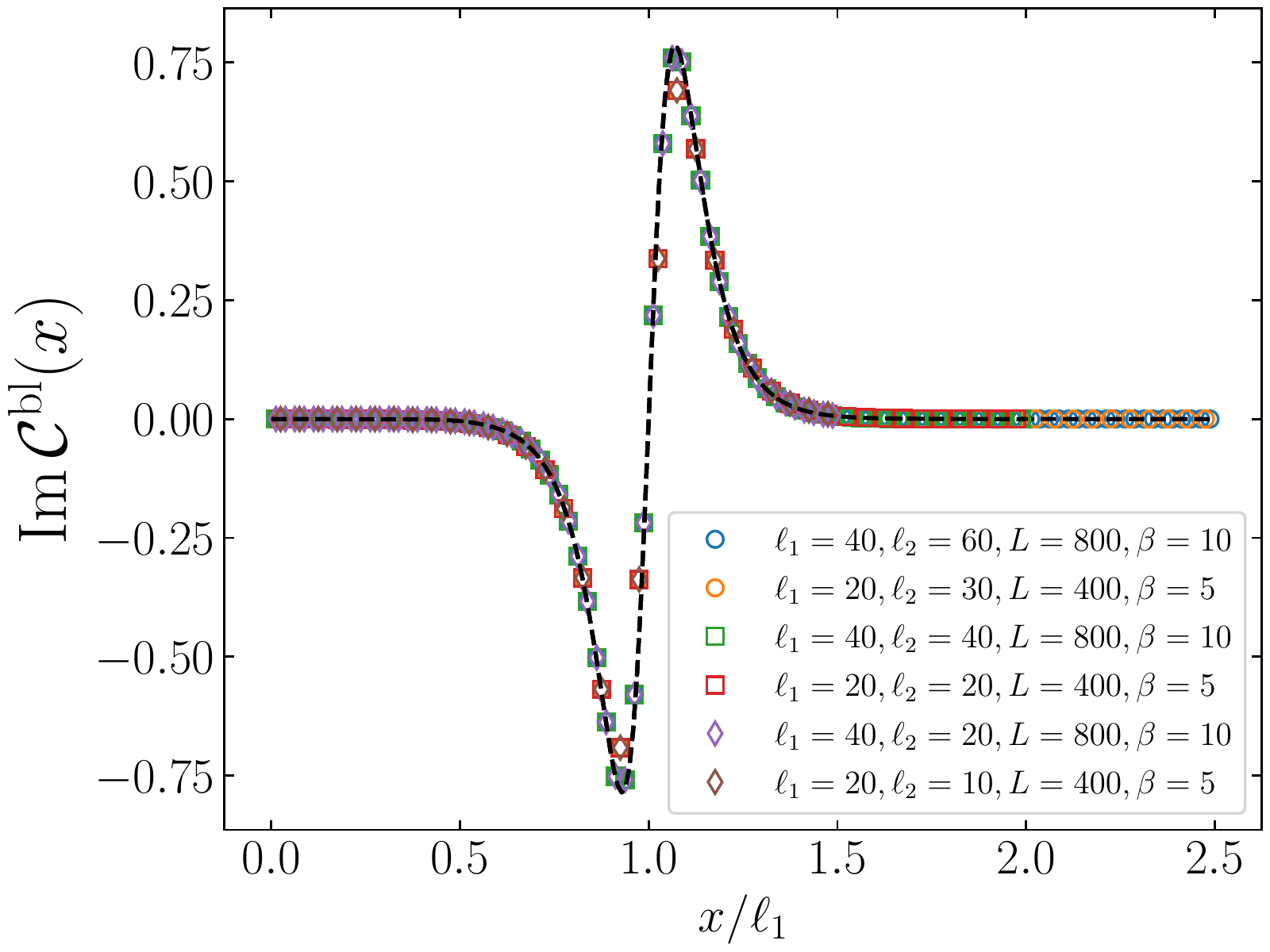}}
    \caption{Same benchmark of the analytic prediction for the twisted negativity Hamiltonian $\widetilde{\mathcal{N}}_A$ as in Fig.~\ref{fig:twisted_plane_intervalliuguali} but at finite temperature. The geometry we consider is $A=[1,\ell_1] \cup [\ell_1+1,\ell_1+\ell_2]$ for different values of the ratio $\ell_1/\ell_2 = 0.5, 1, 1.5$. The system size is fixed as $L/\ell_1=20$ and we rescale the inverse temperature $\beta$ such that $\beta/\ell_1=1/4$. The analytical predictions have been obtained by doing a conformal mapping from the plane to an infinite cylinder of circumference $\beta$ in Eq.~\eqref{eq:NHtildegeneric}.}   
    \label{fig:twisted_thermal}
\end{figure}

Let us first consider the twisted negativity Hamiltonian of the ground state on the infinite line, i.e, on the plane. The geometry under analysis $A = A_1 \cup A_2, A_1 = [-\ell, 0], A_2 = [0, \ell]$ consists of two adjacent intervals of identical length $\ell$, and we perform a partial transpose operation on the first one, $A_1$.

\noindent As we did for $\mathcal{N}_A$, the continuum limit of $\widetilde{\mathcal N}_A$ is identical to the one of the entanglement Hamiltonian described in Sec.~\ref{sec:EHlatticeLim}, since it depends only on the expansion of the lattice fermion in Eq.~\eqref{eq:linearFermion}.
However, differently from all the cases considered so far, we have numerically checked that even at half-filling $k_F = \frac{\pi}{2 s}$, the twisted negativity kernel $\widetilde{\eta}$ in Eq.~\eqref{eq:PeschelTwist} does not present a checkerboard structure.
For this reason, also the terms proportional to the sums $\mathcal{C}^\text{loc}(x)$ in Eq.~\eqref{eq:limCosLoc} and $\mathcal{C}^\text{bl}(x)$ in Eq.~\eqref{eq:limCosBiloc} have to be performed. This is the first difference with respect to Ref.~\cite{mvdc-22}, where the study of only the nearest neighbour terms prevented them from finding the operator $\mathcal{C}^\text{bl}(x)$. This also confirms that, in order to recover the continuum limit correctly, a careful treatment of the long-range hoppings has to be taken into account. 
Therefore, besides the energy density $T_{00}(x)$ in Eq.~\eqref{eq:energydensity} and the bi-local operator $T^\text{bl}(x,y)$ in Eq.~\eqref{eq:biloc}, the continuum limit will contain also an imaginary local chemical potential term proportional to the number operator $N(x)$ in Eq.~\eqref{eq:number} and a term proportional to the operator $j^\text{bl}(x, y)$ defined in Eq.~\eqref{eq:bilocalNumber}. Although we cannot derive the form of the weight functions of these operators explicitly, we provide a conjecture that very accurately matches numerical data on the lattice. Indeed, the twisted negativity Hamiltonian reads 
\begin{equation}\begin{split}\label{eq:NH-ftilde}
   \widetilde{\mathcal{N}}_A &= \int \dd x\, \beta^R_\text{loc}(x)\, T_{00}(x) + \ii \int \dd x\, \tilde{\mu}(x)\, N(x) \\
    &\hspace{2cm} + \int \dd x\, \tilde{\beta}_\text{bl}(x)\, T^\text{bl}(x, \tilde{x}^R_{p}) + \ii \int \dd x\, \tilde{\mu}_\text{bl}(x)\, j^\text{bl}(x, \tilde{x}^R)
\end{split}\end{equation}
where the inverse negativity temperature $\beta^R_\text{loc}(x)$ is given by 
\begin{equation}\label{eq:betaloc_twisted}
    \beta^R_\text{loc}(x)=\frac{1}{z^{R'}(x)},
\end{equation}
with $z^{R}(x)$ given in Eq.~\eqref{eq:CHZetaRev}, i.e. its functional form is the same as for $\mathcal{N}_A$. Despite being localised around the same conjugate point $\tilde{x}^R$ in Eq.~\eqref{eq:conjnegTzero} as the negativity Hamiltonian $\mathcal{N}_A$, the other weight functions are different and we report them here
\begin{equation}\label{eq:betaloc_chem}
    \tilde{\mu}(x) = \frac{1}{4} \left ( 1 - \frac{x}{\ell} \right ),
\end{equation}
\begin{equation}\label{eq:betabiloc_twisted}
    \tilde{\beta}_\text{bl}(x) = - \frac{1}{4} \sqrt{1 - \frac{x^2}{\ell^2}},
\end{equation}
\begin{equation}\label{eq:betabiloc_chem}
    \tilde{\mu}_\text{bl}(x) = \frac{1}{4} \frac{x}{\ell} \sqrt{1 - \frac{x^2}{\ell^2}}.
\end{equation}
The weight function of the number operator $N(x)$ is the same that was conjectured in \cite{mvdc-22}, while the weight functions for $T^\text{bl}(x, \tilde{x}^R)$ and $ j^\text{bl}(x, \tilde{x}^R)$ are different and, we stress again, in order to recover them, it is important to sum over all the elements of the kernel of the negativity Hamiltonian, as done in Eq.~\eqref{eq:limSinBiloc}. 
We also benchmark the analytical predictions from Eq.~\eqref{eq:NH-ftilde} in Fig.~\ref{fig:twisted_plane_intervalliuguali}. The good agreement between the lattice computations and Eq.~\eqref{eq:NH-ftilde} supports our conjecture.

\noindent Our prediction for equal intervals can be mapped into a geometry with adjacent intervals of different length using a M\"obius transformation. For $A = A_1 \cup A_2, A_1 = [a, b], A_2 = [b, c]$, the M\"obius transformation
\begin{equation}
    \xi(z) = \frac{(z-b) (c-a) \ell}{(z-b) (a-2 b+c)+2 (b-a) (c-b)}\, ,
\end{equation}
maps $A$ into the subsystem $\xi(A_1) = [-\ell, 0], \xi(A_2) = [0, \ell]$, for which Eq.~\eqref{eq:NH-ftilde} is valid.
In order to properly apply the transformation, we also need to consider the Jacobians arising from the transformation of the fields.
As discussed in Sec.~\ref{sec:EH}, to understand the transformations of the fields it is convenient to pass to Euclidean time and consider, for example, only the holomorphic component. Under this conformal mapping, the operators appearing in Eq.~\eqref{eq:NH-ftilde} transform as
\begin{equation}\label{eq:transf_operators}
\begin{split}
    N(z) &= \xi'(z)\, N(\xi(z)),\\
    T^\text{bl}(z, w) &= \xi'(z)^{1/2}\, \xi'(w)^{1/2}\, T^\text{bl}(\xi(z), \xi(w)),\\
    j^\text{bl}(z, w) &= \xi'(z)^{1/2}\, \xi'(w)^{1/2}\, j^\text{bl}(\xi(z), \xi(w)),
\end{split}
\end{equation}
where we have used that the fermions $\psi, \psi^\dagger$ transform as $\psi(z) = \left ( \frac{\partial \xi}{\partial z}\right )^{1/2} \psi(\xi(z))$ (and analogously for the anti-holomorphic part). Therefore, taking into account Eq.~\eqref{eq:transf_operators} and the Jacobians of the transformation, we obtain the following expression for the twisted negativity Hamiltonian of two intervals of arbitrary length on the infinite line
\begin{equation}\begin{split}\label{eq:NHtildegeneric}
    \widetilde{\mathcal{N}}_A &= \int \dd x\, \beta^R_\text{loc}(x)\, T_{00}(x) + \ii \int \dd x\, \tilde{\mu}(\xi(x))\, N(x) \\
    &\hspace{1.5cm} + \int \dd x\, \tilde{\beta}_\text{bl}(\xi(x))\sqrt{\frac{\xi(x)}{\xi(\tilde{x}^R)}} \, T^\text{bl}(x, \tilde{x}^R) + \ii \int \dd x\, \tilde{\mu}_\text{bl}(x)\sqrt{\frac{\xi(x)}{\xi(\tilde{x}^R)}}\, j^\text{bl}(x, \tilde{x}^R),
\end{split}\end{equation}
where $\beta^R_\text{loc}(x) = 1/\partial_x z^R(\xi(x))$ with $z^R$ given by Eq.~\eqref{eq:CHZetaRev}.
By doing another conformal mapping $\xi(x) \to  e^{\frac{2 \pi}{\beta} x}$ in Eq.~\eqref{eq:NHtildegeneric}, we can obtain the result for two intervals on the infinite line at finite temperature, as shown in Fig.~\ref{fig:mapping}. We report a check of our conjecture in Fig.~\ref{fig:twisted_thermal} for different ratios of the length $\ell_2/\ell_1 = 0.5, 1, 1.5$, with $\beta/\ell_1=1/4$ and $L=20\, \ell_1$. Beyond the good agreement, we observe that the weight function of the number operator $N(x)$ drastically changes: the linear behaviour in $x$ found at $T=0$ becomes a kink interpolating from $\pi$ for $x<\ell_1$ to $0$ for larger $x$.
To summarise, starting from our conjecture for the twisted negativity Hamiltonian for two intervals of equal size on the infinite line, through a series of conformal mappings, we are able to find an expression also for the finite temperature case, which is a concrete example of a global mixed state. 

\section{Conclusions}
\label{concl}
In this manuscript we have continued the analysis initiated in Refs.~\cite{mvdc-22,rmtc-23} about the study of the negativity Hamiltonian, i.e. an operatorial characterisation of entanglement in mixed states. The most relevant novelty introduced here is the study of the entanglement in thermal states, which represent genuine examples of globally mixed states. Until now, the only configurations considered were non complementary subsystems at zero temperature. Here, we studied the negativity Hamiltonian of free massless Dirac fermions on a torus, for an arbitrary set of
disjoint intervals at generic temperature. The structure of the negativity Hamiltonian exhibits a pattern similar to the entanglement Hamiltonian found in the same geometry in Ref.~\cite{fr-19,bpn-19}: in addition to a local term, each
point is non-locally coupled to an infinite but discrete set of other points. However, contrarily to what happens for the entanglement Hamiltonian, when the reversed and non-reversed subsystems have the same length, the bi-local solutions collapse on each other and we find only a finite number of bi-local terms, which couple each point only to another one in each other interval. \\
We also analysed in detail the negativity Hamiltonian in a bipartite configuration. If the state is pure, the relation between the entanglement entropy and the negativity is well-known \cite{cct-neg-1} and we retrieve it here. If the temperature is different from zero, a bipartite system is the first non-trivial example in which the negativity becomes essential to proper detect the quantum correlations. Also in this case, we found an infinite number of bi-local contributions, which reduce to one single bi-local solution only in the case of infinite  system size. Our analytical findings are supported by exact numerical computations in a free-fermion chain. \\
Another main result of this manuscript is the negativity Hamiltonian computed from the twisted partial transpose, cf. Eq~\eqref{eq:rhotilde}. 
Through a careful numerical analysis, we identified the local and bi-local operators and their weight functions for two intervals on the infinite line both at zero and finite temperature. It would be interesting to derive analytically the conjectured formulae for the twisted negativity Hamiltonian, e.g. using the methods discussed in Appendix \ref{app:resolvent}.\\
This study about the negativity Hamiltonian adds an important contribution to the operatorial characterisation of the mixed state entanglement, but there is still much work to do. For example, a challenging task is to exploit the mild non-locality of the negativity Hamiltonian together with the Hamiltonian reconstruction methods already used in \cite{dvz-17,kbevz-21,kszev-21} to reconstruct the negativity spectrum. 
Similarly, it is still an open problem to derive the conformal negativity spectrum \cite{rac-16b} from the negativity Hamiltonian, as instead done for the entanglement spectrum in Ref. \cite{act-17}. 
Another interesting direction is the study of the negativity Hamiltonian in higher dimensional systems, following what has been done for the entanglement Hamiltonian \cite{jt-21}.
Finally, it would be also interesting to study whether one can define a notion of modular flow \cite{haagbook, h-21,efrs-20} for the partial transpose reduced density matrix and its eventual connections with the negativity Hamiltonian.

\section*{Acknowledgements}
We thank Filiberto Ares, Pierre-Antoine Bernard, Michele Fossati and Francesco Gentile for useful discussions.
FR and PC acknowledge support from ERC under Consolidator grant number 771536 (NEMO). SM thanks support from Caltech Institute for Quantum Information and Matter and the Walter Burke Institute for Theoretical Physics at Caltech.

\appendix

\section{The resolvent method for the negativity Hamiltonian}\label{app:resolvent}

In \cite{ch-09}, the field-theoretical prediction for the kernel $H_A$ of the entanglement Hamiltonian on the plane in Eq.~\eqref{eq:CHEH} was obtained from the knowledge of the \emph{resolvent} of the Green function $C_A$ restricted to the subsystem (see also \cite{kvw-17, kvw-18, fr-19, bpn-19, bgpn-19, efrs-20}). In this appendix we show how to generalise the resolvent method of \cite{ch-09} to the negativity Hamiltonian in the case of multiple intervals on the plane, confirming the validity of the construction of Refs, \cite{mvdc-22, rmtc-23} that we have used in Sec.~\ref{sec:NH}.

\begin{figure}\centering
    \begin{tikzpicture}
        \draw[->]   (-7,0)  --  (7,0);
        \draw   (7,-.3) node    {$\text{Re }z$};
        \draw[->]   (0,-4)  --  (0,4);
        \draw   (.6,4) node    {$\text{Im }z$};
        \draw[dashed]   (0,0)   circle  (2);
        \draw   (1.4,2.)   node    {$\mathcal{C}$};
        \filldraw   (-1.6,.4)   circle  (.04);
        \filldraw   (-1.6,-.4)  circle  (.04);
        \filldraw   (-.3,1.2)    circle  (.04);
        \filldraw   (-.3,-1.2)   circle  (.04);
        \filldraw   (.3,1.2)    circle  (.04);
        \filldraw   (.3,-1.2)   circle  (.04);
        \filldraw   (1.6,.4)   circle  (.04);
        \filldraw   (1.6,-.4)  circle  (.04);
        \draw[decorate, decoration=snake]    (-7,0)  --  (-2,0);
    \end{tikzpicture}
    \caption{Representation of the contour of integration in Eqs.~\eqref{eq:derivationResolvent1}. The dashed line represents the contour $\mathcal{C}$ around the poles (small black dots), while the wavy line denotes the branch cut of $\log(1+z)$.}
    \label{fig:resolvent}
\end{figure}
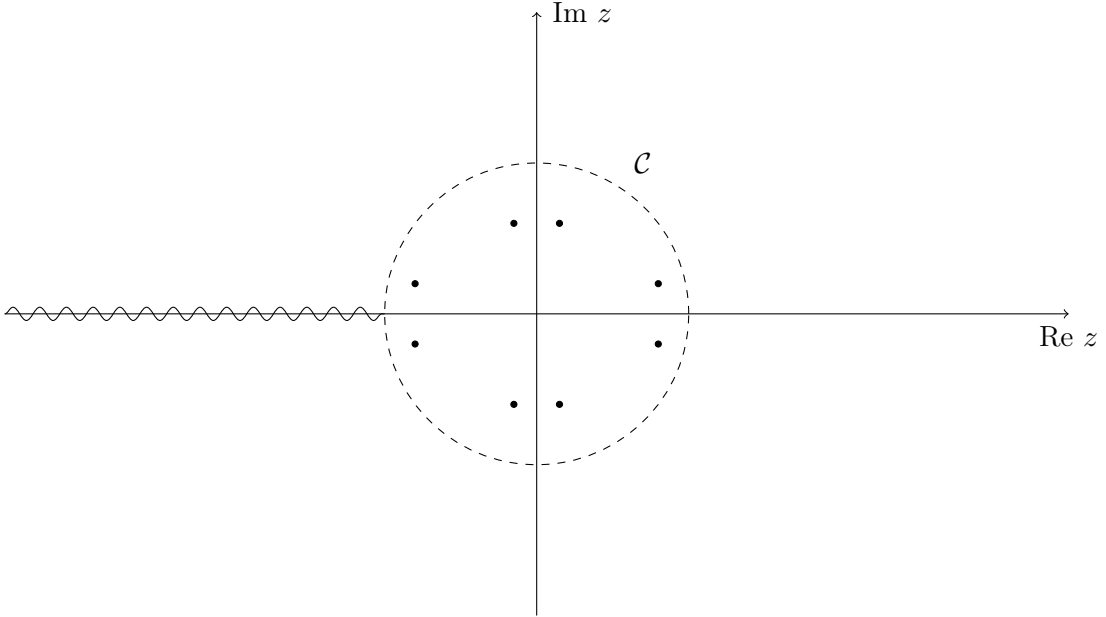

For our purposes, we recast the resolvent method in terms of the partially reversed covariance matrix $\Gamma_A^{R_1}$. To fix the ideas, we present the calculation for chiral fermions. Applying the partial reversal procedure in Eq.~\eqref{eq:defCovarianceRev} to the Green function we find
\begin{equation} \label{eq:CovarianceRevDiracFT}
    \Gamma_A^{R_1}(x,y) = - \frac{1}{\ii \pi} \mathcal{P} \frac{1}{x-y}\,\ii^{\Theta_1(x)}\, \ii^{\Theta_1(y)}\, ,
\end{equation}
where the function $\Theta_1(x)$, defined in Eq.~\eqref{eq:theta1}, is equal to $1$ only for $x \in A_1$, $0$ otherwise and $\mathcal{P}$ denotes Cauchy's principal value.
Recall from the main text that the kernel of the negativity Hamiltonian can be related via Peschel's formula in Eq.~\eqref{eq:PeschelNeg} to the reversed covariance matrix $\Gamma_A^{R_1}$. To apply Eq.~\eqref{eq:PeschelNeg} in the continuum theory, we first consider a single eigenvalue $g$ of $\Gamma_A^{R_1}$.
For the entanglement Hamiltonian, in \cite{ch-09} it was used the fact that the spectrum of the Green function is real and contained in $[0, 1]$.
In the case of the negativity Hamiltonian, we can use the knowledge that the eigenvalues of $\Gamma_A^{R_1}$ are contained in the unit complex disc $|g| < 1$ \cite{shapourian-19}, as depicted in Fig.~\ref{fig:resolvent}.
Then, Peschel's formula for the single eigenvalue can be rewritten using Cauchy's theorem as
\begin{equation} \label{eq:derivationResolvent1} \begin{split}
    \log\!\left[ 1 + g \right ] - \log\!\left[ 1 - g \right ] = \frac{1}{2\pi \ii} \oint_{\mathcal{C}} \dd z \left [\frac{1}{z-g} - \frac{1}{z+g} \right ] \log (1 + z)\, , 
\end{split}\end{equation}
where the branch cut of the logarithm is taken to go from $-\infty$ to $-1$. Since $|g|<1$, the contour of integration $\mathcal{C}$ in Eq.~\eqref{eq:derivationResolvent1} can always be taken to avoid the branch cut (see Fig.~\ref{fig:resolvent}) and therefore can be deformed continuously to integrate along the branch cut and on a small circle at infinity.
Denoting the upper and lower branches of the complex logarithm as $\log^+$ and $\log^-$ respectively, and using the fact that the difference of the two branches is $\log^+ - \log^- = 2\pi \ii$ we find for every eigenvalue $g$ of $\Gamma_A^{R_1}$
\begin{equation} \label{eq:derivationResolvent2} \begin{split}
    \log\!\left[ 1 + g \right ] - \log\!\left[ 1 - g \right ] &= \frac{1}{2\pi \ii} \int_{-\infty}^{-1} \dd z \left [\frac{1}{z-g} - \frac{1}{z+g} \right ] \left [\log^+(1+z) - \log^-(1+z)\right ] \\
    &= - \int_{1}^{\infty} \dd z \left [\frac{1}{g-z} + \frac{1}{g+z}\right ]\, .
\end{split}\end{equation}
Since this holds for every eigenvalue, it holds also for the operator, leading finally to the expression for the kernel of the negativity Hamiltonian
\begin{equation}\label{eq:relationResolventKernel}
    N_A(x, y) = \frac{1}{2\pi} \log\!\left [ \frac{\mathbb{I}_A + \Gamma_A^{R_1}}{\mathbb{I}_A - \Gamma_A^{R_1}} \right ] = - \frac{1}{2 \pi} \int_1^\infty \dd \zeta\, \big [ R(\zeta;x,y) + R(-\zeta;x,y) \big ]\,,
\end{equation}
where we have introduced the resolvent of the partially reversed covariance matrix of Eq.~\eqref{eq:CovarianceRevDiracFT}
\begin{equation}\label{eq:defResolventNeg}
    R(\zeta;x, y) = \frac{1}{\Gamma_A^{R_1} - \zeta \mathbb{I}_A} = \left [- \frac{1}{\ii \pi} \mathcal{P}\frac{\ii^{\Theta_1(x)}\, \ii^{\Theta_1(y)}}{x-y} - \zeta \delta(x-y) \right ]^{-1}.
\end{equation}
Note that throughout this appendix, $2\pi N_A(x,y)$ corresponds to the continuum limit of $\eta$ defined in Eq.~\eqref{eq:latticeNH}.

In order to find the explicit form of the resolvent in Eq.~\eqref{eq:defResolventNeg}, we need to solve a singular integral equation. By construction, the resolvent \eqref{eq:defResolventNeg} satisfies
\begin{equation}
    - \zeta R(\zeta;x, y) -\frac{\ii^{\Theta_1(y)}}{\ii \pi} \mathcal{P}\!\int \dd z\, \frac{  R(\zeta;x, z)\, \ii^{\Theta_1(z)}}{z-y}  = \delta(x-y)\,.
\end{equation}
Multiplying both sides by $(-)\ii^{\Theta_1(y)}$
\begin{equation}\label{eq:resolventEquation}
    \zeta R(\zeta;x, y)\, \ii^{\Theta_1(y)} + \frac{(-1)^{\Theta_1(y)}}{\ii \pi} \mathcal{P}\!\int \dd z\, \frac{R(\zeta;x, z)\, \ii^{\Theta_1(z)}}{z-y} = (-) \ii^{\Theta_1(y)}\, \delta(x-y)\, ,
\end{equation}
we see that Eq.~\eqref{eq:resolventEquation} has the form of a \emph{characteristic singular integral equation} \cite{musk-book}
\begin{equation}\label{eq:characteristicSingularEq}
    a(y) \phi(y) + \frac{b(y)}{\ii \pi} \mathcal{P}\!\int \dd z\, \frac{\phi(z)}{z-y} = f(y)\, ,
\end{equation}
in the unknown function $\phi(y) = \ii^{\Theta_1(y)} R(\zeta;x, y)$, with the identification $a = \zeta$, $b(y) = (-1)^{\Theta_1(y)}$ and $f(y) = (-) \ii^{\Theta_1(y)}\, \delta(x-y)$. 
Comparing Eq.~\eqref{eq:resolventEquation} with the analogous one for the entanglement Hamiltonian in \cite{ch-09}, we see that the most important difference is the presence of the function $b(y) = (-1)^{\Theta_1(y)}$ in front of the Cauchy kernel, which changes sign if the interval is reversed.
Now, we show that this function is precisely responsible for the inversion of the extrema $a_j, b_j$ of the partially reversed intervals in the expression of the negativity Hamiltonian.\\
To solve Eq.~\eqref{eq:resolventEquation}, we introduce \cite{musk-book}
\begin{equation}\label{eq:negG}
    G(y) = \frac{a(y) - b(y)}{a(y) + b(y)} = \frac{\zeta - (-1)^{\Theta_1(y)}}{\zeta + (-1)^{\Theta_1(y)}} =  \left[ \frac{\zeta - 1}{\zeta + 1} \right ]^{(-1)^{\Theta_1(y)}}\,, 
\end{equation}
and the solution of Eq.~\eqref{eq:characteristicSingularEq} will be expressed in terms of the function \cite{musk-book}
\begin{equation}\label{eq:negOmega}\begin{split}
    \omega(y) &= \sqrt{a^2(y)-b^2(y)}\,  \exp\!\left \{ \frac{1}{2\pi i} \mathcal{P}\!\int \dd z\, \frac{\log G(z)}{z-y} \right \}\\
    &= \sqrt{\zeta^2 - 1}\,  \exp\!\left \{ \frac{1}{2\pi i} \log \frac{\zeta-1}{\zeta+1}\, \mathcal{P}\!\int \dd z\, \frac{(-1)^{\Theta_1(z)}}{z-y} \right \}\\
    &= \sqrt{\zeta^2 - 1}\,  \exp\!\left \{- \frac{1}{2\pi i} \log \frac{\zeta-1}{\zeta+1}\, \left [ \sum_{i \in A_2} \log\left|\frac{y-a_i}{y-b_i} \right | - \sum_{j \in A_1} \log\left|\frac{y-a_j}{y-b_j} \right | \right ] \right \}\\
    &= \sqrt{\zeta^2 - 1}\,  \exp\!\left \{- \frac{z^R(y)}{2\pi i} \log \frac{\zeta-1}{\zeta+1} \right \}\, ,
\end{split}\end{equation}
where $z^R$ is precisely the function in Eq.~\eqref{eq:CHZetaRev}, obtained by exchanging the extrema $a_j, b_j$ of the reversed intervals in the expression of Eq.~\eqref{eq:CHZeta}. As we can see, the factor $(-1)^{\Theta_1(z)}$ in the second row of Eq.~\eqref{eq:negOmega} is responsible for the exchange of the extrema in Eq.~\eqref{eq:CHZetaRev}.\\
The general solution of the characteristic singular equation \eqref{eq:characteristicSingularEq} is  \cite{musk-book}
\begin{equation}\label{eq:generalSolutionSingEq}
    \phi(y) = \frac{1}{a^2(y) - b^2(y)} \left [ a(y) f(y) - \frac{b(y)\omega(y)}{\ii \pi}\, \mathcal{P}\! \int \dd z\, \frac{f(z)}{(z-y)\,  \omega(z)} \right ]\, ,
\end{equation}
which specialised to our Eq.~\eqref{eq:resolventEquation} gives
\begin{equation}\label{eq:negResolvent}\begin{split}
    &R(\zeta;x, y) = \frac{(-\ii)^{\Theta_1(y)}}{\zeta^2-1} \left [ - \zeta\, \delta(x-y)\, \ii^{\Theta_1(y)} - \frac{(-1)^{\Theta_1(y)} \omega(y)}{\ii \pi}\, \mathcal{P}\! \int \dd z\, \frac{\delta(x-z)\, (-)\ii^{\Theta_1(z)}}{(z-y)\,  \omega(z)} \right ] \\
    & = \frac{1}{1-\zeta^2} \left [ \zeta\, \delta(x-y) - \frac{1 }{\ii \pi}\,\frac{\omega(y)}{\omega(x)}\, \mathcal{P} \frac{1}{(x-y)}  \, \ii^{\Theta_1(x)} \ii^{\Theta_1(y)} \right ]\\
    &= \frac{1}{1-\zeta^2} \left [ \zeta\, \delta(x-y) - \frac{1 }{\ii \pi}\, \mathcal{P} \frac{1}{(x-y)}\, \ii^{\Theta_1(x)} \ii^{\Theta_1(y)}\, \exp\!\left \{ \frac{1}{2\pi i} \log \frac{\zeta-1}{\zeta+1} \left [z^R(x) -z^R(y)\right ]\right \} \right ].
\end{split}\end{equation}
If we compare the resolvent for the negativity Hamiltonian on the plane in Eq.~\eqref{eq:resolventEquation} with the one obtained in the context of the entanglement Hamiltonian in \cite{ch-09}, we see that the main differences are the presence of the imaginary factors $\ii^{\Theta_1(x)} \ii^{\Theta_1(y)}$ and the substitution of the function \eqref{eq:CHZeta} with the one in Eq.~\eqref{eq:CHZetaRev} where the extrema of the reversed intervals are exchanged.

With the knowledge of the resolvent in Eq.~\eqref{eq:negResolvent}, we can finally obtain the kernel of the negativity Hamiltonian by substituting it in Eq.~\eqref{eq:relationResolventKernel}. Changing variables as $s = \frac{1}{2\pi} \log\frac{\zeta-1}{\zeta+1}$ we find, formally
\begin{equation}\label{eq:formalKernel}
    N_A(x, y) = - \frac{\ii}{2\pi} \int_{-\infty}^{+\infty} \dd s\, \frac{ e^{-\ii\, s\, \left [ z^R(x)-z^R(y) \right ]}}{x-y}\, \ii^{\Theta_1(x)} \ii^{\Theta_1(y)} = - \ii\, \frac{\delta\!\left (z^R(x)-z^R(y)\right )}{x-y}\, \ii^{\Theta_1(x)} \ii^{\Theta_1(y)}\, .
\end{equation}
In the formal expression of the kernel $N_A (x, y)$, the Dirac delta is calculated in the solution of the equation $z^R(x) = z^R(y)$.
However, when dealing with the trivial solution $y=x$ which corresponds to the local part of the kernel, Eq.~\eqref{eq:formalKernel} is proportional to the product of distributions $\delta(x-y)/(x-y)$ with coincident singular support. As discussed in \cite{ch-09}, such an expression is ambiguous and it is necessary to regularise it.
Following \cite{ch-09}, the product is the distribution $T$ that satisfies the algebraic distributional equation $(x-y) T = \delta(x-y)$, whose solution is $T = -\partial_x \delta(x-y) + \kappa\, \delta(x-y)$, where $\kappa$ is an arbitrary constant which is fixed by requiring that the local part of $N_A$ is hermitian \cite{ch-09}.
For this reason, we find it more convenient to explicitly antisymmetrise the kernel in the variables $x$ and $y$, which cancels the $\kappa\, \delta(x-y)$ contribution.\\ 
We also use the fact that the function $z^R$ in Eq.~\eqref{eq:CHZetaRev} has the property that it is monotonically decreasing in the reversed intervals $A_1$ and monotonically increasing outside, which implies for its derivative
\begin{equation}\label{eq:zprimoR}
    \left | \left (z^{R}(x) \right )' \right | = (-1)^{\Theta_1(x)} \left (z^{R}(x) \right )' \equiv \frac{(-1)^{\Theta_1(x)}}{\beta_\text{loc}^R(x)}\, .
\end{equation}
Then, by replacing Eq.~\eqref{eq:zprimoR} in the term of Eq.~\eqref{eq:formalKernel} corresponding to the trivial solution $y = x$ we find
\begin{equation}\begin{split}
    N_A^\text{loc}(x, y) &= -\frac{\ii}{2} \left [ \frac{(-1)^{\Theta_1(y)}}{\left | \left ( z^R(y) \right )' \right |} \frac{\delta(x-y)}{x-y} - \frac{(-1)^{\Theta_1(x)}}{\left | \left (z^R(x) \right )' \right |} \frac{\delta(y-x)}{y-x} \right ] \\
    &= \frac{\ii}{2} \Big [\beta_\text{loc}^R(y)\, \partial_x \delta(y-x) - \beta_\text{loc}^R(x)\,\partial_y \delta(x-y) \Big ]\, ,
\end{split}\end{equation}
which, when plugged in the expression for the negativity Hamiltonian reproduces the local part
\begin{equation}\begin{split}
    \mathcal{N}_A^\text{loc} &= \int_A \dd x \int_A \dd y\, \psi^\dagger(x)\, N_A^\text{loc}(x,y)\, \psi(y)\\
    &= \int_A \dd x\, \beta_\text{loc}^R(x) \left [ - \frac{\ii}{2}:\!\!\left ( \partial_x \psi^\dagger(x) \psi(x) - \psi^\dagger(x) \partial_x\psi(x) \right )\!\! : \right ]\, .
\end{split}\end{equation}

The $n-1$ non-trivial solutions $y=\tilde{x}^R_p$ of the equation $z^R(y) = z^R(x)$ instead give rise to the bi-local terms. Explicitly anti-symmetrising the expression in the variables $x, y$ gives 
\begin{equation}\begin{split}
    N_A^\text{bl}(x,y) &= -\frac{\ii}{2}\frac{1}{x - y} \sum_{p=1}^{n-1} \left [ \frac{\delta(y - \tilde{x}^R_p)}{\left | \left (z^R(\tilde{x}^R_p) \right )' \right |} + \frac{\delta(x - \tilde{y}^R_p)}{\left | \left (z^R(\tilde{y}^R_p)  \right )'\right |} \right ] \ii^{\Theta_1(x)} \ii^{\Theta_1(y)} \\
    &=  -\frac{\ii}{2} \sum_{p=1}^{n-1} \Bigg [ \ii^{\Theta_1(x)} (-\ii)^{\Theta_1(\tilde{x}^R_p)} \frac{\beta_\text{loc}^R(\tilde{x}^R_p)}{x-\tilde{x}^R_p}\, \delta(y - \tilde{x}^R_p) \\
    &\hspace{5cm} - \ii^{\Theta_1(y)} (-\ii)^{\Theta_1(\tilde{y}^R_p)} \frac{\beta_\text{loc}^R(\tilde{y}^R_p)}{y-\tilde{y}^R_p}\, \delta(x - \tilde{y}^R_p) \Bigg ] \, ,
\end{split}\end{equation}
leading to the bi-local part of the negativity Hamiltonian
\begin{equation}\begin{split}
    \mathcal{N}_A^\text{bl} &= \int_A \dd x \int_A \dd y\, \psi^\dagger(x)\, N_A^\text{nl}(x,y)\, \psi(y)\\
    &= \sum_{p=1}^{n-1} \int \dd x\, \frac{\beta_\text{loc}^R(\tilde{x}^R_p)}{x-\tilde{x}^R_p}\, \ii^{\Theta_1(x)} (-\ii)^{\Theta_1(\tilde{x}^R_p)}  \, \left [- \frac{\ii}{2}:\!\!\left (\psi^\dagger(x) \psi(\tilde{x}^R_p) - \psi^\dagger(\tilde{x}^R_p) \psi(x) \right )\!\! : \right ]\, .
\end{split}\end{equation}
This resolvent procedure could be analogously extended to the case on the cylinder or on the torus considered in Sec.~\ref{sec:NH}. Therefore, we can formally justify not only the construction introduced in \cite{mvdc-22} to compute the negativity Hamiltonian on the plane, but also at finite temperature or size, proving the correctness of the results found in this manuscript.

\section{Mathematical identities}\label{app:tools}
We report here the main mathematical tools we have used throughout the manuscript. 
The Weierstrass zeta function is defined by \cite{whittaker}
\begin{equation}
    \zeta(x)=\frac{1}{z}+\sum_{\lambda \neq 0}\left(\frac{1}{z+\lambda}-\frac{1}{\lambda}+\frac{z}{\lambda^2}\right).
\end{equation}
It enters in the class of elliptic functions and it is quasiperiodic, i.e. it satisfies
\begin{equation}
    \zeta(x+P_i)=\zeta(x)+2\zeta(P_i/2),
\end{equation}
where $P_i$, $i=1,2$, are the fundamental periods. In the case of interest for us, $P_1=L$ and $P_2=i\beta$. 
In order to prove the equality in Eq.~\eqref{eq:TorusZetabeta}, we have used the following representation of the Weierstrass zeta function through Jacobi functions
\begin{equation}
\zeta(x)=\frac{2x\pi}{L\beta}-\ii\frac{2x \zeta(\ii\beta/2)}{\beta}+\frac{\pi}{L}\frac{\vartheta'_1\!\left ( \frac{\pi}{L} x \big | q\right )}{\vartheta_1\!\left ( \frac{\pi}{L} x \big | q\right )}.
\end{equation}
For completeness, we also report here the definition of the Weierstrass sigma function used in Eq.~\eqref{eq:TorusZeta} 
\begin{equation}
    \sigma(x)=x\prod_{\lambda \neq 0}\left[\left(1+\frac{x}{\lambda}\right)e^{-\frac{x}{\lambda}+\frac{1}{2}(\frac{x}{\lambda})^2}\right].
\end{equation}
Also the equality in Eq.~\eqref{eq:TorusZeta} can be proven by using the following property
\begin{equation}
    \sigma(x)=\frac{L}{\pi}e^{\zeta(L/2)\frac{x^2}{L}}\frac{\vartheta_1\!\left ( \frac{\pi}{L} x \big | q\right )}{\vartheta'_1\!\left ( 0 \big | q\right )}.
\end{equation}
We also define the Jacobi theta functions $\theta_1(z|u)$ \cite{whittaker}
\begin{equation}
\begin{split}\label{Theta3Def}
\theta_1(u|q)=
&=
\sum_{k=-\infty}^{\infty} 
(-1)^{k-1/2}q^{\left(k+\frac{1}{2}\right)^2}\, 
 e^{ \ii(2k+1) u},
 \end{split}
\end{equation}
which satisfies the following asymptotic behaviour in the limit $\tau  \to 0$
\begin{equation}\label{eq:thetaLimit}\begin{split}
    \vartheta_1\!\left ( u \big | q \right ) \sim \frac{2\, \ii}{\sqrt{-\ii \tau}}\, e^{-\ii \left (\pi^2 + 4 u^2 \right )/4\pi\tau}\sin\!\left ( \frac{u}{\tau} \right ) 
    = 2 \left ( \frac{L}{\beta} \right )^{1/2} e^{-\frac{L}{4 \pi \beta}\left ( \pi^2 + 4 u^2  \right ) } \sinh\!\left ( \frac{u L}{\beta} \right ).
\end{split}\end{equation}
This expansion turns out to be useful to recover Eq.~\eqref{eq:limitEH}.
Finally, we remind here the definition for the $q-$digamma function \cite{qdigamma} used in Eqs.~\eqref{eq:ramond} and \eqref{eq:NS}
\begin{equation}
\begin{split}
    \psi_q(x)=-\log(1-q)+\log q\sum_{n=1}^{\infty}\frac{q^{nx}}{1-q^n}, \quad 0<q<1\\
    \psi_q(x)=-\log(q-1)+\log q\left(x-\frac{1}{2}-\sum_{n=1}^{\infty}\frac{q^{-nx}}{1-q^{-n}}\right). \quad q>1
    \end{split}
\end{equation}

\end{document}